\documentclass[12pt]{article}
\usepackage{amsmath,epsf}
\usepackage[T1]{fontenc}
\usepackage[latin1]{inputenc}
\usepackage{epsf}
\usepackage{amsfonts,amssymb}
\usepackage[usenames,dvipsnames]{color}
\newcommand{\al}{\alpha}

\newcommand{\de}{\delta}
\newcommand{\De}{\Delta}
\newcommand{\vep}{\varepsilon}

\newcommand{\ka}{\kappa}
\newcommand{\la}{\lambda}
\newcommand{\si}{\sigma}

\newcommand{\om}{\omega}

\newcommand{\vp}{\varphi}
\newcommand{\La}{\Lambda}

\newcommand{\Lao}{\Lambda_0}

\newcommand{\xv}{\vec x}

\newcommand{\cH}{{\cal H}}

\newcommand{\cL}{{\cal L}}

\newcommand{\pa}{\partial}

\newcommand{\ti}[1]{\tilde{#1}}

\newcommand{\qed}{\hfill \rule {1ex}{1ex}\\ }

\newcommand{\eq}{\begin{equation}}
\newcommand{\eqe}{\end{equation}}
\newcommand{\nom}{|\,\omega(x_1)|}
\newcounter{saveeqn}

\begin{document}
\message{reelletc.tex (Version 1.0): Befehle zur Darstellung |R  |N, Aufruf=
z.B. \string\bbbr}
%
%
%  Sonderzeichen
\message{reelletc.tex (Version 1.0): Befehle zur Darstellung |R  |N, Aufruf=
z.B. \string\bbbr}
\font \smallescriptscriptfont = cmr5
\font \smallescriptfont       = cmr5 at 7pt
\font \smalletextfont         = cmr5 at 10pt
\font \tensans                = cmss10
\font \fivesans               = cmss10 at 5pt
\font \sixsans                = cmss10 at 6pt
\font \sevensans              = cmss10 at 7pt
\font \ninesans               = cmss10 at 9pt
\newfam\sansfam
\textfont\sansfam=\tensans\scriptfont\sansfam=\sevensans
\scriptscriptfont\sansfam=\fivesans
\def\sans{\fam\sansfam\tensans}
%----------------------------------------------------------
\def\bbbr{{\rm I\!R}} %reelle Zahlen
\def\bbbn{{\rm I\!N}} %natuerliche Zahlen
\def\bbbE{{\rm I\!E}} %Einheitsmatrix by I. Zoller
\def\bbbm{{\rm I\!M}}
\def\bbbh{{\rm I\!H}}
\def\bbbk{{\rm I\!K}}
\def\bbbd{{\rm I\!D}}
\def\bbbp{{\rm I\!P}}
\def\bbbone{{\mathchoice {\rm 1\mskip-4mu l} {\rm 1\mskip-4mu l}
{\rm 1\mskip-4.5mu l} {\rm 1\mskip-5mu l}}}
\def\bbbc{{\mathchoice {\setbox0=\hbox{$\displaystyle\rm C$}\hbox{\hbox
to0pt{\kern0.4\wd0\vrule height0.9\ht0\hss}\box0}}
{\setbox0=\hbox{$\textstyle\rm C$}\hbox{\hbox
to0pt{\kern0.4\wd0\vrule height0.9\ht0\hss}\box0}}
{\setbox0=\hbox{$\scriptstyle\rm C$}\hbox{\hbox
to0pt{\kern0.4\wd0\vrule height0.9\ht0\hss}\box0}}
{\setbox0=\hbox{$\scriptscriptstyle\rm C$}\hbox{\hbox
to0pt{\kern0.4\wd0\vrule height0.9\ht0\hss}\box0}}}}

\def\bbbe{{\mathchoice {\setbox0=\hbox{\smalletextfont e}\hbox{\raise
0.1\ht0\hbox to0pt{\kern0.4\wd0\vrule width0.3pt height0.7\ht0\hss}\box0}}
{\setbox0=\hbox{\smalletextfont e}\hbox{\raise
0.1\ht0\hbox to0pt{\kern0.4\wd0\vrule width0.3pt height0.7\ht0\hss}\box0}}
{\setbox0=\hbox{\smallescriptfont e}\hbox{\raise
0.1\ht0\hbox to0pt{\kern0.5\wd0\vrule width0.2pt height0.7\ht0\hss}\box0}}
{\setbox0=\hbox{\smallescriptscriptfont e}\hbox{\raise
0.1\ht0\hbox to0pt{\kern0.4\wd0\vrule width0.2pt height0.7\ht0\hss}\box0}}}}

\def\bbbq{{\mathchoice {\setbox0=\hbox{$\displaystyle\rm Q$}\hbox{\raise
0.15\ht0\hbox to0pt{\kern0.4\wd0\vrule height0.8\ht0\hss}\box0}}
{\setbox0=\hbox{$\textstyle\rm Q$}\hbox{\raise
0.15\ht0\hbox to0pt{\kern0.4\wd0\vrule height0.8\ht0\hss}\box0}}
{\setbox0=\hbox{$\scriptstyle\rm Q$}\hbox{\raise
0.15\ht0\hbox to0pt{\kern0.4\wd0\vrule height0.7\ht0\hss}\box0}}
{\setbox0=\hbox{$\scriptscriptstyle\rm Q$}\hbox{\raise
0.15\ht0\hbox to0pt{\kern0.4\wd0\vrule height0.7\ht0\hss}\box0}}}}

\def\bbbt{{\mathchoice {\setbox0=\hbox{$\displaystyle\rm
T$}\hbox{\hbox to0pt{\kern0.3\wd0\vrule height0.9\ht0\hss}\box0}}
{\setbox0=\hbox{$\textstyle\rm T$}\hbox{\hbox
to0pt{\kern0.3\wd0\vrule height0.9\ht0\hss}\box0}}
{\setbox0=\hbox{$\scriptstyle\rm T$}\hbox{\hbox
to0pt{\kern0.3\wd0\vrule height0.9\ht0\hss}\box0}}
{\setbox0=\hbox{$\scriptscriptstyle\rm T$}\hbox{\hbox
to0pt{\kern0.3\wd0\vrule height0.9\ht0\hss}\box0}}}}

\def\bbbs{{\mathchoice
{\setbox0=\hbox{$\displaystyle     \rm S$}\hbox{\raise0.5\ht0\hbox
to0pt{\kern0.35\wd0\vrule height0.45\ht0\hss}\hbox
to0pt{\kern0.55\wd0\vrule height0.5\ht0\hss}\box0}}
{\setbox0=\hbox{$\textstyle        \rm S$}\hbox{\raise0.5\ht0\hbox
to0pt{\kern0.35\wd0\vrule height0.45\ht0\hss}\hbox
to0pt{\kern0.55\wd0\vrule height0.5\ht0\hss}\box0}}
{\setbox0=\hbox{$\scriptstyle      \rm S$}\hbox{\raise0.5\ht0\hbox
to0pt{\kern0.35\wd0\vrule height0.45\ht0\hss}\raise0.05\ht0\hbox
to0pt{\kern0.5\wd0\vrule height0.45\ht0\hss}\box0}}
{\setbox0=\hbox{$\scriptscriptstyle\rm S$}\hbox{\raise0.5\ht0\hbox
to0pt{\kern0.4\wd0\vrule height0.45\ht0\hss}\raise0.05\ht0\hbox
to0pt{\kern0.55\wd0\vrule height0.45\ht0\hss}\box0}}}}

\def\bbbz{{\mathchoice {\hbox{$\sans\textstyle Z\kern-0.4em Z$}}
{\hbox{$\sans\textstyle Z\kern-0.4em Z$}}
{\hbox{$\sans\scriptstyle Z\kern-0.3em Z$}}
{\hbox{$\sans\scriptscriptstyle Z\kern-0.2em Z$}}}}
\noindent

\title{ Renormalization  Proof for  Massive $\vp_4^4$ Theory 
on Riemannian Manifolds}

\author{Christoph Kopper\footnote{\ kopper@cpht.polytechnique.fr} \\
Centre de Physique Th{\'e}orique, CNRS, UMR 7644\\
Ecole Polytechnique\\
F-91128 Palaiseau, France
\and 
Volkhard F. M{\"u}ller\footnote{\ vfm@physik.uni-kl.de}\\
Fachbereich Physik, Technische Universit{\"a}t Kaiserslautern\\
D-67653 Kaiserslautern, Germany
} 

\date{}

\maketitle

\begin{abstract}
In this paper we present an inductive renormalizability proof
for massive $\vp_4^4$ theory on Riemannian manifolds, 
based on the Wegner-Wilson flow equations of the Wilson 
renormalization group, adapted to perturbation theory. 
The proof goes in hand with bounds on the perturbative Schwinger functions
which imply tree decay between their position arguments.
An essential prerequisite are precise bounds on the short and long distance
behaviour of the heat kernel on the manifold. With the aid of a
regularity assumption (often taken for granted) we also show, that 
for suitable renormalization conditions
the bare action takes the minimal form, that is to say, there appear the
same counter terms as in flat space, apart from a logarithmically
divergent one which is proportional to the scalar curvature.

\end{abstract}
\newpage

\noindent
\section{Introduction }

Among the different schemes deviced to prove the perturbative
renormalizability of a local quantum field theory, the one based
on the Wegner-Wilson differential flow equations of the Wilson 
renormalization group shows a distinctive
characteristic: it circumvents completely the combinatoric
complexity of generating Feynman diagrams and the subsequent
cumbersome analysis of Feynman integrals with in general 
overlapping divergences. Initiated by Polchinski [Pol],
this approach to renormalization has now been adapted
 to a wide variety of physically interesting instances.
Partial reviews of the rigorous work which started from [KKS] 
may be found in [Kop1], [Sal], [Kop2], [M\"u].
It is tempting to extend the approach via Wilson's 
flow equation further to prove the perturbative renormalizability
of a quantum field theory defined on curved spacetime.
 There is a caveat, however.
Using functional integration, one actually deals with a quantum
field theory defined on a "Euclidean section" of curved 
spacetime, i.e. on a Riemannian manifold. In contrast to 
flat space there is no Wick rotation of Lorentzian curved
spacetime, in general. Nevertheless, beyond static spacetimes,
 on particular nonstatic ones the analytic continuation of a
quantum field theory to a corresponding Euclidean formulation
has been rigorously shown recently:
 Bros, Epstein and Moschella [BEM] considered
a quantum field theory on the anti-de Sitter (AdS) spacetime within a
 Wightman-type approach. As a consequence of certain spectral assumptions
 they show that the $n$-point correlation functions admit an analytic
 continuation to tuboidal domains (of $n$ copies) of the complexified
 covering space of the AdS spacetime. This continuation includes the
 Euclidean AdS spacetime and satisfies there Osterwalder-Schrader positivity.
  Euclidean AdS spacetime is a Riemannian manifold with constant negative
  curvature.\footnote{It is called 'hyperbolic space' 
   in the mathematical literature.} Moreover, Birke and Fr\"ohlich
 [BiFr], establishing in an algebraic approach Wick rotation
of quantum field theories at finite temperature, also presented
 a reconstruction of quantum field theories on specific
 curved spacetimes from corresponding imaginary-time
 formulations, using group-theoretical techniques.\\
Our work starts straight considering a Riemannian manifold as
 given "spacetime". This manifold is assumed to be geodesically
complete and to have all its sectional curvatures confined by
a negative lower and a positive upper bound.
 We then study perturbative renormalizability of massive
 $\vp_4^4\,$-theory defined on such a manifold  
by analy\-sing the generating functional
$L^{\La,\Lao}$ of connected (free propagator) amputated 
Schwinger functions (CAS).
From the physical point of view it seems justified to restrict to
this class of manifolds, since
in situations where curvature becomes
large or where singularities appear the treatment of gravity as a
classical background effect becomes questionable anyway. 
As there is no translation symmetry, the CAS and thus the
system of flow equations relating them have to be dealt with
in position space. Establishing bounds involving these CAS
we heavily rely on global lower and upper bounds for the heat
kernel on the manifold, found in the mathematical literature.

Around the beginning of the eighties a considerable amount
 of work was carried out to formulate   quantum field theory
perturbatively  on curved spacetime. 
Based on the intuition that ultraviolet
divergences involve arbitrarily short wavelenghts, an 
approximating local momentum space representation of the Feynman
 propagator in curved spacetime was developed in
[Bir],[BPP],[BuPn] for the $\phi^4$-theory and generalized
in [BuPr],[Bun1]. Combined with dimensional regularization,
the euclideanized  $\phi^4$-theory was then shown to be
renormalizable with local counterterms in one- and two-loop
order. Furthermore, choosing the same general approach,
Bunch [Bun2] has demonstrated the BPHZ renormalization of the
 $\phi^4$-theory on euclideanized curved spacetime,
by taking into account the power counting singular contributions
in the asymptotic expansion of the propagator around its
euclidean form.
A different kind of generally applicable dimensional
regularization scheme has been given by L\"uscher [L\"u],
who applies it to the $\phi^4$-theory on an arbitrary compact
four-dimensional manifold with positive metric and to the
Yang-Mills gauge theory on $ S^4$. He also shows the
renormalizability of the $\phi^4$-theory by local
counterterms at the one- and two-loop levels.
Further references on work before 1982 can be found in the monograph [BiDa].
More recently, the perturbative construction of the $\phi^4$-theory
has been performed in an algebraic setting by Brunetti,
Fredenhagen, Hollands and Wald [BrFr],[HoWa1],[HoWa2].
These authors adapted the renormalization method of Epstein
 and Glaser to construct the algebra of local, covariant
quantum fields of the  $\phi^4$-model on a globally
hyperbolic curved spacetime to any order of the perturbative
expansion, making use of techniques from microlocal analysis. 
The crucial notion of a local and covariant quantum field
 introduced in [HoWa1,2] has been further formalized by
Brunetti, Fredenhagen and Verch [BFV]. \\
 This paper is organized as follows~: In Section 2 we collect
and slightly adapt global bounds on the heat kernel found
 in the mathematical literature, which are pertinent to our
treatment. The action considered and the system of
perturbative flow equations satisfied by the CAS is set up
in Section 3. To establish bounds on the CAS, being
distributions, they have to be folded first with test
functions. In Section 4 a suitable class of 
test functions is introduced, together with tree structures
with the aid of which the bounds to be derived on the CAS
are going to be expressed. In Section 5 we state the boundary 
and the renormalization conditions used to integrate the
flow equations of the irrelevant and relevant terms, respectively.
The flow equations permit to be quite general in this respect,
englobing basically all situations of interest.
Section 6 is the central one of this paper. We state and prove
inductive bounds on the Schwinger functions which, being uniform 
in the cutoff, directly lead to renormalizability. Beyond they imply 
tree decay of the Schwinger functions between their
external points. The last section is devoted to the proof that the bare
action of the theory may be chosen minimally, i.e. with position 
independent counter terms apart from one (logarithmically divergent) 
term which is proportional to the scalar curvature of the manifold. 
Here we have to make the assumption that geometric
quantities on the manifold have a smooth expansion (to lowest orders)
w.r.t. contributions of curvature terms of increasing mass dimension.

%%%%%%%%%%%%%%%%%%%%%%%%%%%%%%%%%%%%%%%%%%%%%%%%%%%%%%
\section{ The heat kernel}

We consider geodesically complete simply connected  Riemannian manifolds 
$\cal M\,$ of dimension $n$ 
without boundary,
whose sectional curvatures are bounded between two constants
$-k^2$ and $\kappa^2\,$.
The related heat kernel then  has the following properties~:
\eq 
K(t,x,y) \in C^{\infty}( (0,\infty) \times {\cal M} \times {\cal M})\ ,
\label{hk0}
\eqe
\eq
0 < \ K(t,x,y) < \infty
\ ,
\label{hk1}
\eqe
\eq
K(t,x,y) = \ K(t,y,x)\ ,
\label{hk2}
\eqe
\eq 
 \int_{\cal M} K(t,x,y)\ dV(y)\ =\ 1\ ,
\label{hk3}
\eqe  
\eq
K(t_1+t_2,x,y) =\ \int_{\cal M}
 K(t_1,x,z) \ K(t_2,z,y)\ dV(z)\ . 
\label{hk4}
\eqe
Stochastic completeness (\ref{hk3}) holds due to the assumed
bounded curvature, cf. [Tay, ch.6, Prop.2.3 ].
Mathematicians have established quite
sharp pointwise bounds on the respective heat kernels
 of various classes of manifolds. We are going to explicit now 
 some of these bounds, because we will rely on them in the subsequent 
construction. Some bounds are known to hold for 
$0 < t < T\,$, others for  $0 < t \,$ or $ T < t $.
We will write $t_{\de} =t(1+\de)\,$ where the parameter $\de\,$ satisfies
$\,0 < \de <1\,$  and may be chosen arbitrarily small. 
Furthermore  $c,\ C,$ collectively 
denote constants which depend 
 on $\de, n\,$ and - if involved in the claim - on $k^2, T\,$.\\
On complete Riemannian
manifolds of dimension $n$ with nonnegative Ricci curvature 
 the heat kernel satifies the lower and upper 
bounds [LiYa, Dav1]
\eq
\frac{c}{\sqrt{|{\cal B}(x,t^{1/2})|\, |{\cal B}(y,t^{1/2})|}}\,
e ^{-\frac{d^2(x,y)}{4 t(1-\de)}} \ \le \ 
K(t,x,y)\ \le \ 
\frac{C}{\sqrt{|{\cal B}(x,t^{1/2})| \, |{\cal B}(y,t^{1/2})|}}\,
e ^{-\frac{d^2(x,y)}{4 t(1+\de)}}  
\label{hk5}
\eqe
valid for all $x,y \in \cal{M}$ and $ t >0\, $.\\
In the case of negative Ricci curvature bounded below let
 $ E \geq 0 $ denote the bottom of the spectrum of the operator
  $ - \Delta\, $. Then it holds, see [Dav2, Theorems 16 and 17 ]~:\\
 If $ \delta > 0 $, there exists a constant $ c_{\delta } $ such that
\begin{equation} \label{f20}
         K(t,x,y) \leq  c_\delta
   \big (\,|{\cal B}(x,t^{1/2})| \, |{\cal B}(y,t^{1/2})|\big )^{-\frac{1}{2}}
             \exp \Big( - \frac{d^2(x,y)}{4t(1+\de)} \Big)
\end{equation}   
for $ 0 < t < 1$ and for all $ x,y \in \mathcal{M} $, whereas \\
\begin{equation} \label{f21}
     K(t,x,y) \leq  c_{\delta} 
 \Big (|{\cal B}(x,1)| \, |{\cal B}(y,1)|  \Big )^{- \frac{1}{2}}
  \exp \Big \{ ( \delta - E ) t - \frac{d^{\, 2}(x,y)}{4t(1+\de)} \Big \}
\end{equation}
for $ 1 \leq  t < \infty $ and for all $x,y \in \mathcal{M} $. \\
Moreover, a lower bound  of the form appearing in (\ref{hk5})
holds here, too, however restricted to $ 0 < t < T$, [Var].\\
In addition, given bounded sectional curvature
 $ - k^2 \leq Sec_{\cal{M}} \leq \kappa^2 $, there is the
lower bound
\eq
K(t,x,y) \geq c\,  \exp\bigg ( - \tilde{E} t- 
   C\, \frac{ d^{\,2}(x,y)}{t} \bigg ) 
\label{F22}
\eqe
for all  $ x,y \in \mathcal{M} $ and for $ t > T $, with
constants $ c,C > 0 $ and $\tilde{E} > E $, possibly much
larger, [Gri, ch. 7.5 ]. \\
On a Cartan-Hadamard manifold of dimension $n\,$, 
i.e. a geodesically complete simply connected noncompact 
Riemannian manifold with nonpositive
sectional curvature, and assuming that the sectional curvature
is bounded below by $- K^2 $, we also have for all
 $ x,y \in \mathcal{M} $, and for  $0 <t $, [Gri, ch. 7.4 ],
\eq
K(t,x,y)\ \le \ 
\frac{C}{\min(1, t^{n/2})}\,
\exp \bigg( - \frac{(n-1)^2}{4}K^2 t
    -\frac{d^{\,2}(x,y)}{4 t_{\de}}\bigg)\ .
\label{hk8}
\eqe
For later use we extract from these bounds particular
versions valid for four-dimensional complete Riemannian
 manifolds whose sectional curvatures may range between two
constants $ - k^2 $ and $ \ka^2 $.\\
 From volume comparison, cf. e.g. [Cha, sect.3.4], follows:\\
i) If all sectional curvatures of $\cal{M}$ have values
 in $ [- k^2, 0],\,k > 0 $, fixed, then \\ 
\eq
\frac{\pi^2}{2} t^2 \leq |{\cal B}(x,t^{1/2})|
  \leq \frac{\pi^2}{2} t^2\, h_4(k\, t^{1/2})
\label{comp}
\eqe
with the positive increasing function
$$ h_4(r)=\ \frac{\cosh(3r)-9\cosh r +8}{3r^4}, 
         \quad h_4(0) = 1 \, ,$$ 
ii) if all sectional curvatures of $\cal{M}$ have values
 in $ [0, \ka^2],\, \ka > 0 $, fixed, then 
\footnote{ The restriction on $ t $ accounts for
  the injectivity radius of the manifold.}\\ 
\eq
\frac{\pi^2}{2} t^2 \, s_4(\ka \, t^{1/2})
\leq |{\cal B}(x,t^{1/2})|
  \leq \frac{\pi^2}{2} t^2 ,\quad \mbox{for}\quad \ka\, t^{1/2} < \pi,
\label{comp'}
\eqe
with the positive decreasing function, $ 0 < r < \pi$,
$$ s_4(r)=\ \frac{\cos(3r)-9\cos r +8}{3r^4}, 
         \quad s_4(0) = 1 \, .$$ 
Taking (\ref{hk5}) together with (\ref{comp'}), as well as
taking (\ref{hk8}) or (\ref{f20}) and the lower bound
from (\ref{hk5}) \footnote{Remember the statement after (\ref{f21}).}
 together with (\ref{comp}), we obtain, restricting 
\footnote{The restriction is necessary both for the upper and lower
bounds.}
 to $\, 0 < t < T $:
\eq
\frac{c}{t^2}\ \exp(-\frac{d^2(x,y)}{4 t(1-\de)}) \ \le \ 
K(t,x,y)\ \le \ 
\frac{C}{t^2} \ \exp(-\frac{d^2(x,y)}{4 t(1+\de)})\ . 
\label{hk10}
\eqe
The constants $c,\  C $ depend on $ k^2, \ka^2, \de, T $,
but do not depend on $ t $.
As a consequence of this lower and upper bound  
 we obtain under the same conditions 
\eq
 d^s(x,y) \,K(t,x,y) \ \le \ c\,'\  t^{\, s/2}\ K(t_{\,\de\, '},x,y)\ 
 \quad \mbox{for} \,\, t \leq T ,
\label{d}
\eqe
with $\de\,' > \de $.   
For $\,1\le s \le 3\,$ we also need the bound
\eq
|\nabla^s \,K(t,x,y)| \ \le \ C \ t^{-s/2}\ K(t_{\de},x,y)\ 
\label{D}
\eqe
based on [CLY], [Dav3] and valid for $ 0 < t < T $.
Here $\,\nabla^s\,$ denotes a covariant derivative
of order $s\,$ w.r.t. $x\,$ and the norm is that of (\ref{nor}).
The constant $C$ here also depends on the norm of the covariant
derivatives of the curvature tensor up to order $s-1\,$.
From (\ref{D}) and the heat equation it follows directly that
\eq
|\pa_t \,K(t,x,y)| \ \le \ C \ t^{-1}\ K(t_{\de},x,y)\ .
\label{pat}
\eqe
Finally we note the following recently proven bound on the
logarithmic derivative of the heat kernel [SoZh] which holds
for $Ric_{\cal M} \ge -k^2\,$ and for $\,0 < t < T\,$:
\eq
\frac{|\nabla \,K(t,x,y)|}{K(t,x,y)} \ \le \
O(1)\ \frac{1}{t^{1/2}} \ (1+\frac{d^2(x,y)}{t})\ .  
\label{logD}
\eqe

In closing this section we remark that the restriction to manifolds
$\cal M$ of the kind considered is not dictated by the validity of
our methods of proof. It rather seems to be a choice which is
reasonable and interesting on physical grounds.

%%%%%%%%%%%%%%%%%%%%%%%%%%%%%%%%%%%%%%%%%%%%%%%%%%%%%%%%%%%%%%%%
\section{The Action and the Flow Equations}

The regularized (free) propagator is
 given in terms of the heat kernel by
\eq
{C}^{\vep,t}(x,y)\,=\, \int_{\vep}^{t}
dt' \ e ^{-m^2\,t'}\ K(t',x,y)\ .
\label{propa}
\eqe
Its derivative  w.r.t. $t\,$ is denoted as
\eq
 C_t(x,y):=\pa_{t}C^{\vep,t}(x,y)=\
e ^{-m^2\,t}\ K(t,x,y)\ .
\label{dpropa}
\eqe
We assume $\,0 < \vep  \le t < \infty\,$
so that the  flow parameter $t$ takes the role of a 
long distance cutoff, whereas  $\vep$ is 
a short distance regularization. The full propagator is recovered for 
$\vep=0$ and $t \to \infty\,$.
For finite $\vep$ and in finite volume 
the positivity and regularity properties of ${C}^{\vep,t}\,$ 
permit to define the theory rigorously 
from the functional integral
\eq
e^{- {1\over \hbar} (L^{\vep,t}(\vp)+ I^{\vep,t})}
\,=\, 
\int \, d\mu_{\vep,t}(\phi) \; 
e^{- {1\over\hbar} L^{\vep,\vep}(\phi\,+\,\vp)} \ ,
\quad L^{\vep,t}(0):=0\ ,
\label{funcin}
\eqe
where the factors of $\hbar\,$ have been introduced to allow for a consistent
loop expansion in the sequel.
In (\ref{funcin}) $\,d\mu_{\vep,t}(\phi) $ denotes the
 Gaussian measure with covariance $\hbar C^{\vep,t}(x,y)$.
The test function $\vp\,$ here is supposed  to be in the 
support of the Gaussian measures $\,d\mu_{t',t''}(\phi)\, $,
$\vep \le t' \le t'' < \infty\,$, which implies in particular
that it is in $C^{\infty}(\cal M)\,$. 
The normalization factor $\,e^{-{1\over\hbar} I^{\vep,t}} $ is
due to vacuum contributions. It
diverges in infinite volume so that we can take the infinite volume
limit only when it has been eliminated. 
We do not make the finite
volume explicit here since it plays no role in the 
sequel. 

The functional 
$\,L^{\vep}(\vp)~:=\,L^{\vep,\vep}(\vp)\,$ is the bare (inter)action including
counterterms, viewed as a formal power series in $\,\hbar\,$. The
superscript $\vep\,$ indicates the UV cutoff.
For shortness we will pose in the following, with $x,y \in \cal{M}$,
 $\vec x = (x_1,\cdots, x_n)\in {\cal{M}}^{\times n} $,
\[
\int_x\ :=\ \int_{\cal M} dV(x)\, , \qquad
\int_{\vec x}\ :=\ \prod_{i=1}^{n}\int_{\cal M} dV(x_i)\, ,
\]
and
\[
\ti \de(x,y)\,:=\ |g|^{-1/2}(x) \ \de(x,y)\ .
\]
 As is known from lowest order calculations [Bir],[BPP],[L\"u],
in curved spacetime there will appear an additional counterterm
of the type $\int_x R(x)\, \vp^2(x)\,$ which is proportional to the 
scalar curvature  $\,R(x)\,$  
of the spacetime manifold $\cal M\,$ considered. So the bare
interaction for the symmetric  $\vp_4^4$ theory would be 
\eq
   L^{\vep}(\vp) = {\lambda \over 4!}  \int_x \vp^4(x)  
   \, +\, {1 \over 2} \int_x\ \{ (a ^{\vep} +\xi ^{\vep} R(x))\vp^2(x) +
    b ^{\vep} g^{\mu \nu}(x)\  
 \partial_{\mu}\vp(x)\cdot\partial_{\nu} \vp(x)  +
    {2 \over 4!}\, c ^{\vep} \vp^4(x)\} 
\label{nawi}
\eqe  
where $\la >0$ is the renormalized coupling, and   
the cutoff dependent parameters $ a ^{\vep},\ \xi ^{\vep},\  b
^{\vep},\ c ^{\vep} $
- which remain to be fixed and which are directly related 
to the mass, curvature, wave function, and
coupling constant counterterms \footnote{Since it is not necessary  
in the flow equation framework
to introduce
bare fields in distinction from renormalized ones 
(our field is the renormalized one in this language), there is a
slight difference, which is to be kept in mind only 
when comparing to other schemes.} - will fulfill
\eq
a ^{\vep},\ \xi ^{\vep},\ b ^{\vep},\ c ^{\vep} =O(\hbar)\,.
\label{coef}
\eqe
It seems to us that there is no a priori reason to restrict
to bare interactions of this form. In fact, since 
there is no translation invariance in curved space time,
{\it all} counter terms and even the coupling $\la$ itself 
may be position dependent.
Quite generally the bare action is not a directly observable
 physical object, 
and the constraints on  its form stem from the symmetry properties
of the theory which are imposed, on its field content 
and on the form of the propagator. The symmetry properties
depend in particular on the renormalization conditions which fix the 
physical (relevant) parameters of the theory. 
%In curved space time they should reflect intrinsic 
%geometrical properties.
They might be position dependent~: e.g. a local scattering
experiment performed at different places at the same external
momenta might give different cross sections and, as a consequence 
of this, the renormalized coupling, fixed in terms of the
cross section, would be position dependent. 
It is therefore natural to  admit more general bare 
interactions \footnote{One could of
  course be even more general.}
\[
   L^{\vep}(\vp) = \int_x   {\la(x) \over 4!} \,\vp^4(x)  
   \ + \ 
\]
\[
{1 \over 2} \int_x\ \{a^{\vep}(x) \,\vp^2(x) +
    u^{\,\mu,{\vep}}(x) \,\vp(x)\,\nabla_{\mu}\vp(x) + 
      \hat b^{\vep}(x) g^{\mu \nu}(x)\  
 \partial_{\mu}\vp(x)\cdot\partial_{\nu} \vp(x) 
 +     {2 \over 4!}\, c ^{\vep}(x) \vp^4(x)\}\, .
\] 
Here $\,  \la(x),\  a ^{\vep}(x),\  \hat
b ^{\vep}(x),\ c ^{\vep}(x)\,$
are general scalars and $u^{\,\mu,\vep}(x)$
is a general vector, all functions are  supposed to be smooth,
and $|\la(x)|\,$  (of course)
uniformly bounded on $\cal M\,$.
When calculating the two point function
${\cal L}^{\vep}(x_1,x_2)=\
\de/\de_{\vp(x_1)}\,\de/\de_{\vp(x_2 )}\, L^{\vep}(\vp)|_{\vp \equiv
      0}\,$ 
from this bare action one obtains
\begin{eqnarray} 
\mathcal{L}_{2}^{\,\epsilon}(x_1, x_2) &= &  a^{\epsilon}(x_1) 
                       \, \tilde{\delta}(x_2,x_1)
     - \frac{1}{2}( \nabla_\mu u^{\,\mu, \epsilon})(x_1)
         \, \tilde{\delta}(x_2,x_1) \nonumber \\ 
  &-& |g(x_2)|^{-{1 \over 2}}\,
   \pa_{\mu}^{(2)}\,\hat{b}^{\vep}(x_2) \,
   g^{\mu \nu}(x_2)\,|g(x_2)|^{1 \over 2}\,\pa_{\nu}^{(2)}
             \, \tilde{\delta}(x_2,x_1)\ .  
\label{gen2}
\end{eqnarray}
This means that $u^{\,\mu, \epsilon}\,$ only contributes via
its divergence, and that the contribution  $\sim  \nabla_\mu
    u^{\,\mu,\epsilon} \,$  can be absorbed in $a^{\epsilon}\,$. 
On the other hand the particular
 tensor coupling $ \hat b^{\vep}(x)g^{\mu \nu}(x)\, $ 
can be generalized - without changing symmetry properties -
into a smooth symmetric $ (2,0)$-tensor 
field $ b^{\,\mu \nu, \vep}(x)\,$. In (\ref{gen2})
the  product $ \hat b^{\vep}(x)g^{\mu \nu}(x)$ is then  replaced by 
$ b^{\,\mu \nu, \vep}(x)\,$, and we  recognise that in
 (\ref{gen2}) the generalisation 
\eq
 \Delta^{(b)}\, := \, |g(x)|^{- {1 \over 2}}\,\pa_{\mu} \,
      b^{\,\mu \nu}(x)\,|g(x)|^{1 \over 2}\,\pa_{\nu} 
\label{gede}
\eqe
of the Laplace-Beltrami operator $\De $, (\ref{a4}), appears.
We thus adopt a bare interaction 
of the form 
\eq
L^{\vep}(\vp) = \int_x {\la(x) \over 4!} \,  \vp^4(x)  
 +
{1 \over 2}\int_x\ \{\, {\tilde a}^{\vep}(x) \,\vp^2(x) + 
   \, b^{\,\mu \nu, \vep}(x)\  
 \partial_{\mu}\vp(x)\cdot\partial_{\nu} \vp(x) 
 +     {2 \over 4!}\, c ^{\vep}(x) \vp^4(x)\}
\label{nawig}
\eqe 
with smooth scalar functions $\,{\tilde a}^{\vep}(x),\ c ^{\vep}(x)\,$
and a smooth symmetric tensor field $\  b ^{\mu \nu,\vep}(x)\,$
- which remain to be fixed and which are of (at least) order $\,\hbar\,$.

The flow equation (FE)  
is obtained from (\ref{funcin}) on differentiating 
w.r.t. $t\,$. It is a differential equation for the functional
$L^{\vep,t}\,$~: 
\eq
\partial_{t}(L^{\vep,t} + I^{\vep,t} ) \,=\,\frac{\hbar}{2}\,
\langle\frac{\delta}{\delta \vp}, C_t\,
\frac{\delta}{\delta \vp}\rangle L^{\vep,t}
\,-\,
\frac{1}{2}\, \langle \frac{\delta}{\delta
  \vp}L^{\vep,t}, C_t\,
\frac{\delta}{\delta \vp} L^{\vep,t}\rangle \;\,.
\label{feq}
\eqe
By $\langle\ ,\  \rangle$ we denote the standard inner product in 
$L^2({\cal M}, dV( x))\,$. The FE can also be stated in integrated form
\eq
e^{- {1\over \hbar} (L^{\vep,t}(\vp)+ I^{\vep,t})}
\,=\, 
e^{\hbar\De_{\cal F}(\vep,t)}\ 
e^{- {1\over\hbar} L^{\vep,\vep}(\vp)} \ .
\label{intfe}
\eqe
The functional Laplace operator $\De_{\cal F}(\vep,t)\,$ is given by
\[
\De_{\cal F}(\vep,t)=\,\frac12\
\langle \de_{\vp},\ C^{\vep,t} \ \de_{\vp} \rangle 
\]
using the notation 
$\delta_{\vp(x)}=\delta/\delta\vp(x)$.
We may expand $L^{\vep,t}(\vp)\,$ w.r.t. the number of fields
$\vp\,$ setting
\eq
L^{\vep,t}_{n}(\vp)~:= \
\frac{1}{n!}\,
{\pa ^n\over \pa \ka ^n}\ L^{\vep,t}(\ka \vp)|_{\ka =0}\ .
\eqe
The functional $L^{\vep,t}(\vp)\,$ 
can also be expanded in a formal powers series w.r.t. 
$\hbar$, and in a double series w.r.t. 
$\hbar$ and the number of fields
\eq
L^{\vep,t}(\vp)\,=\,\sum_{l=0}^{\infty}
\hbar^l\,L^{\vep,t}_{l}(\vp)\,=\, \sum_{l=0}^{\infty}\hbar^l\,\sum_{n
 =2}^{\infty}L^{\vep,t}_{n,l} (\vp) \ ,\quad
L^{\vep,t}_{2,0}(\vp)\equiv 0\ .
\label{3.3}
\eqe
Corresponding expansions for $ \ti a ^{\vep}(x),\ 
 b^{\,\mu \nu, \vep}(x),\ c ^{\vep}(x)\,$ are 
$ \ti a ^{\vep}(x)=\sum_{l \ge 1} \ti a_l ^{\vep}(x) \hbar^l$ etc. 
We can then rewrite (\ref{feq}) in loop order $l\ $ as
\[
\partial_{t}L^{\vep,t}_{n,l}(\vp ) \,=\,
\]
\eq\frac{1}{2}\,
\int_{x,y} C_{t}(x,y)\ \Bigl[\de_{\vp(x)}\
\de_{\vp(y)}\,L^{\vep,t}_{n+2,l-1}(\vp )
\,-\
\sum_{n_1+n_2 =n+2 \atop 
l_1+l_2=l }
(\de_{\vp(x)}\,L^{\vep,t}_{n_1,l_1}(\vp ))\
\de_{\vp(y)}\,L^{\vep,t}_{n_2,l_2}(\vp )
\Bigr]\ .
\label{nfeq}
\eqe
From $L^{\vep,t}_{l}(\vp)$ we obtain the connected amputated
Schwinger functions of loop order $l$
as 
\eq
 {\cal L}^{\vep,t}_{n,l}(x_1,\ldots,x_{n}) 
\,:=\ 
\de_{\vp(x_1)} \ldots \de_{\vp(x_n)}
L^{\vep,t}_l|_{\vp \equiv 0}\ .
\label{CAS}
\eqe
It is straightforward to realize that the    
${\cL}^{\vep,t}_{n,l}\,$ are distributions \\  
1) which are completely {\it symmetric}
w.r.t. permutations of the arguments 
$(x_1 ,\ldots, x_{n})\,$\\
2) and which  {\it fall off rapidly} with the distances $d(x_i,x_j)\,$.\\
These facts follow
from (\ref{nawig}), (\ref{intfe}) and the properties 
of the regularized propagator (\ref{hk5}), (\ref{hk10}).
The distributional character of the  ${\cL}^{\vep,t}_{n,l}\,$ 
is related to the fact that we consider {\it amputated}
Schwinger functions. Thus there is associated a factor of 
$\ti\de(x_i,z)\,$
to the external line  joining  
$x_i$ to an internal vertex at $z$
which is integrated over.
Therefore the distributional character is
 different according to whether one or
several (up to three) external lines end in a given $z$-vertex.
From the point of view of the FE the distributional character  is a
consequence of the boundary conditions, see (\ref{nawig}) and
(\ref{bo2}), (\ref{bo3}). Note that   
by  (\ref{hk0}) the propagators $C^{\vep,t}\,$ which join different vertices 
are smooth functions of their position arguments for 
$\vep >0\,$. 
The two-point function is the most divergent object as regards its
flow for  small $t$. But the distributional structure 
of its regularized version is particularly
simple. 
In fact it follows
from (\ref{nawig}) and (\ref{intfe}) 
that $\,{\cal L}^{\vep,t}_{2,l} (x_1,x_2) \,$ can be written as a
sum over regularized Feynman amplitudes with vertices from   
$\, L^{\vep} \,$ and with regularized propagators
which satisfy (\ref{hk0}), (\ref{hk5}). Thus the only distributional
singularities appearing are of the form 
${\ti \de}(x_2,x_1)\,$ and ${\De^{(b_l^{\vep,\vep})}}_{x_2}\ {\ti
  \de}(x_2,x_1)\,$. Thus $\,{\cal L}^{\vep,t}_{2,l} (x_1,x_2) \,$ 
can be written as a linear combination of these two contributions and a
smooth function $f(x_1,x_2)\,$ of rapid decrease in $\,d(x_1,x_2)\,$.\\

The FE for the Schwinger functions derived from (\ref{nfeq})
takes the following form~:
\eq
\pa_{t}
{\cL}^{\vep,t}_{n,l}(x_1,\ldots,x_{n})=\,
\frac12\,\int_{x,y} C_t(x,y)\ \Biggl\{
{\cL}^{\vep,t}_{n+2,l-1}(x_1,\ldots,x_{n},x,y)\  -
\label{fequ}
\eqe
\[
\sum_{l_1+l_2=l,\atop
n_1+n_2=n}\Bigl[
 {\cal L}^{\vep,t}_{n_1+1,l_1}(x_1,\ldots,x_{n_1},x)\,
\,
{\cal L}^{\vep,t}_{n_2+1,l_2}(y,x_{n_1+1},
\ldots,x_{n})\Bigr]_{sym}\Biggr\}\, .
\]
Here $sym$ means symmetrization - i.e. summing over all
permutations \footnote{by our choice of 
normalization there is no  normalization factor to be divided out}
  of  $(x_1,\ldots, x_{n})$ {\it modulo those} 
which only rearrange the arguments of a factor.
 \footnote{This  
may be implemented by counting only the configurations
  in which the permuted position variables appearing in 
${\cL}^{\vep,t}_{n_1+1}\,$ and ${\cL}^{\vep,t}_{n_2+1}\,$ 
appear in lexicographic order.} \\
For the renormalization proof we also need the FE 
for the Schwinger functions derived w.r.t. the UV cutoff $\vep$. 
Integrating the FE over $t'\,$ between $\vep$ and $t$ and then
deriving w.r.t. $\vep$ we obtain 
\[
\pa_{\vep} \,{\cL}^{\vep,t}_{n,l}(x_1,\ldots,x_{n})=\
 \pa_{\vep}{\cL}^{\vep,\vep}_{n,l}(x_1,\ldots,x_{n})\, -\,
\frac12\,\int_{x,y} C_{\vep}(x,y)\ \Biggl\{
{\cL}^{\vep,\vep}_{n+2,l-1}(x_1,\ldots,x_{n},x,y)\  -
\]
\eq
\sum_{l_1+l_2=l,\atop
n_1+n_2=n}\Bigl[
 {\cal L}^{\vep,\vep}_{n_1+1,l_1}(x_1,\ldots,x_{n_1},x)\,
\,
{\cal L}^{\vep,\vep}_{n_2+1,l_2}(y,x_{n_1+1},
\ldots,x_{n})\Bigr]_{sym}\Biggr\}\ +
\label{fequvep}
\eqe
\[
\frac12\,\int_{x,y} \int_{\vep}^t dt'\ C_{t'}(x,y)\ \Biggl\{
\pa_{\vep}
{\cL}^{\vep,t'}_{n+2,l-1}(x_1,\ldots,x_{n},x,y)\  -
\]
\[
\sum_{l_1+l_2=l,\atop
n_1+n_2=n}\Bigl[
 \pa_{\vep}
\Bigl({\cal L}^{\vep,t'}_{n_1+1,l_1}(x_1,\ldots,x_{n_1},x)\,
\,
{\cal L}^{\vep,t'}_{n_2+1,l_2}(y,x_{n_1+1},
\ldots,x_{n})\Bigr)\Bigr]_{sym}\Biggr\}\ .
\]
Integrating the FE instead from $t<1$ to $t=1$ 
and then deriving w.r.t. $\vep\,$  we get 
\eq
\pa_{\vep} \,{\cL}^{\vep,t}_{n,l}(x_1,\ldots,x_{n})=\
 \pa_{\vep}{\cL}^{\vep,1}_{n,l}(x_1,\ldots,x_{n})\ - 
\label{fequvep2}\eqe
\[
\frac12\,\int_{x,y} \int_{t}^1 dt' \ C_{t'}(x,y)\ \Biggl\{
\pa_{\vep}
{\cL}^{\vep,t'}_{n+2,l-1}(x_1,\ldots,x_{n},x,y)\  -
\]
\[
\sum_{l_1+l_2=l,\atop
n_1+n_2=n}\Bigl[
 \pa_{\vep}
\Bigl({\cal L}^{\vep,t'}_{n_1+1,l_1}(x_1,\ldots,x_{n_1},x)\,
\,
{\cal L}^{\vep,t'}_{n_2+1,l_2}(y,x_{n_1+1},
\ldots,x_{n})\Bigr)\Bigr]_{sym}\Biggr\}\ .
\]

%%%%%%%%%%%%%%%%%%%%%%%%%%%%%%%%%%%%%%%%%%%%%%%%%%%%%%%%%%%%%%%%
\section{Test functions and Tree structures}

The distributional character of the 
${\cal L}^{\vep,t}_{n,l}\, $ necessitates
the introduction of test functions against which they will be
integrated. Later on we will only use a subclass of the test functions 
introduced in the subsequent definition, see (\ref{phic}).
\\[.1cm]  
{\it Definition 1}~: For $\,   n \in \bbbn\,$ we set
\[
\cH_n: =\{ \vp(\xv_n)= \vp_1(x_1)\ldots   \vp_n(x_n)\ 
| \ \vp_i \in C^{\infty}({\cal M})\cap   
L^{\infty}({\cal M})\} \  .
\]
We wrote $\,\xv_n=\ (x_1,\ldots,x_n)\,$
and we shall write \footnote{By the bosonic symmetry of the 
${\cal L}^{\vep,t}_{n,l}\,$ all 
bounds are  independent of the particular role assigned to
the coordinate $x_1\,$, which can be exchanged with any other coordinate.} 
$\,x_{2,n}=\, (x_2,\ldots,x_n) \in {\cal M}^{\times (n-1)} \,$.\\
For $\vp\, \in \cH_{n-1}\,$ we set
\eq 
{\cal L}^{\vep,t}_{n,l}(x_1,\vp) :=\, 
\int_{x_{2,n}} 
 {\cal L}^{\vep,t}_{n,l} (\xv_n)\  \vp(x_{2,n})\ .
\label{lin}
\eqe
The regularized Schwinger functions are obviously linear
w.r.t. the test functions:
\eq 
{\cal L}^{\vep,t}_{n,l}(x_1,a \,\vp_1+ b\,\vp_2) =\, 
a\ {\cal L}^{\vep,t}_{n,l}(x_1,\vp_1) \,+\,
b\ {\cal L}^{\vep,t}_{n,l}(x_1,\vp_2) \ ,\quad a,b \in \bbbc\ ,
\label{lin1}
\eqe
and it also follows from from
(\ref{nawig}) and (\ref{intfe}) and the properties 
of the regularized propagator (\ref{hk0}), (\ref{hk8})
that they satisfy
\[
{\cal L}^{\vep,t}_{n,l}(x_1,\vp)\in \cH_{1}\ .
\]
We will also consider Schwinger functions multiplied by 
products of factors $\si(x_j,x_1) ^{\mu}$, (\ref{sig}).\\[.1cm]
{\it Definition 2}~:  
We introduce a smooth (external) covector field 
$ \om_\mu (x)$ and form the bi-scalar insertions
\eq
 E_{(i)}\equiv E(x_i, x_1~;\om) := \si(x_i, x_1)^\mu\, \om_{\mu}(x_1),
  \qquad  i=2, \cdots, n\ ,
\label{ino}  
\eqe
and, more generally, for  $r\in \bbbn\,$,
\eq
 E ^{(r)}_{(i)}\equiv E(x_i, x_1~;\om^{(r)}) 
:= \si(x_i, x_1)^{\mu_1}\ldots \si(x_i, x_1)^{\mu_r}\, \om^{(r)}_{\mu_1
  \ldots\mu_r }(x_1) 
\label{inor}
\eqe
with a smooth (external) symmetric covariant tensor field 
$\om^{(r)}_{\mu_1 \ldots\mu_r } (x)$ of rank $r$. 
We have, because of (\ref{nsi}),
\eq
 | E(x_i, x_1~;\om^{(r)}) | \leq |\,\om^{(r)}(x_1)| \, d^{\,r}(x_i,x_1)
\label{nino}
\eqe
with the norm $ |\,\om^{(r)}(x_1)| $ according to (\ref{nor}).
 For  $r\,\in \bbbn\,$ we then pose
\eq
 {\cal L}^{\vep,t}_{n,l}(x_1, E^{(r)}_{(i)}\, \vp):=
\int_{x_{2,n}} E(x_{i},x_{1}~;\om^{(r)})\,
 {\cal L}^{\vep,t}_{n,l} (\xv_n)\ \vp(x_{2,n})\ .  
\label{ins}
\eqe
Mostly we will suppress $\om$ in the notation
as we did in (\ref{ins}).
Furthermore, for given $x_1, x_2\, \in \cal{ M}$
 we consider the products
\eq
 F^{(r)}_{(12)} {\cal L}^{\vep,t}_{n,l}(x_1, x_2, \vp) :=
d^{\,3-r}(x_1, x_2)E(x_{2},x_{1}~;\om^{(r)})\,
  \int_{x_{3,n}}{\cal L}^{\vep,t}_{n,l} (\xv_n)\ \vp(x_{3,n})  
\label{2ins}
\eqe
for $ r = 0,1,2\, $, with $ E \equiv 1\mbox{ if}\,\, r  = 0\, $,
 and with $\vp(x_{3,n})\equiv 1\,$ (and no integration)
   \mbox {if} $n <3\,$.\\[.1cm]
{\it Definition 3}~:  
i) A graph $G(V,{\cal P})\,$ is defined as a set of vertices $V\,$  
and a set ${\cal P}\,$ of unordered pairs $p$ of vertices called lines/edges. 
Two lines are connected if they share a vertex in common.
A graph is connected if for each pair of vertices
$(i,j)\,$ there exists a path of connected lines connecting $i\,$ to
$j\,$. A tree is a connected graph $G(V,{\cal P})\,$ with $|{\cal P}|=|V|-1\,$.
For a tree one can prove that the  path of connected lines connecting 
$i\,$ to $j\,$ is {\it unique}. A rooted tree is a tree where one vertex
in $V$ has been chosen to be its root. The incidence number $c_i\,$ of
the vertex $i\,$ in a tree is the number of distinct lines containing
$i\,$. The subset $V_e \subset V\,$, containing the vertices $i\,$ with 
 $c_i=1\,$, excluding the root (if it has $c=1\,$), is called the set of
external vertices. All other vertices  are called internal
vertices. We denote by $\mathcal{T}^s $ the set of all trees 
such that $|V_e|=s-1\,,\ s\ge 2\,$. Subsequently we will consider
trees where the set of vertices is identified\footnote{In
mathematically straight notation a vertex
should be viewed as the image of an element of a discrete set
under a mapping from this set into $\cal M$.} with a set of points
in the manifold $\cal M\,$.  For a  tree $T^s \in \mathcal{T}^s\,$
we will call $x_1\in \cal M\,$ 
its root vertex, and $Y=\{y_2,\ldots,y_s\}\,$
the  set of points in $\cal M\,$ to be identified with its external
vertices. Likewise we call $Z=\{ z_1,\ldots, z_r\}\,$ with $r \ge 0\,$
the set of internal vertices of $T^s\,$.\\ 
ii) For $y_i \in Y\,$ there exists exactly one $p \in \cal P$
such that $y_i \in p\,$. 
For $x_1$ there exist $p_1,\ldots,p_{c_1}\in\cal P$ with $1 \le c_1 \le s-1$
such that $x_1 \in p_1,\ldots,x_1 \in p_{c_1}$. For $z_j \in Z$ there
exist $p^{(z_j)}_1,\ldots,p^{(z_j)}_{c_j}\in\cal E$ with $2 \le c_j \le s \,$
such that  $z_j \in  p^{(z_j)}_1,\ldots, z_j \in  p^{(z_j)}_{c_j}\,$. 
We call $c_1=c(x_1)$ the
incidence number of the root vertex and $c(z_j)\,$ the
incidence number of the internal vertex $z_j$ of the tree.\\ 
We call a line $p \in \cal P$ an external line of the tree 
if there exists $y_i$ such that $y_i \in p\,$.
 The set of external lines is denoted
$\cal J$. The remaining lines are called internal lines of the tree 
and are denoted by $\cal I$, hence $\cal P = \cal J \cup \cal I $.\\
iii) Denoting by $ v_c $ the number of vertices having incidence number  
$c$, it follows from the definition that $\sum_{c\ge 2} (c-2)\, v_c  = s-3+
\delta_{c_1,1}\,$.
By $ T_l^s $ we denote a tree $ T^s \in \mathcal{T}^s$ 
satisfying $ v_2+\de_{c_1,1} \leq \, 3l-2+s/2\, $ for $l\ge 1\,$
and satisfying $ v_2=0\,$ for $l =0\,$. Then 
  $ \mathcal{T}_l^s $ denotes the set of all trees $ T_l^s $. 
We indicate the external vertices and internal vertices of the tree
by writing $T_l^{s}(x_1, y_{2, \, s}, \vec z \,)$
with $y_{2, \, s}=(y_2,\ldots,y_s)\,,\ \vec z =(z_1,\ldots,z_r)\,$.\\
iv) We also define for $i \le s$ the set 
of {\it twice rooted trees} denoted
as $\mathcal{T}^{s,(12)}_l\,$. The trees
${T}^{s,(12)}_l\,\in \mathcal{T}^{s,(12)}_l\,$ are defined
exactly as the trees $\,T^{s}_l\, $ apart from the fact that
they have two root vertices $x_1, \ x_2\,$ with the properties of
ii) above, and $s-2$ external vertices.
\\[.1cm]
{\it Definition 4}~: 
For a  tree $T_l^{s+2}(x_1, y_{2, \,s+2},\vec z \,)$
we define the {\it reduced tree}\\
$ T_{l,y_i,y_j}^{s}(x_1,y_2, \ldots, y_{i-1},y_{i+1},
\ldots,y_{j-1},y_{j+1},\ldots, y_{s+2}, \vec z_{ij})\,$ to be the unique
tree to be obtained from $T_l^{s+2}(x_1, y_{2, \,s+2},\vec z \,)$ through
the following procedure~:\\
i) By taking off the two external vertices $y_i, y_j\,$ together
 with  the  external lines attached to them.\\
ii) By taking off the internal vertices  - if any -
which have acquired incidence number $c=1\,$ through the previous 
process,   and by also taking off the lines attached  to them.\\
iii) If the process ii) has produced a new vertex of incidence
number 1 go back to ii).\\[.1cm]
In the sequel we shall bound the CAS folded with test functions.  
Here we restrict to test functions of the following form~:
Let $1\le s \le n\,$ and 
$\,\tau= \tau_{2,s}= (\tau_2,\ldots,\tau_{s})\,$ with 
$\,0< \tau_i \,$
\eq
\vp_{\tau,y_{2,s} }( x_{2,n})~:=\ 
\prod_{i=2}^{s}
K(\tau_{i}, x_i,y_i)\  \prod_{i=s+1}^n
\mathbf{1}(x_i)\ .
\label{phic}
\eqe
Here $\mathbf{1}(x)=1\ \ \, \forall x \in \cal M\,$.
These test functions are factorized\footnote{The function $\mathbf{1}(x)\,$
is obtained on integrating 
$K(\tau,x,y)\, $ over $y\,$. This could be used to unify the
notation.}. The nonconstant functions
are smooth, globally defined and rapidly decreasing on $\cal M\,$.
The pair $ \tau_j , y_j $ determines the width 
  and the center of localisation of the test function. 
This definition can be generalized by choosing any other subset
of $s$ coordinates among $x_2, \ldots, x_n\,$.
We also define \footnote{Note that $\vp^{(j)}_{\tau,y_{2,s} }$ depends
  on $x_1$ which is not indicated.}
 for $2 \le j \le s\,$
\eq
\vp^{(j)}_{\tau,y_{2,s} }( x_{2,n})~:=\ 
K^{(1)}(\tau_{j}, x_j,x_1;y_j)\  \prod_{i=2,i\neq j}^{s}
\ K(\tau_{i}, x_i,y_i)\  \prod_{i=s+1}^n
\mathbf{1}(x_i)
\label{phid}
\eqe
with 
\eq
K^{(1)}(\tau_{j}, x_j,x_1;y_j) =\
 K(\tau_j, x_j,y_j)-\ K(\tau_j, x_1,y_j)\ .
\label{k1}
\eqe
\noindent
{\it Definition 5:}
Given $\tau$, $y_{2,s}\,$, $\de >0\,$, and a set of internal vertices
$\vec z =(z_1,\ldots, z_r)\,$, and attributing positive 
 parameters  $t_{{\cal I}}= \{ t_I| I \in {\cal I}\}\,$
to the internal lines, the {\it weight factor 
$ {\cal F}(t_{{\cal I}},\tau;T_l^{s}( x_1, y_{2,s},\vec z))$ 
of a tree $T_l^{s}( x_1, y_{2,s},\vec z)\,$ 
at scales $t_{{\cal I}}$ } 
is defined as a product of heat kernels associated with
the internal and external lines of the tree. 
We set
\eq
{\cal F}(t_{{\cal I}},\tau;T_l^{s}(x_1, y_{2,s},\vec z))
:=\ 
\prod_{I\in {\cal I}}C_{t_{I,\de}}(I)\
 \prod_{J \in {{\cal J}}}\ K(\tau_{J,\de}, J)\ .
\label{f}
\eqe
Here we denote by $\tau_{J}\,$ the
 entry $\tau_i\,$ in $\tau\,$ carrying the index of the external
coordinate $y_i$ in which the external line  $J\,$ ends. 
For $I =\{a,b\}\,$ the notation
$C_{t_I}(I)\,$ stands  for $C_{t_I}(a,b)\,$.
\\
We then also define the {\it integrated weight factor} of a tree by
\eq
{\cal F}(t,\tau;T_l^{s}~; x_1, y_{2,s})~:= 
   \sup_{\{t_I| I \in {\cal I},\, \vep \le t_I \le t\}} 
  \int_{\vec z} {\cal F}(t_{{\cal I}},\tau;T_l^{s}( x_1, y_{2,s},\vec
  z))\ .
\label{ff}
\eqe
It depends on $\vep$, but note that its limit for $\vep \to 0\,$ exists,
and that typically the $\sup\,$ is expected to be taken for the 
maximal values of $t\,$ admitted. Therefore we suppress the dependence
on $\vep\,$ in the notation.   
Finally we introduce the shorthand
notation for the {\it global weight factor 
 $\,{\cal F}_{s,l}(t,\tau;x_1, y_{2,s})\,$ or
 more shortly $\,{\cal F}_{s,l}(t,\tau)\,$ } which is defined as follows
\eq
 {\cal F}_{s,l}(t,\tau)\,\equiv\,{\cal F}_{s,l}(t,\tau;x_1,
 y_{2,s})\,:=\, 
\sum_{T_l^{s} \in {\cal T}_l^s}
{\cal F}(t,\tau;T_l^{s}~;x_1, y_{2,s})\ .
\label{ges}
\eqe
In complete analogy we define the weight factors and global
weight factors for twice rooted trees
which we denote as
${\cal F}(t,\tau;T_l^{s,(12)}~;x_1,x_2, y_{3,s})\,$
resp. 
${\cal F}^{(12)}_{s,l}(t,\tau;x_1,x_2,y_{3,s})\,$
or  $\,{\cal F}^{(12)}_{s,l}(t,\tau)\,$.
Following the definitions (\ref{f})-(\ref{ges}) we also define 
for $t \ge 1\,$
\eq
{\cal F}^{\,t}(\tau;T_l^{s}(x_1, y_{2,s},\vec z))
:=\ 
 \sup_{\{t_I| I \in {\cal I},\, \vep \le t_I \le 1\}} 
\prod_{I\in {\cal I}}[(C_{t_{I,\de}}(I)
+\int_{1}^{t}C_{t'}(I)\ dt')]
\
 \prod_{J \in {{\cal J}}}\ K(\tau_{J,\de}, J)\ ,
\label{fi}
\eqe
\eq
{\cal F}^{\, t}(\tau;T_l^{s}~; x_1, y_{2,s})~:= 
    \int_{\vec z} {\cal F}^{\,t}(\tau;T_l^{s}( x_1, y_{2,s},\vec z))\ ,
\label{iff}
\eqe
and  
\eq
 {\cal F}^{\,t}_{s,l}(\tau)\,:=\, 
\sum_{T_l^{s} \in {\cal T}_l^s}
{\cal F}^{\,t}(\tau;T_l^{s}~;x_1, y_{2,s})\ .
\label{iges}
\eqe
For $s=1\,$ we set $\, {\cal F}_{1,l}(t,\tau)\,\equiv\,1 \ .$\\
We give more explicitly the form of ${\cal F}_{2,l}(t,\tau;x,y)\,$.
It is by definition given through
\[
{\cal F}_{2,l}(t,\tau;x,y)\,=\,
\sum_{T_l^2} {\cal F}_{2,l}(t,\tau;T_l^2~;x,y)\ = 
 \ K(\tau_{\de}, x,y)\ +\
\]
\[
\sum_{n = 1}^{3l-2}\, \,
\sup_{\{t_{I_i}| \vep\, \leq\, t_{I_i}\, \leq\, t,\, i=1.\cdots,n\}} 
\, \,[\prod_{1 \le i \le n } \int_{z_i}]\
C_{t_{I_1,\de}}(x,z_1) \ldots  C_{t_{I_n,\de}}(z_{n-1},z_n)\ 
K({\tau}_{\de}, z_n,y)\ .
\]
Using (\ref{hk4}) we get 
\eq
{\cal F}_{2,l}(t,\tau;x,y)\ = 
\sum_{n = 0}^{3l-2}\, \,
\sup_{\{t_{I_i}| \vep\, \leq\, t_{I_i}\, \leq\, t,\, i=1.\cdots,n\}} 
C_{\tau_{\de}+ \sum_1^n t_{I_i,\de}}(x,y) \  e ^{m^2\tau_{\de}}\ .
\label{f2r}
\eqe
\\ 

Let us shortly comment on why we are led to introduce tree structures
and weight factors in our context. In fact we will establish
bounds for the CAS (\ref{lin}) inductively by concluding
from the CAS appearing on the r.h.s. of the FE   (\ref{fequ}) on the CAS 
appearing on the l.h.s. Assume we have bounds in terms of weight factors 
of trees for the $\cal L$'s on the r.h.s. The second contribution  on
the r.h.s. of  (\ref{fequ})  then lends itself immediately to
reproduce such a bound if the factor, associated in our bound 
with a line of the tree, is a bound on $C_t(x,y)\,$, and if the vertices
$x, \, y\,$ appearing in the bound are integrated over. This is the
case for our definition of weight factors since in particular internal
vertices are integrated over. For the first contribution on
the r.h.s. of  (\ref{fequ}) we would like to pass from a tree
associated with ${\cal L}_{n+2}$ to a tree associated with ${\cal L}_{n}\,$.
This requires the bound to be such that its integration 
over $x, \, y $ against the factor  $C_t(x,y)\,$ finally leads to
an expression bounded by a tree bound on  ${\cal L}_{n}\,$. 
This is at the origin of the notion of reduced tree introduced above,
where two external points have disappeared.  

For simplicity we choose the test functions appearing in the weight 
factors to be heat kernels themselves. These form a
sufficiently large set. However, to get inductive control of the local 
counter terms, we also have to admit the situation where some of the  
external coordinates are just integrated over all of $\cal M\,$. This
leads to the general form of the admitted test functions 
(\ref{phic}).

%%%%%%%%%%%%%%%%%%%%%%%%%%%%%%%%%%%%%%%%%%%%%%%%%%%%%%%%%%%%
\section{Boundary and renormalization conditions}

From the mathematical point of view the renormalization 
problem in the FE framework appears as a mixed boundary value
problem. The relevant terms are fixed by renorma\-lization conditions at
a large value \,$t_R$\, of the flow parameter $t$, 
 all other boundary terms are fixed at the short-distance 
 cutoff $ t=\vep $.\\[.2cm]
To extract the relevant terms - contained in 
 ${\cal L}^{\vep,t}_{2,l}(x_1,\vp)\,$  and
 ${\cal L}^{\vep,t}_{l,4}(x_1,\vp)\,$-
 a covariant Taylor expansion
with remainder term (\ref{sloe}), (\ref{Rn})
of the test function $ \, \vp \,$ is used, $\vep \le t$~: 
\eq
{\cal L}^{\vep,t}_{2,l}(x_1,\vp)\,=\,
  a^{\vep,t}_l(x_1)\ \vp(x_1)
\,  - \, f^{\mu,\vep,t}_l(x_1)\ (\nabla _{\mu} \vp)(x_1)
\,  - \,
   b^{\,\mu \nu,\vep,t}_l(x_1) (\nabla_\mu \nabla_\nu  \vp)(x_1)  
\,+\,  
{\ell}^{\vep,t}_{2,l}\ (x_1, \vp)\ ,
\label{2l}
\eqe
\eq
{\cal L}^{\vep,t}_{4,l}(x_1,\vp)=\
 c^{\vep,t}_l (x_1)\ \vp_2(x_1)\  \vp_3(x_1) \vp_4(x_1) + \ 
{\ell}^{\vep,t}_{4,l} (x_1, \vp)\,.
\label{4l}
\eqe   Then the relevant terms appear as
$$
a^{\vep,t}_l(x_1) = \int_{x_2}\! {\cal L}^{\vep,t}_{2,l} (x_1,x_2)\ ,\quad
f ^{\mu,\vep,t}_l(x_1) =
\int_{x_2}\! \si(x_2,x_1)^{\mu} \  
 {\cal L}^{\vep,t}_{2,l} (x_1,x_2)\ ,  
$$
\eq
b^{\mu\nu,\vep,t}_l(x_1)=-\,
\frac12 \, \int_{x_2}\! \si(x_2,x_1)^{\mu} \  \si(x_2,x_1)^{\nu} 
   {\cal L}^{\vep,t}_{2,l} (x_1,x_2) \, ,
\label{coeff}
\eqe
\eq
c^{\vep,t}_l(x_1)=
\int_{x_2,x_3,x_4}\!\!\! 
{\cal L}^{\vep,t}_{4,l} (x_1, \ldots  ,x_4)\, ,
\label{coeff4}
\eqe
and the `remainders' ${\ell}^{\vep,t}_{2,l}$ and ${\ell}^{\vep,t}_{4,l}$
have the respective forms
\eq
 {\ell}^{\vep,t}_{2,l} (x_1, \vp)\ =\ \int_{x_2} 
{\cal L}^{\vep,t}_{2,l} (x_1,x_2)
\int_{0}^{s} dr\, \frac{ (s-r)^2}{2!} \,\,
 {\dot x}_{12}^{\nu_{3}} (r)\, {\dot x}_{12}^{\nu_{2}} (r) 
         \, {\dot x}_{12}^{\nu_1} (r) 
        \big( \nabla_{\nu_{3}}  \nabla_{\nu_{2}} 
\label{re2}         \nabla_{\nu_{1}}\vp \big)(x_{12}(r)) 
\eqe
where $ s = d(x_1, x_2)$\, and $x_{12}(r)\, $ 
is the point on the geodesic segment from
 $x_1$ to $x_2\,$ at arc length $r\,$; and
\[
{\ell}^{\vep,t}_{4,l} (x_1, \vp)\ =\ \int_{x_2,x_3,x_4} 
{\cal L}^{\vep,t}_{4,l} (x_1,\ldots,x_4)
\Bigl[ \int_{0}^{s_{12}} dr\, \,\,
       {\dot x}^{\nu}_{12} (r)\  \big( \nabla_{\nu}
        \vp_2\big)(x_{12}(r)) \ \vp_3(x_3)\vp_4(x_4) \  +\
\]
\[
 \vp_2(x_1)  \int_{0}^{s_{13}} dr\, \,\,
      {\dot x}^{\nu}_{13} (r)\  \big( \nabla_{\nu}
        \vp_3\big)(x_{13}(r)) \ \vp_4(x_4) \ +
\]
\eq
  \vp_2(x_1) \   \vp_3(x_1)     \int_{0}^{s_{14}} dr\, \,\,
    {\dot x}^{\nu}_{14} (r)\  \big( \nabla_{\nu}
        \vp_4\big)(x_{14}(r))\Bigr]\, .
\label{re4}
\eqe 
Reparametrizing the geodesic segment $ x_{12}(r) = X(\rho),\,
r = d(x_1, x_2) \rho\, ,\, 0 \leq \rho \leq 1 $, we can
 rewrite the remainder (\ref{re2})  
employing (\ref{Rna})
 $$   {\ell}^{\vep,t}_{2,l} (x_1, \vp) = \int_{x_2}\!\!\!
 d^{\,3}(x_1, x_2) {\cal L}^{\vep,t}_{2,l} (x_1,x_2)
  \int_{0}^{1}\!\!\! d\rho \frac{ (1-\rho)^{\,2}}{2 !\,d^{3}(x_1, x_2)}  
   {\dot X}^{\nu_{3}}(\rho)\, {\dot X}^{\nu_{2}}(\rho) 
         \, {\dot X}^{\nu_1}(\rho)
    \,  \om^{(3)}_{{\nu}_3 {\nu}_2 {\nu}_1}(X(\rho))   $$                    
\eq
 \mbox{where}\qquad  \om^{(3)}_{{\nu}_3 {\nu}_2 {\nu}_1}(x) \, = \, 
    \big( \nabla_{\nu_{3}}  \nabla_{\nu_{2}} 
        \nabla_{\nu_{1}}\vp \big)(x)\ . 
 \label{re22} 
\eqe        
1) {\bf Boundary conditions} at $\,t =\vep\,$~:\\
 The bare interaction (\ref{nawig}) 
implies that at   $t =\vep\ $  
- with $ {\cal L}^{\vep}\equiv  {\cal L}^{\vep,\vep}$ -
\eq
{\cal L}^{\vep}_{n,l} (x_1,\ldots x_{n}) \equiv 0\ 
\mbox{ for}\quad n >4\ , \quad {\cal L}^{\vep}_{2,0} \equiv 0
\label{bo1}
\eqe
\eq 
{\cal L}^{\vep}_{2,l} (x_1,x_2) 
=  {\tilde a}_l^{\epsilon}(x_1)\, \tilde{\delta}(x_2,x_1)
- \ \Delta^{(b)}_2 \, \tilde{\delta}(x_2,x_1)\,,
 \,b =  b_l^{\,\mu \nu,\, \vep} (x_2) \,, 
\label{bo2}
\eqe
\eq
{\cal L}^{\vep}_{4,l} (x_1,\ldots x_{4}) =
({ \de}_{l,0}\ \la(x_1) \,+\,(1-{ \de}_{l,0})\,c_l^{\vep}(x_1))\
{\ti \de}(x_2,x_1)\,{\ti \de}(x_3,x_1)\,{\ti \de}(x_4,x_1)\ .
\label{bo3}
\eqe
To cope with the relevant part of the expansion (\ref{2l})
we consider a corresponding bare part
\eq
{\cal L}^{\,\vep}_{2,\,l}(x_1,\vp)\,=\,
  a^{\,\vep}_l(x_1)\ \vp(x_1)
\,  - \, f^{\mu,\,\vep}_l(x_1)\ (\nabla _{\mu} \vp)(x_1)
\,  - \, b^{\,\mu \nu,\,\vep}_l(x_1) (\nabla_\mu \nabla_\nu  \vp)(x_1)\,.  
\label{baex}
\eqe
The identity
\eq
 -\, b^{\,\mu \nu}_l(x) (\nabla_\mu \nabla_\nu  \vp)(x)  
\,= \, \nabla_\nu\, b^{\,\mu \nu}_l (x)\cdot
  \big (\nabla_\mu \vp\big )(x)\, -\, 
    \Delta ^{(b)} \vp  (x)\,, 
   \quad b = b^{\,\mu \nu}_l(x)
\label{cov2}
\eqe    
suggests to decompose the bare vector coefficient 
appearing in (\ref{baex}) as
\eq
 f^{\mu,\,\vep}_l(x_1)\, = \, {\tilde f}^{\mu,\,\vep}_l(x_1)\, 
   + \, \nabla_\nu\, b^{\,\mu \nu,\,\vep}_l (x_1)\, .
\label{bave}
\eqe
By folding (\ref{baex}) with a test function $ \vp $
we obtain after partial integration \footnote{\,Here  $ \vp \,$
is assumed to be smooth, and to decrease sufficiently rapidly 
if $ \cal M $ is noncompact. Apart from the present consideration
and from the general analysis of the effective action  after
(\ref{funcin}),  we do not introduce test functions 
against which the first (root) vertex  is integrated.} 
\eq
\int_{x} \,\vp(x)\,{\cal L}^{\,\vep}_{2,\,l}(x,\vp)\,=\,   
\int_{x}\Big \{ \big ( a^{\vep}_l(x)+ \frac{1}{2}\,
     \nabla_{\mu} {\tilde f}^{\mu,\,\vep}_l(x)\big )\, \vp^2(x) + 
    \, b^{\,\mu \nu, \vep}_l(x)\  
 \partial_{\mu}\vp(x)\cdot\partial_{\nu} \vp(x) \Big \} \,.
\label{baac}
\eqe 
This agrees in form with the corresponding content of 
the bare action (\ref{nawig}).
From the boundary conditions  (\ref{bo1})-(\ref{bo3}) 
we deduce
\eq
{\ell}^{\vep,\vep}_{2,l}\  (x_1,\,\vp)\ =\  0\ ,\quad
{\ell}^{\vep,\vep}_{4,l}\  (x_1,\,\vp)\ =\  0\ .      
\label{bell}
\eqe
    
The renormalization problem is related to the behaviour of the heat
kernel at small values of $t$. Therefore this problem is
essentially solved if we can integrate the flow equations up to some
finite value $t_R$ of $t$. For shortness we choose units such that 
$t_R =1$. We will come to the limit $t\to \infty\,$ later, see
Proposition 3. The positive mass $m>0\,$ only plays a role when this limit
is taken. We pose\\
2)  {\bf Renormalization conditions} at $t = t_R := 1\,$ ~:
\footnote{The scale $t_R\,$ is related to the scale $T$ appearing in
  the bounds on the heat kernel (\ref{hk10}), (\ref{d}), (\ref{D}) .}   
\eq
a ^{\vep, 1}_l(x_1):=\, a ^R_l(x_1) ,\quad
f^{\mu,\vep, 1}_l(x_1):=\, f^{\mu,R}_l(x_1),\quad
 b^{\mu\nu, \vep, 1}_l(x_1):=\, b^{\mu\nu, R}_l(x_1)\, ,
\label{renbed}
\eqe
\eq
c ^{\vep, 1}_l(x_1) :=\, c ^R_l(x_1)\, ,\qquad
\label{renbed4}
\eqe
where $b^{\mu\nu, R}_l(x)\,$ is  a smooth 
 symmetric tensor of type
$(2,0)\,$, $f^{\mu, R}_l(x)\,$ is  a smooth vector
 and $a^R_l(x)\,,\ c ^R_l(x)\,$ are smooth  scalars on 
$\cal M\,$, all uniformly bounded in the norm (\ref{nor}).
Typically the renormalization conditions  are assumed
 to be cutoff-independent.
To be able to analyse the relation between the bare (inter)action and the
renormalization conditions in more detail later on,
 we shall be more general in also 
admitting weakly $\vep$-dependent renormalization functions
satisfying
\eq
|\pa_{\vep}\,a_l^R(x)| \,< \,
O(\vep^{-\eta}) \  ,\quad
|a_l^R(x)| \,< \, const\,+\,
O(\vep^{1-\eta})\ , \quad \eta \le 1/2\ ,
\label{vepdep}
\eqe
with analogous expressions for the other renormalization functions.

In the particular case of $\, {\cal M}\,$ having constant curvature,
i.e. where all sectional curvatures of  $\, {\cal M}\,$ have a
 constant value $\, \rho \, $, a transitive isometry group
 $ \, G \,$ acts on  $\, {\cal M}\,$. There are three types of
 such manifolds: The sphere $\, {\cal S}^4 \,$ with $\, \rho = k^2 $
 \, and \,$G = SO(5)$\,, the flat space  $ {\mathbf R}^4\,$
 with $\, \rho = 0 $ \,and \,$G = SO(4)\otimes_s {\mathbf R}^4 $\,,
 the hyperbolic space \,$ {\cal H}^4$\, with  $\, \rho = - k^2 $\, 
 and \, $ G = SO_0(4,1)$,\, the subscript denoting the component
 connected to the identity. \\
Requiring the Schwinger functions to show this symmetry $\,G\,$,
results in the following restrictions on the relevant terms: \\
i)$\quad a^{\vep,t}_l(x), \,  c^{\vep,t}_l (x)\quad $ do not
   depend on\, $ x \in {\cal M} $,  \\  
ii)$\quad f^{\mu,\vep,t}_l(x)\equiv 0\, , \quad 
   b^{\,\mu \nu,\vep,t}_l(x) = g^{\,\mu \nu}(x)\, 
      b^{\,\vep,t}_l $\,,\,\, hence 
 \, $\nabla_{\nu}\, b^{\,\mu \nu,\vep,t}_l(x)\equiv 0 $. \\ 
However, there is a further (dimensionless) parameter
 \, $ \zeta = k^2/ m^2 $\, on which 
$a^{\vep,t}_l,\ b^{\vep,t}_l,\ c^{\vep,t}_l\, $ may
depend,  in general.
   
%%%%%%%%%%%%%%%%%%%%%%%%%%%%%%%%%%%%%%%%%%%%%%%%%%%%%%%%%%%%%%%
\section{Renormalizability}

The subsequent proposition is proven for test functions 
of the form
$\,\vp_{\tau_{2,s},y_{2,s}}( x_{2,n})$, (\ref{phic}).
In the end of this section we join some remarks on possible
extensions of the class of test functions.
By Bose symmetry the bounds stay unaltered if any permuted
subset of external coordinates (and not $\,x_2, \ldots, x_s\,$)
is folded with test functions. 
\\
\noindent
{\bf Proposition 1}: \\ 
{\it 
We consider $0 < \vep \le t \le 1\,$
 and $\vep < \tau_i \,$, furthermore $1 \le s  \le n\,$,  $2 \le i  \le
 n\,$, $2 \le j \le s \,$  and $0 \le r \le 3\,$.
We consider test functions either of the form
$\vp_{\tau_{2,s},y_{2,s}}( x_{2,n})\,$
or $\vp^{(j)}_{\tau_{2,s},y_{2,s}}( x_{2,n})\,$, which are also
denoted 
in shorthand as $\vp_s \,$ resp. $\vp^{(j)}_s\,$.\\
In all subsequent bounds  we understand ${\cal P}_{l}$ to denote 
a polynomial of degree $\le \sup(l,0)\,$ - each time it appears
possibly a new one - with nonnegative coefficients which 
may depend on $l,n,\de$ \footnote{We suppose that
$\de>0$ may be chosen arbitrarily small in the definition of $\cal F$.
The constants in ${\cal P}_{l}$ then depend on the choice of $\de$.},
on $\sup_{\cal M}|\la(x)|\,$,  
as well as on $k^2\,$, $\ka ^2\,$ 
and the bounds on the first and second
 covariant derivatives of the curvature tensor (see (\ref{hk10}) -
(\ref{D})), {\it but not on $\,\vep,\,t,\,m\, $} and $\tau$. 
Constants $O(1)\,$ in the subsequent proof
are to be understood in the same 
way. By $(t,\tau)\,$ we denote $\inf \{\tau_2,\ldots,\tau_s, t\}$, by 
  $(t,\tau)_{i}\,$ we denote 
$\inf \{\tau_2,\ldots, \not\!\! \tau_i,\ldots, \tau_s, t\}$.

\noindent
Then we claim the following bounds - using the shorthand (\ref{ges}) -
\eq
|\,{\cal L}^{\vep,t}_{n,l} (x_1,\vp_{\tau,y_{2,s}})|\,\le\ 
t^{\frac{n-4}{2}}\
{\cal P}_l\log(t,\tau)^{-1}\ 
{\cal F}_{s,l}(t,\tau) \ , \quad n \ge 4  
\label{prop1}
\eqe
\eq
|\, {\cal L}^{\vep,t}_{n,l} (x_1, E_{(i)}^{(r)} \,\vp_{\tau,y_{2,s}})|\,
\le\ |\,\om^{(r)}(x_1) | \ t^{\frac{n+r-4}{2}}\
{\cal P}_l\log(t,\tau)_{i}^{-1}\ 
{\cal F}_{s,l}(t,\tau) \ , \quad n > 4, \ r>0  
\label{prop1r}
\eqe
\eq
|\, {\cal L}^{\vep,t}_{n,l} (x_1, E_{(i)}^{(r)} \vp_{\tau,y_{2,s}})|\,
\le\  |\,\om^{(r)}(x_1) |  \  t^{\frac{n+r-4}{2}}\
{\cal P}_{l-1}\log(t,\tau)_{i}^{-1}\ 
{\cal F}_{s,l}(t,\tau) 
\label{prop3}
\eqe
\centerline{$\, n=2, r = 3\,$ or $\, n=4,\ r>0\,$ }  
\eq
|\, {\cal L}^{\vep,t}_{2,l} (x_1,\vp_{{\tau},y_{2}})|\,\le\ 
(t,\tau)^{-1}\
{\cal P}_{l-1}\log (t,\tau)^{-1}\  {\cal F}_{2,l}(t,\tau)
\label{prop20}
\eqe
\eq
|\, {\cal L}^{\vep,t}_{2,l} (x_1,E_{(2)}\, \vp_{{\tau},y_{2}})|\,
\le\ \nom  \ (t , \tau)^{-1/2}\
{\cal P}_{l-1}\log (t,\tau)^{-1}\  {\cal F}_{2,l}(t,\tau)
\label{prop21}
\eqe
\eq
|\, {\cal L}^{\vep,t}_{2,l} (x_1,E_{(2)}^{(2)}\, \vp_{{\tau},y_{2}})|\,
\le\  |\,\om^{(2)}(x_1) | \ {\cal P}_{l-1}\log (t,\tau)^{-1}\ 
 {\cal F}_{2,l}(t,\tau)
\label{prop22}
\eqe
\eq
|\,{\cal L}^{\vep,t}_{n,l} (x_1,\vp^{(j)}_{\tau,y_{2,s}})|\,\le\ 
t^{\frac{n-4}{2}}\ \bigg(\frac{t}{\tau_j}\bigg)^{1/2}\
{\cal P}_l\log(t,\tau)^{-1}\ 
{\cal F}_{s,l}(t,\tau)   \ ,\quad n>2
\label{prop5}
\eqe
\eq
|\, {\cal L}^{\vep,t}_{2,l} (x_1,\vp^{(2)}_{{\tau},
  y_{2}})|\,\le\ 
 (\frac{t}{\tau})^{1/2} (t, \tau)^{-1}\
{\cal P}_{l-1}\log(t,\tau)^{-1}\  {\cal F}_{2,l}(t,\tau)
\label{prop2j}
\eqe
\eq
|\,F^{(r)}_{(12)}{\cal L}^{\vep,t}_{n,l}(x_1, x_2, \vp)| \,\le\
 |\,\om^{(r)}(x_1) | \ t^{\frac{n-1}{2}}\ {\cal P}_{l-1}\log(t,\tau)^{-1}\  
 {\cal F}^{(12)}_{s,l}(t,\tau)\ . 
\label{prop2r}
\eqe
$$   r = 0,1,2 \quad and  \, \,  |\,\om^{(0)}(x_1) | \equiv 1 \, .$$
}
{\sl Remark~:} The full series of the previous bounds is needed to
close the inductive argument in the subsequent proof. The reader who
only wants to know what the bounds are can restrict to (\ref{prop1}), 
(\ref{prop20}).

\noindent
{\it Proof}~:
The bounds  stated in the proposition are proven inductively
using the (standard) inductive scheme which proceeds upwards in $n+2l$,
and for given $n+2l$ upwards in $l$.\\ 
Thus the induction starts with the pair $(4,0)\,$. For this term
the r.h.s. of the FE vanishes so that
${\cL}^{\vep,t}_{4,0}(x_1,\ldots,x_4) = \la(x_1)\,{\ti \de}(x_2,x_1)\,
{\ti \de}(x_3,x_1)\,{\ti \de}(x_4,x_1)\, $ which is compatible with
our bounds (after folding with suitable $\vp\,$). Generally 
it is important to note that the boundary conditions are
 compatible with the bounds of the proposition.\\
We  will first derive bounds for the derivatives 
$\pa_t\, {\cal L}^{\vep,t}_{n,l} (x_1, E_{(i)}^{(r)} \vp_s)\,$,
  where $\, E_{(i)}^{(0)}\equiv 1 $, and 
$\pa_{t}\, {\cal L}^{\vep,t}_{n,l} (x_1,\vp^{(j)}_s)\,$. 
Afterwards these bounds are integrated over w.r.t. $t$. \\
A) We start  considering the case $r=0\,$ and test functions
$\vp_s\,$.\\ 
A1) Here we first treat the first term on the r.h.s. of 
(\ref{fequ})
\[
  R_1 := \, \int_{x_{2,n},  x,y}
 \vp_s( x_{2,n})\  C_{t}(x,y)\ 
{\cL}^{\vep,t}_{n+2,l-1}(\xv_n,x,y) \ .
\] 
We may rewrite this expression as
\[
  R_1 =  \int_v \int_{x_2,\ldots,x_n,  x,y}
\vp_s(x_{2,n})\   C_{t/2}(x,v)\  C_{t/2}(v,y) \
{\cL}^{\vep,t}_{n+2,l-1}(\xv_n,x,y)\ =\
\]
\[ 
 \int_v {\cL}^{\vep,t}_{n+2,l-1}(x_1,\vp_s  
\times C_{t/2}(\cdot,v)\times C_{t/2}(v,\cdot))\ .
\]
Applying the induction hypothesis to 
${\cL}^{\vep,t}_{n+2,l-1}(x_1,\vp_s \times  
C_{t/2}(\cdot,v)\times C_{t/2}(v,\cdot))\,$
we thus obtain the bound 
\eq
 |R_1| \le t^{\frac{n+2-4}{2}} \ {\cal P}_{l-1}\log(t,\tau)^{-1}
\int_v \int_{\vec z}
\sum_{T_{l-1}^{s+2}(x_1,y_{2,s},v,v)}
{\cal F}(t,\tau,\frac t2,\frac t2;
T_{l-1}^{s+2}(x_1,y_{2,s},v,v,\vec z))\ .
\label{1st1}
\eqe
For any contribution to 
(\ref{1st1}) we denote by $z',\ z''\,$ the vertices in
 the respective tree $\,T_{l-1}^{s+2}( x_1, y_{2,s},v,v,\vec z)$
to which the  test functions
$  C_{t/2}(\cdot,v),\   C_{t/2}(v,\cdot)\,$ are attached.
Interchanging $\int_{\vec z}$ (see (\ref{ff}))
 and $\int_v\,$ and 
performing the integral over $v\,$ using
 (\ref{hk4}), (\ref{dpropa}), we get a contribution
\eq
\int_{v}  
C_{t/2}(z',v)\ C_{t/2}(v,z'')\ = \
\ C_t (z',z'')\  \le \ O(1)\  t^{-2}\ .
\label{rom}
\eqe
Using this bound we can majorize 
$\,\int_v 
{\cal F}(t,\tau,\frac t2,\frac t2;
T_{l-1}^{s+2}~;x_1,y_{2,s},v,v)\,$ by
\[
O(1)\ t^{-2}\  
{\cal F}(t,\tau;T_{l}^{s}~;x_1,y_{2,s})
\]
where the tree $\,T_{l}^{s}\,$ is the 
{\it reduced tree}
obtained from $\,T_{l-1}^{s+2}\,$ by taking away the two external
lines ending in $v\,$, see Definition 4.
 The reduction process for each tree fixes uniquely
the set of internal vertices  of $\,T_{l}^{s}\,$ 
in terms of those of 
 $\,T_{l-1}^{s+2}\,$. 
Note that the elimination of 
vertices of incidence number 1 together with their adjacent
line is justified by the fact that $\int_{z'}  C_{t_I,\de_l}(z,z') \le 1\,$.
Note also that in the tree
$T^{s}_l\,$ the number $v_2\,$ of vertices of incidence number
$2$  may have increased by at most $2\,$, 
as compared to $T_{l-1}^{s+2}\,$,
so that $T^{s}_l\,$  is indeed an element from  $\,{\cal T}_l^{s}\,$.
We keep track of this lower index $l$ in the
tree basically to show that the number of vertices always stays
finite (in fact does not grow faster than linearly in $l\,$ for $n$ fixed).\\
The final bound for the first term on the r.h.s. of the FE is thus 
\eq
 |R_1| \le t^{\frac{n-6}{2}}\ 
\ {\cal P}_{l-1}\log(t,\tau)^{-1}\ 
\sum_{T_{l}^{s}(x_1,y_{2,s})}
{\cal F}(t,\tau;T_{l}^{s}~;x_1,y_{2,s})
\label{1st}
\eqe  
where constants have been absorbed in $\,{\cal P}_{l-1}\log\,$.\\

\noindent
A2) We now  consider the second term in (\ref{fequ}) for\\ 
i) $n > 4 $\\
Picking a generic term
from the  symmetrized sum and arguing as in A1) 
we have to bound 
\[
 R_2 := \int_{x_{2,n}, x,y} \vp_s(x_{2,n})\  
  C_{t}(x,y)\ 
{\cL}^{\vep,t}_{n_1+1,l_1}(x_1,\ldots,x_{n_1},x)
\
{\cL}^{\vep,t}_{n_2+1,l_2}(y,x_{n_1+1},\ldots,x_{n})
\]
which we rewrite similarly as in A1)
\eq
R_2 = \int_v \int_{x_2,\ldots,x_n,  x,y} \vp_s(x_{2,n})\  
  C_{t/2}(x,v)\   C_{t/2}(v,y)\ 
{\cL}^{\vep,t}_{n_1+1,l_1}(x_1,\ldots,x)
\
{\cL}^{\vep,t}_{n_2+1,l_2}(y,\ldots,x_{n})\ .
\label{nuj}
\eqe
Denoting
\[
\vp'_{s_1}(x_{2,n_1})= \prod^{n_1}_{r=2}\vp_r(x_r)\ , \quad
\vp_{s_2}''(x_{n_1+1,n-1})= \prod_{r= n_1+1}^{n-1} \vp_r(x_{r}) 
\]
we identify the two terms
\[
\int_{x_2,\ldots,x_{n_1},  x} \vp'_{s_1}(x_{2,n_1})\  
  C_{t/2}(x,v)\   
{\cL}^{\vep,t}_{n_1+1,l_1}(x_1,\ldots,x)
\]
and 
\[
\int_{x_{n_1+1},\ldots,x_n, y} \vp_{s_2}''(x_{n_1+1,n-1})\  
 C_{t/2}(v,y)\ 
{\cL}^{\vep,t}_{n_2+1,l_2}(y,\ldots,x_{n})
\]
and thus write (\ref{nuj}) as
\eq
 R_2 =   \int_{x_n} \int_{v}
{\cL}^{\vep,t}_{n_1+1,l_1}(x_1,\vp'_{s_1}\times  
C_{t/2}(\cdot ,v))  
\   
{\cL}^{\vep,t}_{n_2+1,l_2}(x_n,\,C_{t/2}(v,\cdot)\times \vp_{s_2}'' )
\ \, \vp_n(x_n)\ .
\label{bose2}
\eqe
On applying the
induction hypothesis to both terms in (\ref{bose2}), restricting  
first to $ n_1, n_2 > 1,$ we obtain
the bound
\[
 |R_2| \le \, t^{\frac{n+2-8 }{2}}\ 
{\cal P}_{l_1}\log(t,\tau)^{-1}\ 
 \int_{x_n}\int_v 
\sum_{T_{l_1}^{s_1+1},\
  T_{l_2}^{s_2+1}}
{\cal F}(t,\tau',t/2; T_{l_1}^{s_1+1}~;x_1,y_2,\ldots,y_{s_1},v)\,  \cdot 
\]
\eq 
 \cdot
\ {\cal P}_{l_2}\log(t,\tau)^{-1}\ 
{\cal F}(t,\tau'',t/2;T_{l_2}^{s_2+1}~;x_n, v,y_{s_1+1},\ldots
\ldots, y_{s(n)}) \ \, \vp_n(x_n) \ .
\label{gensi}
\eqe
Here $s(n)=s$ if $s<n$, and  $s(n)=s-1$ if $s=n$. 
Interchanging the integral over $v\,$
with the sum over trees we obtain
\eq
   |R_2| \le \, t^{\frac{n -6 }{2}}\  
\ {\cal P}_{l}\log(t,\tau)^{-1} 
\sum_{T_l^{s}(T_{l_1}^{s_1+1},\
  T_{l_2}^{s_2+1})(x_1,y_2\ldots,y_{s})} \int_{x_n}
{\cal F}(t,\tau_{2,s};T_l^{s}~;x_1,y_2\ldots,y_{s})
\label{2nd}
\eqe
with the following explanations~:\\
Any contribution in the sum over trees $\,T_l^{s}(T_{l_1}^{s_1+1},\
  T_{l_2}^{s_2+1})(x_1,y_2,\ldots,y_{s},  \vec z)\,$
is obtained from  
$\,T_{l_1}^{s_1+1}(x_1,y_2\ldots,y_{s_1},v ,  \vec z^{\,'})\,$ and
$\,T_{l_2}^{s_2+1}(x_n ,v,y_{s_1+1},\ldots,y_{s},  \vec z^{\,''})\,$
by joining these  two  trees via the lines going from the vertices
$z'$ and $z''$ to $ v$, where $z'$ and $z''$ are
 the vertices attached to  $v$ in
the two trees. These two lines have parameters $t/2\,$. 
We use the equality 
\eq
\int_v C_{t/2}(z',v)\  C_{t/2}(v,z'')\ =\ C_{t}(z',z'')
\label{falt}
\eqe
so that the new internal line has 
a $t$-parameter in the interval $[\vep,t]\,$ over which the $\sup$
is taken in the definition of $\cal F$. \footnote{We can of course majorize
$ C_{t}(z',z'') \le O(1)\  C_{t_{\de}}(z',z'')$.}
\\
 When performing the integral over $x_n$ in
(\ref{gensi}) we remember that $x_n\,$ is the root vertex of
$T_{l_2}^{s_2+1}(x_n, v,y_{s_1+1},\ldots \ldots, y_{s},\vec z )\,$.
If $s=n$ we have $\vp_n(x_n)= C_{\tau_n}(x_n,y_n)\,$, and 
$x_n$ becomes an internal vertex,
 and $y_n$ an
external vertex, of $T_l^s\,$.
If $s<n$, then  $\vp_n(x_n)\equiv 1\,$, and the vertex $x_n\,$
becomes an internal vertex of $T_l^s\,$
unless $c(x_n)=1\,$. In this last case 
integration over $x_n$ together with (\ref{hk3}) permits to take away
the vertex $x_n$ and the internal line joining it to
an internal vertex $ \,z_j\,$ of the tree $ T_{l_2}^{s_2+1}\,$
\footnote{If $x_n$ has turned into a
  vertex of incidence number 2 for $T_l^s\,$,
the bound $v_2 +\de_{c_1,1} \le
 3l-2+s/2\,$ is easily verified.}. If (originally) 
$ \, c(z_j) = 2 \,$ this elimination process continues.
Thus the final bound for $R_2$, and hence for the second
 term in (\ref{fequ}) is the same
as (\ref{1st}), apart from
  changing $ {\cal P}_{l-1}\log(t,\tau)^{-1} \to 
 {\cal P}_{l}\log(t,\tau)^{-1}\,$.
Note that this bound is established in the same way if
${\cL}^{\vep,t}_{n_1+1,l_1}(x_1,\ldots,x )$ or 
${\cL}^{\vep,t}_{n_2+1,l_2}(y,\ldots,x_{n})$ are two-point
functions~: In this case the parameter $\tau$ appearing in
(\ref{prop20}) equals $t/2$ so that
 $(t, \tau)^{-1}\,$ can be replaced
by $2/t\,$.

\noindent
Taking both contributions from the r.h.s. of the FE together
and summing over all trees we have established the bounds
\eq
|\pa_t  {\cal L}^{\vep,t}_{n,l} (x_1,\vp_s)| 
\ \le \ t^{\frac{n-6}{2}} \ {\cal P}_{l}\log  (t,\tau)^{-1}
\ {\cal F}_{s,l}(t,\tau)
\ , \quad n> 4 \ .
\label{formeln1}
\eqe

\noindent
ii) $n\le 4$ \\ 
In this case we have $\,n_1+1 =2\,$ and/or $\,n_2+1 =2\,$.
Thus at least one of the polynomials 
$\, {\cal P}_{l_i}\log(t,\tau)^{-1}\,$ 
appearing in the bounds (\ref{gensi}) 
can by induction be replaced by
 $\, {\cal P}_{l_i-1}\log t^{-1}\,$.  
Therefore proceeding exactly as in the previous case 
we obtain the bounds 
\eq
|\pa_t {\cal L}_{n,l}^{\vep,t}(x_1,\vp_s)| 
\ \le \  t^{\frac{n-6}{2}}\ {\cal P}_{l-1}\log   (t,\tau)^{-1}\  
{\cal F}_{s,l}(t,\tau)
\ , \quad n \le 4 \ .
\label{formeln2}
\eqe
\\
\noindent
B) $\,r >0\,, \,$ cf.(\ref{ins})\\
For the first term on the r.h.s. of the flow
equation resulting from (\ref{fequ}) the bounds for $1 \le r \le 3\,$
 are proven exactly as in A1).  
For the second term  we proceed similarly  as in A2). 
We pick a
generic term on the r.h.s. 
\[
 \int_{x_{2,n}, x,y} \vp_s(x_{2,n})\  
  C_{t}(x,y)\  E(x_k,x_1~; \om^{(r)}) 
{\cL}^{\vep,t}_{n_1+1,l_1}(x_1,\ldots,x_{n_1},x)
\
{\cL}^{\vep,t}_{n_2+1,l_2}(y,x_{n_1+1},\ldots,x_{n})\ .
\]
In the case where $k \le n_1\,$ 
the proof is the same as for $r=0$, up to
inserting the modified induction hypothesis for
\[
 \int_{x_{2,n_1}, x} \vp_{s_1}(x_{2,n_1})\  
  C_{t/2}(x,v)\ E(x_k,x_1~; \om^{(r)}) 
{\cL}^{\vep,t}_{n_1+1,l_1}(x_1,\ldots,x_{n_1},x)\ .
\]
If $k > n_1\,$
we assume without restriction $k = n\,$ and proceed 
 again as in A2) to obtain the bound
\[
t^{\frac{n-6 }{2}}\ 
{\cal P}_{l_1}\log(t,\tau)_{n}^{-1}\ 
 \int_{x_n}\int_v | E(x_n,x_1~; \om^{(r)})| \ 
\sum_{T_{l_1}^{s_1+1},\
  T_{l_2}^{s_2+1}}
{\cal F}(t,\tau,t/2; T_{l_1}^{s_1+1}~;x_1,y_2,\ldots,y_{s_1},v)\ 
\]
\eq 
 \cdot
\ {\cal P}_{l_2}\log(t,\tau)_{n}^{-1}\ 
{\cal F}(t,\tau,t/2;T_{l_2}^{s_2+1}~;x_n, v,y_{s_1+1},\ldots
\ldots, y_{s(n)}) \ \vp_n(x_n)\ .
\label{gensis}
\eqe
Observing the inequality (\ref{nino}) together with
\eq
d(x_n,x_1) \le \ \sum_{a=1}^q d(v_a,v_{a-1})
\label{sid}
\eqe
where $\{v_a\}\,$ are the positions of the internal 
vertices in the tree
$T_l^{s}(T_{l_1}^{s_1+1},\,   T_{l_2}^{s_2+1})$ defined as in A2),
on the path joining $x_1=v_0$ and $x_n = v_q\,$, we then use
the bound (\ref{d}).  
The cases $s=n$ and $s<n$ are treated as in A2), using once more
 the bound (\ref{d}).\\
The previous reasoning holds a fortiori for $\pa_t\
 F^{(r)}_{(12)}{\cal L}^{\vep,t}_{n,l}(x_1, x_2, \vp)$\,,
since in these cases we have $ |\, F^{(r)}_{(12)}\,|
  \leq d^{\,3}(x_1, x_2)  |\, \om^{(r)}(x_1) | \,,\,
 |\, \om^{(0)}(x_1)\,| \equiv 1 .$ 
 Here then $x_2$ takes the role of $x_n\, $.

\noindent
Proceeding as before we thus obtain for $r \neq 0\,$
(after absorbing again  all constants in ${\cal P}_l$)
\eq
| \pa_t\
 {\cal L}^{\vep,t}_{n,l} (x_1, E_{(i)}^{(r)}\, \vp_s)|\,\le\ 
  | \om^{(r)}(x_1) | \,
  t^{\frac{n+r-6}{2}}\ {\cal P}_l\log  (t,\tau)_i^{-1} \
 {\cal F}_{s,l }(t,\tau)\ ,\quad
 n > 4
\label{bdprop2}
\eqe
\eq
|\pa_t \ {\cal L}^{\vep,t}_{4,l}(x_1, E_{(i)}^{(r)}\, \vp_s)|\,\le\ 
  | \om^{(r)}(x_1) | \, t^{\frac{r-2}{2}}\
 {\cal P}_{l-1}\log(t,\tau)_{i}^{-1}\
 {\cal F}_{s,l }(t,\tau) 
\label{bdprop3}
\eqe
\eq
|\pa_t \ {\cal L}^{\vep,t}_{2,l}(x_1, E_{(2)}^{(r)}\ \vp_2)|\,\le\ 
   \, |\,\om^{(r)}(x_1) | \   t^{\frac{r-4}{2}}\ {\cal P}_{l-2}\log t^{-1}\
 {\cal F}_{2,l }(t,\tau)
\label{bdprop22}
\eqe
\eq
|\,\pa_t \ F^{(r)}_{(12)} {\cal L}^{\vep,t}_{n,l}(x_1, x_2,\vp)|
 \,\le\  |\,\om^{(r)}(x_1) | \
    t^{\frac{n-3}{2}}\ {\cal P}_{l-1}\log(t,\tau)^{-1}\  
 {\cal F}^{(12)}_{s,l}(t,\tau)\ . 
\label{prop2rr}
\eqe
In (\ref{prop2rr}) $\tau\,$ stands for 
$\,(\tau_3,\ldots,\tau_s)\,$, furthermore   
$\,  r = 0,1,2 \, \mbox{ and } \, |\,\om^{(0)}(x_1) | \equiv
1\, $.\\[.2cm] 
The bounds for (\ref{coeff})-(\ref{coeff4}), 
\eq
|\,\pa_{t}\, c^{\vep,t}_{l}(x_1)\,| \le \
{1 \over t}\ {\cal P}_{l-1}\log \frac{1}{ t}  \ ,\quad
|\ \pa_{t}\,a^{\vep,t}_{l}(x_1)\,| \le \ t^{-2} \ {\cal P}_{l-1}\log
\frac{1}{ t}\ ,  
\label{ca}
\eqe
\begin{eqnarray}
|\,\pa_{t}\, f^{\mu,\vep,t}_{l}(x_1) \,\om_\mu (x_1) |
& \le & \nom \,\  t^{-3/2} \ {\cal P}_{l-2}\log
\frac{1}{  t}\ ,  \label{b2pt1}\\
|\,\pa_{t}\, \, b^{\mu\nu,\vep,t}_{l}(x_1) \,\om_{\mu \nu}^{(2)} (x_1)\,|
 & \le & \ |\, \om^{(2)}(x_1) | \, {1 \over t}\ 
{\cal P}_{l-2}\log \frac{1}{  t} 
\label{b2pt}
\end{eqnarray}
are obtained on restricting  the previous considerations to the case $s=1\,$,
in which all external coordinates are integrated over, e.g.
\[
\pa_{t}\,a^{\vep,t}_{l}(x_1)\,=\, 
\frac12\,\int_{x_2,x,y} C_{t}(x,y)
\Biggl\{
{\cL}^{\vep,t}_{4,l-1}(x_1,x_{2},x,y)\,  -
\sum_{l_1+l_2=l}
\Bigl[
 {\cal L}^{\vep, t}_{2,l_1}(x_1,x)\,
{\cal L}^{\vep, t}_{2,l_2}(y,x_{2})\Bigr]_{sym}\Biggr\} \ .
\]
The polynomials appearing in (\ref{b2pt1}), (\ref{b2pt}) 
are of degree $\le \,l-2\,$, corresponding 
to the fact that on the r.h.s. of the FE (\ref{fequ}) for 
these terms, there appear 
$ {\cal L}^{\vep,t}_{l-1,4}$ and
$ {\cal L}^{\vep,t}_{l_1,2}\, {\cal L}^{\vep,t}_{l_2,2}\,$
with insertions $ E_{(2)}^{(r)}\, , \,r = 1, 2\,$. Both
are bounded inductively by  
polynomials of total degree $\le \sup (l-2,0)\,$.\\

\noindent
C) We come to the bound on 
$\pa_t {\cal L}^{\vep,t}_{n,l} (x_1,\vp^{(j)}_{s})\,$,
cf. (\ref{prop5}).
As compared to  B) the only case which requires new analysis 
is the bound on the second term from the r.h.s. of the FE (\ref{fequ}), 
in the case  $j >s_1\,$. Then we assume without
restriction, similarly as in B),
 that $j=s\,$. The term to be bounded corresponding to
(\ref{gensi}) is then
\[
t^{\frac{n-6 }{2}}\ 
{\cal P}_{l_1}\log(t,\tau)^{-1}\ 
 \int_v \sum_{T_{l_1}^{s_1+1},\
  T_{l_2}^{s_2+1}}
{\cal F}(t,\tau',t/2; T_{l_1}^{s_1+1}~;x_1,y_2,\ldots,y_{s_1},v)\ \cdot 
\]
\eq 
\ \cdot\ {\cal P}_{l_2}\log(t,\tau)^{-1}\  \int_{x_s}
{\cal F}(t,\tau'',t/2;T_{l_2}^{s_2+1}~;x_s, v,y_{s_1+1},\ldots
\ldots, y_{s-1}) \ |K^{(1)}(\tau_s, x_s,x_1;y_s)| \ .
\label{gensik}
\eqe
To bound this expression we telescope the difference  
$\,K^{(1)}(\tau_s, x_s,x_1;y_s)$, cf.(\ref{k1}),
along the tree $T_l^s( T_{l_1}^{s_1+1},  T_{l_2}^{s_2+1})\,$
obtained from the two initial trees by joining them via $v$
as in A2) and proceeding similarly as in (\ref{sid}).
We then have to bound expressions of the type
\[
 C_{t_{I,\de}}(v_{a-1},v_a)\
|K(\tau_{s}, v_a,y_s)\ -\ K(\tau_{s}, v_{a-1},y_s)|
\]
where $v_{a-1},v_a\,$ are adjacent internal vertices in 
$T_l^s( T_{l_1}^{s_1+1},  T_{l_2}^{s_2+1})\,$ 
on the unique path from $\,x_1\,$ to $\,y_s\,$.
Making use of the covariant Schl\"omilch formula (\ref{sloe})-(\ref{Rnb})
for the difference $\,K^{(1)}(\tau_s, v_a,v_{a-1};y_s)$,
 we obtain
$$ |K(\tau_{s}, v_a,y_s)\ -\ K(\tau_{s}, v_{a-1},y_s)|
  \leq  \int_{0}^{s} d r
     \, |\, \nabla_{(1)} K(\tau_s, z(r),y_s)\,| $$
$$  =  d(v_{a-1},v_a) \int_{0}^{1} d \rho
 \, |\, \nabla_{(1)} K(\tau_s, v_{a-1, a}(\rho),y_s)\,| $$
\eq 
  \leq  O(1)\, \frac{ d(v_{a-1},v_{a})}{\sqrt{\tau_s}}\,
      \int_{0}^{1} d \rho
    \, K(\tau_{s,\de '}, v_{a-1, a}(\rho),y_s)
\label{tel}
\eqe
where $ z(r)=v_{a-1, a}(\rho)$ lies at distance
$ r = \rho\, d(v_{a-1},v_{a}),
 \, 0 \leq \rho \leq 1, $ from $v_{a-1}\,$
 on the reparametrized geodesic segment
from $ v_{a-1}$ to $v_a$. The last inequality results 
from (\ref{D}). 
Introducing for 
\eq
  3\, \de < 1 :\qquad
 b = 2\,\frac{1+3\de}{1-3\de}\ , 
\label{tb}
\eqe
we then bound, with $ \de' > 0$ to be fixed later,
\begin{eqnarray}
 C_{t_{I,\de}}(v_{a-1},v_{a})\
|\,K(\tau_{s}, v_a,y_s)\ -\ K(\tau_{s}, v_{a-1},y_s)|
\ \le{\qquad\qquad\qquad\qquad\qquad }
\nonumber 
\\
\le \ \left\{ \begin{array}{r@{\quad \quad}l} 
 C_{t_{I,\de}}(v_{a-1},v_{a})\,K(\tau_{s}, v_a,y_s)\ +
C_{t_{I,\de}}(v_{a-1},v_{a})\, K(\tau_{s}, v_{a-1},y_s)
\ ,\quad\,b \,t \ge \de '\tau_s
\\   
O(1)\  C_{t_{I,2\de}}(v_{a-1},v_{a})\,(\frac{t_I}{\tau_s})^{1/2}\
 \int_{0}^{1} d\rho \ K(\tau_{s,\de '}, v_{a-1,a}(\rho),y_s)\,
\ ,\quad b\,t < \de '\tau_s \,\, .
\end{array}\right .
\label{arr}
\end{eqnarray}
The last line follows using (\ref{tel}) and
 absorbing the factor $d(v_{a-1},v_a)$
 in  $ C_{t_{I,\de}}(v_{a-1},v_a)\,$ with the aid of (\ref{d}),
 by changing $\de$ to $ 2\de$ .\\ 
The last line in (\ref{arr})  has to be bounded in
such a way as to reproduce a contribution compatible with the
induction hypothesis. To this end we use the (upper) bound (\ref{hk10})
\[
 C_{t_{2\de}}(v_{1},v_{2})\
 K(\tau_{\de '}, v_{1,2}(\rho),y)\ \le \ O(1)\,
\frac{1}{t^2}\ \frac{1}{\tau^2}\
\exp \bigg(-\frac{d^2(v_1,v_2)}{4t(1+3 \de )}\
 -\frac{d^2(v_{1,2}(\rho),y)}{4\tau(1+{\de '})^2}\,\bigg)\ .
\]
Noting that $d(v_{1},v_2)=d(v_1,v_{1,2}(\rho)) + 
d(v_{1,2}(\rho),v_2)\,$ we deduce
\[
\frac{1}{\de '}\ d^2(v_{1},v_2)+d^2(v_{1,2}(\rho),y) \ge
\frac{1}{\de '}\ d^2(v_1,v_{1,2}(\rho)) + d^2(v_{1,2}(\rho),y) \ge
\]
\[
\frac{1}{1+\de '}\ \left(d (v_1,v_{1,2}(\rho)) + d(v_{1,2}(\rho),y)\right)^2
\ge \frac{1}{1+\de '}\  d^2(v_{1},y)\ .
\]
Hence, observing (\ref{tb}), we find for $  b\,t  < \de '\tau \,$ 
\begin{eqnarray}
 \frac{d^2(v_1,v_2)}{4 t(1+3\de)} 
    + \frac{d^2(v_{1,2}(\rho),y)}{4 \tau (1+\de')^2} 
 & = & \frac{d^2(v_1,v_2)}{8 t}
  +  \frac{d^2(v_1,v_2)}{4 b t} 
+ \frac{d^2(v_{1,2}(\rho),y)}{4 \tau(1+\de ')^2 }
             \nonumber \\
 & \geq & \frac{d^2(v_1,v_2)}{8 t} \
         +\ \frac{d^2(v_1,y)}{4\tau(1+\de ')^3 } \ . \nonumber
\end{eqnarray}
With the aid of the
lower bound (\ref{hk10}) we then arrive at
\eq
 C_{t_{I,2\de}}(v_{1},v_{2})\!
 \int_{0}^{1}\!\!\! d\rho \ K(\tau_{s,\de '}, v_{1,2}(\rho),y_s)
 \le O(1)\,
 C_{2 t_{I},\de}(v_{1},v_{2})\, 
 K(\tau_s (1+\de ')^4, v_{1},y)\ . 
\label{abs}
\eqe
 Taking into account (\ref{hk4}) and choosing $\de ' $ such that
$(1+\de ')^4 = 1+\de\,$, i.e.
$\de ' =\de/4+O(\de ^2)\,$,   
 we may thus bound the last line in  (\ref{arr}) by
\eq
O(1)\ (\frac{t}{\tau_s})^{1/2}\ K(\tau_{s,\de}, v_{a-1},y_s)
\ \int_v C_{t_{I},\,\de}(v_{a-1},v)\,
C_{t_{I},\,\de}(v,v_{a})\ , \quad b\,t < \de ' \ \tau\ .
\label{arr2}
\eqe
Note that the addition of a  new internal  vertex $v\,$
of incidence number 2 in
(\ref{arr2}) is compatible with the inequality 
$v_2+\de_{c_1,1} \le 3l - 2 +s/2\,$. \\
Using these bounds and going back to (\ref{gensik})
we realize that the two terms in the first line
of  (\ref{arr}) - case $b\,t \geq \de' \tau_s $ - 
correspond to two new trees  of type ${\cal T}_l^{s}\,$, where 
in comparison to  $T_l^s( T_{l_1}^{s_1+1},  T_{l_2}^{s_2+1})\,$
the incidence number
of the vertex $v_{a-1}$ or $v_{a}$ has increased by one unit.
Similarly  (\ref{arr2})  - case $b\,t < \de ' \tau_s\,$ -
corresponds to a new tree where 
the incidence number
of the vertex $v_{a-1}$ has increased by one unit.
 In (\ref{gensik}) an integral over $x_s\,$ is performed.
If in the new tree
\\
i) $x_s$ has $c(x_s)>1\,$\footnote{remember that the vertex  
$x_s$ in $T_{l_2}^{s_2+1})\,$ is a root vertex by construction}, 
then $x_s$ takes the role of an internal vertex
of the new tree,\\ 
ii) $x_s$ has $c(x_s)=1\,$ we integrate over $x_s$ using (\ref{hk3})
so that the vertex $x_s$ disappears. 

As a consequence of the previous bounds,
on replacing again $s \to j\,$ in 
(\ref{arr}),(\ref{arr2}), we thus obtain for $ n > 4$ - 
see also A2) ii) for (\ref{bd6}),  (\ref{bd2}) -
\begin{eqnarray}
|\pa_t \,{\cal L}^{\vep,t}_{n,l} (x_1,\vp^{(j)}_{\tau_{2,s},
y_{2,s}})|\,\le\ 
\left\{ \begin{array}{r@{\quad \quad}l} 
t^{\frac{n-6}{2}}\ 
{\cal P}_l\log(t,\tau)^{-1}\ 
{\cal F}_{s,l}(t,\tau) 
\ ,\quad \, b\,t \ge  \de '\tau_j  
\\   
t^{\frac{n-6}{2}}\ (\frac{t}{\tau_j})^{1/2}\
{\cal P}_l\log(t,\tau)^{-1}\ 
{\cal F}_{s,l}(t,\tau)
\ ,\quad  b\,t <  \de '\tau_j   
\end{array}\right .
\label{bd5}
\end{eqnarray}
\begin{eqnarray}
|\pa_t \,{\cal L}^{\vep,t}_{4,l} (x_1,\vp^{(j)}_{\tau_{2,s},
y_{2,s}})|\,\le\ 
\left\{ \begin{array}{r@{\quad \quad}l} 
t^{-1}\ 
{\cal P}_{l-1}\log(t,\tau)^{-1}\ 
{\cal F}_{s,l}(t,\tau) 
\ ,\quad  b\,t \ge  \de ' \tau_j 
\\   
t^{-1}\ (\frac{t}{\tau_j})^{1/2}\
{\cal P}_{l-1}\log(t,\tau)^{-1}\ 
{\cal F}_{s,l}(t,\tau)
\ ,\quad b\, t <  \de '\tau_j 
\end{array}\right .
\label{bd6}
\end{eqnarray}
\begin{eqnarray}
|\pa_t \,{\cal L}^{\vep,t}_{2,l} (x_1,\vp^{(2)}_{\tau,y_{2}})|\,\le\ 
\left\{ \begin{array}{r@{\quad \quad}l} 
t^{-2}\ 
{\cal P}_{l-1}\log t^{-1}\ 
{\cal F}_{2,l}(t,\tau) 
\ ,\quad b\,t  \ge   \de '\tau_2 \ \
\\   
t^{-2}\ (\frac{t}{\tau})^{1/2}\
{\cal P}_{l-1}\log t^{-1}\ 
{\cal F}_{2,l}(t,\tau) \ ,\quad b\,t   <  \de '\tau_2 \ .
\end{array}\right .
\label{bd2}
\end{eqnarray}
\\
\noindent
D) From the bounds on the derivatives  $\,\pa_t \,{\cal L}^{\vep,t}_{n,l} \,$
we then verify the induction hypothesis on integrating over
$t\,$. In all cases we need the  bound
\eq
 {\cal F}_{s,l}(t',\tau)\ \le \
 {\cal F}_{s,l}(t,\tau) \quad \mbox{for }\ t' \le t\ ,
\label{bdF}
\eqe
which  follows directly from the definition (\ref{ff}).   

\noindent
a) In the cases \underline{$n+r >4\,$} we have, 
due to the boundary conditions encoded in the 
form of (\ref{nawig}) 
\[
{\cL}^{\vep,t}_{n,l}(x_1,\vp)=\,
\int_{\vep}^{t} dt'\
\pa_{t'}
{\cL}^{\vep,t'}_{n,l}(x_1,\vp)\ .
\]
Then we get from (\ref{formeln1}),
 (\ref{bdprop2})-(\ref{bdprop22}), due to (\ref{bdF}),
\eq
|\, {\cal L}^{\vep,t}_{n,l} (x_1,\vp_s)|\,\le\ 
t^{\frac{n-4}{2}}\
{\cal P}_l\log(t,\tau)^{-1}\  {\cal F}_{s,l}(t,\tau)\ 
\label{res1}
\eqe
\eq
| {\cal L}^{\vep,t}_{n,l} (x_1, E_{(k)}^{(r)}\, \vp_s)|\,\le\ 
   |\, \om^{(r)}(x_1) | \, t^{\frac{n+r-4}{2}}\
{\cal P}_l\log(t,\tau)_k^{-1}\  {\cal F}_{s,l}(t,\tau)\ 
,\quad
 n > 4
\label{res2}
\eqe
\eq
| {\cal L}^{\vep,t}_{4,l} (x_1, E_{(k)}^{(r)}\, \vp_s)|\,\le\ 
    |\, \om^{(r)}(x_1) | \, t^{\frac{r}{2}}\
{\cal P}_{l-1}\log   (t,\tau)_k^{-1}\  {\cal F}_{s,l }(t,\tau) \ 
,\quad
  r > 0 
\label{res3}
\eqe
\eq
| {\cal L}^{\vep,t}_{2,l} (x_1, E_{(2)}^{(3)} \vp_s)|\,\le\ 
 |\,\om^{(3)}(x_1) |  \, t^{\frac{1}{2}}\
 {\cal P}_{l-1}\log t^{-1}\ {\cal F}_{s,l}(t,\tau) \ ,
\label{res4}
\eqe
which proves the proposition for these cases.\\
b) Similarly, for \underline{$n \ge 4\,$}
the boundary conditions (\ref{nawig}) imply that
\[
{\cL}^{\vep,t}_{n,l}(x_1,\vp^{(j)}_s)=\,
\int_{\vep}^{t} dt'\
\pa_{t'}
{\cL}^{\vep,t'}_{n,l}(x_1,\vp^{(j)}_s)\ ,
\]
and we then obtain from
 (\ref{bd5}), (\ref{bd6}) together with (\ref{bdF})
\eq
|{\cal L}^{\vep,t}_{n,l} (x_1,\vp^{(j)}_s)|\,\le\ 
t^{\frac{n-4}{2}}\ \bigg (\frac{t}{\tau_j}\bigg)^{1/2}\ 
{\cal P}_l\log(t,\tau)^{-1}\  {\cal F}_{s,l}(t,\tau)\ .
\label{res5}
\eqe
We note that for $b\,t \ge  \de '\tau_j\,$ 
the integral $\int_{\vep}^{t} dt'\,$
has to be split into $\int_{\vep}^{ \de '\tau_j/b} dt'+
\int_{ \de ' \tau_j/b}^t dt'\,$, and that in the case $n=4$ the polynomial
in logarithms may increase in degree by one unit due to the
logarithmically divergent $t$-integral, 
see (\ref{d242})-(\ref{acht}) below for more details.\\[.1cm]
c) In the case \underline{$n=4\,, \  r=0$} we  
start from the decomposition (\ref{4l})
\[
{\cal L}^{\vep,t}_{4,l} (x_1,\vp)\,=\
c_l^{\vep,t}(x_1)\ \vp(x_1,x_1,x_1)\,+\
{\ell}^{\vep,t}_{4,l} (x_1,\vp)\ .
\]
On integrating the bound for $c_l^{\vep,t}(x_1)$,
(\ref{ca}), from $t\,$ to $1\,$ 
and using the boundary condition (\ref{renbed4})
we get
\eq
|\, c^{\vep,t}_{l}(x_1)\,| \le \
{\cal P}_{l}\log t^{-1}  \ .
\label{4ptt}
\eqe  
Taking together  (\ref{ca})
and (\ref{formeln2}) we verify 
\[
| \pa_{t} {\ell}^{\vep,t}_{4,l} (x_1,\vp_s)|\ \le \
t^{-1}\ 
{\cal P}_{l-1}\log(t,\tau)^{-1}\
{\cal F}_{s,l}(t,\tau)\ . 
\]
A sharper bound for $\, \pa_{t}
{\ell}^{\vep,t}_{4,l} (x_1,\vp_s)  \,$, which when integrated over
$t \ge \vep\,$ stays uniformly bounded in
$\vep$, is
obtained as follows. 
In the case $s=4\,$\footnote{In this case 
$\vp_i(x_i)=\, K(\tau_i, x_i, y_i)\,,\ 1 \le i \le 4\,$.
If $\vp_i =
  \mathbf{1}\,$ for some $i\,$, the corresponding contribution
 to the subsequent 
sum over $i\,$   vanishes.} we decompose the test function
$$ \vp_4(x_2, x_3, x_4) := \prod_{i=2}^{4} K(\tau_i, x_i, y_i)
  = \vp_4(x_1, x_1, x_1) + \psi(x_2, x_3, x_4) \,, $$
 $$  \psi(x_2, x_3, x_4) = \sum_{i=2}^{4} \,
    \prod_{f=2}^{i-1} \, K(\tau_f, x_1, y_f)\,
     K^{(1)}(\tau_i, x_i, x_1; y_i)
      \prod_{j=i+1}^{4} \, K(\tau_j, x_j, y_j)\ . $$
Then ${\ell}^{\vep,t}_{4,l} (x_1,\vp_4)
  =  \mathcal{L}_{4,l}^{\vep, t}(x_1, \psi)\, $, and hence
  the FE (\ref{fequ})  provides 
\eq
\pa_{t}\,{\ell}^{\vep,t}_{4,l} (x_1,\vp_4) \,=\ \frac12
\int_{x_2, x_3, x_4, x,y} \psi(x_2, x_3, x_4)\, 
 C_t(x,y)\ \Bigl\{{\cL}^{\vep,t}_{6,l-1}(x_1,\ldots,x_4, x,y) 
\label{s4t}
\eqe
\[
-\ \sum_{l_1+l_2=l}  
\Bigl[ {\cal L}^{\vep,t}_{4,l_1}(x_1, x_2 ,x_{3},x)\,
{\cal L}^{\vep,t}_{2,l_2}(y,x_{4})
\ +
 {\cal L}^{\vep,t}_{2,l_1}(x_1,x)\,
\,
{\cal L}^{\vep,t}_{4,l_2}(y,x_2,x_3,x_{4})\Bigr]_{sym}\,\Bigr\}\ .
\]
The r.h.s. is a sum over expressions of the same form as the one
 for $\pa_t\mathcal{L}_{4,l}^{\vep, t}(x_1, \vp^{(j)}_{\tau_{2,s},
y_{2,s}})\,$
in  part C. Setting $\tau =\inf_j\{ \tau_j\}$
we obtain, in the same way as there, the bound
\begin{eqnarray}
| \pa_{t} {\ell}^{\vep,t}_{4,l} (x_1,\vp_s)|
\,\le\ 
\left\{ \begin{array}{r@{\quad \quad}l} 
t^{-1}\ 
{\cal P}_{l-1}\log(t,\tau)^{-1}\ 
{\cal F}_{s,l}(t,\tau) 
\ ,\quad b\,t \ge \de ' \tau\ ,
\\   
t^{-1}\ (\frac{t}{\tau})^{1/2}\
{\cal P}_{l-1}\log(t,\tau)^{-1}\ 
{\cal F}_{s,l}(t,\tau)
\ ,\quad b\,t < \de '\tau \ .
\end{array}\right . 
\label{d242}
\end{eqnarray}
Using the boundary condition (\ref{bell}) 
we integrate (\ref{d242}) over $t\,$. This  gives for $b\,t
< \de ' \tau$  (and $\vep\,$
sufficiently small)
\eq
\Big |\int_{\vep}^{t} dt'\ \pa_{t'}\, {\ell}^{\vep,t'}_{4,l}
 (x_1,\vp_s)\Big |\ \le \ 
(\frac{t}{\tau})^{1/2} \
{\cal P}_{l-1}\log{t}^{-1}\ {\cal F}_{s,l}(t,\tau)\, ,
\label{sieben}
\eqe
and for  $b\,t >  \de '\tau \,$, in
which case we may have $t > \tau$ or $ t < \tau\,$,

$$
\Big |\int_{\vep}^{t} dt'\ \pa_{t'}\, 
{\ell}^{\vep,t'}_{4,l} (x_1,\vp_s)\Big|\ \le \
\Big |\int_{\vep}^{ \de '\tau/b} dt'\pa_{t'}\, 
{\ell}^{\vep,t'}_{4,l} (x_1,\vp_s)\Big |+
\Big |\int_{ \de '\tau/b}^t dt'\pa_{t'}\, 
{\ell}^{\vep,t'}_{4,l} (x_1,\vp_s)\Big |
$$
\eq
\le \ 
\bigg(\Big( \frac{\de '}{b}\Big )^{1/2} \, 
      {\cal P}_{l-1}\log \frac{b}{\de '\tau} +\ 
     {\cal P}_l\log \frac{b}{\de '\tau}  
    \bigg)\  {\cal F}_{s,l}(t,\tau ) \ .
\label{acht}
\eqe
Hence, absorbing powers of $\log( \de '/b)\,$ in the coefficients
of ${\cal P}_{l}\log\,$ as usual, 
\eq
| {\ell}^{\vep,t}_{4,l} (x_1,\vp_s)|\, \le \
{\cal P}_{l}\log{\tau}^{-1} \ {\cal   F}_{s,l}(t,\tau) 
\ .
\label{df}
\eqe
From (\ref{4l}), (\ref{4ptt}), (\ref{sieben}) and (\ref{df}) we obtain
\[
| {\cL}^{\vep,t}_{4,l} (x_1,\vp_s)|\ \le\
  {\cal P}_{l}\log(t,\tau)^{-1}\ {\cal F}_{s,l}(t,\tau)\ . 
\]

\noindent 
d)  In the case  \underline{$n=2\,$}
we have the decomposition (\ref{2l}). In addition to the
bounds (\ref{ca})-(\ref{b2pt}) to be integrated from 1 to 
$ t \leq 1 $, we need for 
$\, \pa_t\, {\ell}^{\vep,t}_{2,l}\ (x_1, \vp) $,\ (\ref{re2}),
 a bound, which upon integration from $\vep $  to $t\,$ 
becomes a 
uniformly bounded function on $ \vep \geq 0 $. To this end we
 use the form (\ref{re22}) and choose the test function
 $\vp(x_2) = K(\tau, x_2,y_2) \,$. Taking into account the
 bound (\ref{prop2rr}) for $ n = 2, r=0 $ together with (\ref{Rnb})
yields

 $$  |\, \pa_t\, {\ell}^{\vep,t}_{2,l}\ (x_1, \vp)|  \leq 
  \int_{x_2}  \, t^{- \frac{1}{2}}\
 {\cal P}_{l-1}\log t^{-1}\ {\cal F}_{2,l}^{(12)}(t)
  \int_{0}^{1}d\rho \, \frac{ (1-\rho)^2}{2!}\,
  | \, \nabla_{(1)}^3 K(\tau, X(\rho),y_2) | $$
 \[
\qquad \,\, \leq \, t^{- \frac{1}{2}}\,  \tau^{- \frac{3}{2}}
  \, {\cal P}_{l-1}\log t^{-1}\int_{x_2} {\cal F}_{2,l}^{(12)}(t)
  \int_{0}^{1} d\rho \,
   K(\tau_{ \de '}, X(\rho),y_2)
\]
where (\ref{D}) has been used.
By definition we have 
\[
 {\cal F}^{(12)}_{2,l}(t)\ =\
{\cal F}_{2,l}(t;x_1,x_2)\,=\,
\sum_{T_l^{2,(12)}} {\cal F}_{2,l}(t;T_l^{2,(12)}~;x_1,x_2)
\]
\[
 =\ \sum_{n = 1}^{3l-2}\, \,
\sup_{\{t_{I_{\nu}}| \vep\, \leq\, t_{I_i}\, \leq\, t,\, i=1.\cdots,n\}} 
\, \,[\prod_{1 \le \nu \le n } \int_{z_{\nu}}]\
C_{t_{I_1,\de}}(x_1,z_1) \ldots  C_{t_{I_n,\de}}(z_{n},x_2)
\]     
\[
=\
\sum_{n = 1}^{3l-2}\, \,
\sup_{\{t_{I_{\nu}}| \vep\, \leq\, t_{I_{\nu}}\, \leq\, t,\, \nu=1.\cdots,n\}} 
C_{\sum_1^n t_{I_{\nu},\de}}(x_1,x_2) 
\]
where we used (\ref{hk4}) and (\ref{f2r}). 
We then proceed similarly as in and after (\ref{arr}).
Setting $N=3l-2\,$ we  bound for 
$Nb\,t < \de ' \tau \,$ and for $n \le N\,$ as in (\ref{abs})
\eq
  C_{\sum_1^n t_{I_{\nu},\de}}(x_1,x_2)\
 \int_{0}^{1} d\rho \ K(\tau_{\de '}, X(\rho),y_2)\,
\ \le O(1) \,\,
 C_{2\sum_1^n t_{I_{\nu},\de}}(x_{1},x_{2})\ 
    K(\tau_{\de}, x_{1},y_2) 
\label{rr}
\eqe
so that, observing (\ref{hk3}), 
\eq
 |\, \pa_t\, {\ell}^{\vep,t}_{2,l}\ (x_1, \vp)| 
  \ \le \ \tau^{-3/2}\ t^{-1/2}\ 
{\cal P}_{l-1}\log t^{-1}\ K(\tau_{\de}, x_{1},y_2)\ ,\quad 
N b\,t < \de ' \tau \ . 
\label{rl2}
\eqe
To verify the induction hypothesis (\ref{prop20}) we 
resort to the decomposition (\ref{2l}) and denote the
sum of the first, second and third terms there by
$ {\cal L}^{\vep,t}_{2,l}(x_1,\, \vp)_{rel}$. Integrating
the corresponding bounds (\ref{ca})-(\ref{b2pt}) from $1$
to $t$, and using again (\ref{D}) gives
\eq \label{rel}
|\, {\cal L}^{\vep,t}_{2,l}(x_1,\, \vp)_{rel} | < 
 \Bigl(\, \frac{1}{t}\ {\cal P}_{l-1}\log t^{-1}\ 
 + \frac{1}{(t \tau)^{\frac{1}{2}}}{\cal P}_{l-2}\log  t^{-1}
 + \frac{1}{\tau} {\cal P}_{l-1}\log  t^{-1}\Bigr)\
 K(\tau_{\de}, x_1,y_2)\ .
\eqe
Integrating the remainder (\ref{rl2}) from (small) $\vep$
 with vanishing initial condition
 (\ref{bell}) to $ t < \de '\tau /(bN)$ leads to
\eq
 |\,{\ell}^{\vep,t}_{2,l}\ (x_1, \vp)| 
  \ \le \ \tau^{-3/2}\ t^{1/2}\ 
{\cal P}_{l-1}\log t^{-1}\ K(\tau_{\de}, x_{1},y_2)\ . 
\label{bn2}
\eqe
By way of (\ref{2l}) we obtain from the bounds
(\ref{formeln2}) and (\ref{ca})-(\ref{b2pt})
the bound for $N b\,t  \ge \de ' \tau \,$ 
\eq
 |\, \pa_t\, {\ell}^{\vep,t}_{2,l}\ (x_1, \vp)| 
 \leq \  t^{-2}\ {\cal P}_{l-1}\log   (t,\tau)^{-1}\  
{\cal F}_{2,l}(t,\tau)\ +\
\label{bn22}
\eqe
\[
\Bigl(t^{-2}\ {\cal P}_{l-1}\log  t^{-1}\ 
+\,\sum_{j=1}^2 
t^{\frac{j-4}{2}}\ \tau ^{-j/2}\   
{\cal P}_{l-2}\log  t^{-1}\Bigr)\ K(\tau_{\de}, x_{1},y_2)\ .
\]
Hence, integration and majorization, again observing
both $\tau > t $ and $t  < \tau $,
 gives for $ t > \de '\tau /(bN)$ 
\eq 
|\,{\ell}^{\vep,t}_{2,l}\ (x_1, \vp)|\, \leq \,
\bigg(\, \frac{bN}{\de '\tau}\,
    {\cal P}_{l-1}\log \frac{bN}{\de '\tau} +
  \theta (t - \tau)\frac{1}{\tau} {\cal P}_{l-1}\log  \tau^{-1}
   \bigg )\,{\cal F}_{2,l}(t,\tau)\ +\ 
\label{tgta}
\eqe
$$  \bigg (\, \frac{bN}{\de '\tau}\,
    {\cal P}_{l-1}\log \frac{bN}{\de '\tau} +
 \bigg(\frac{bN}{\de '\tau^2}\bigg)^{1/2}
   {\cal P}_{l-2}\log \frac{bN}{\de '\tau}\, 
   + \frac{1}{\tau} {\cal P}_{l-1}\log \frac{bN}{\de '\tau}
     \bigg )\, K(\tau_{\de}, x_1,y_2)\ .  $$
From (\ref{rel}), (\ref{bn2}) and (\ref{tgta}), 
absorbing constants as usual in $ {\cal P}\log\,$, we then
get
\footnote{The bound (\ref{tgta}) diverges linearly with
$\de '$, whereas in 
(\ref{acht}) the divergence was only logarithmic. This indicates
rapid growth since the bounds then behave as $(\de ') ^{-l}\,$,
a factor of $(\de ')^{-1}\,$ being produced per loop order.
Without trying at all to optimize constants, we still note
that it is possible to choose
for this case $\de =2\,$  in $K(\tau_{\de}, x_{1},y_2)\,$
and bound the two point function 
inductively by ${\cal F}_{2,l}(t,3\tau)$ without changing the bounds on
the other functions. The only place in the proof where there is a
modification due to this factor is in part A2). But here the value of
$\tau$ appearing is $t/2$, see (\ref{falt}), and $3t/2$ can be accommodated
for in the proof by introducing a new vertex of incidence number 2
while respecting the bound on the number of those vertices.
A value $\de =2\,$ then gives for suitable choice of $b$
the value  $b/\de ' \simeq 6\,$.}

\eq
|\, {\cal L}^{\vep,t}_{2,l}(x_1,\, \vp)\,|\ \le \
 (t,\tau)^{-1}\ {\cal P}_{l-1}\log (t,\tau)^{-1}\ 
{\cal F}_{2,l}(t,\tau) 
\eqe
in accord with (\ref{prop20}). 

To establish the bounds on 
$\,  {\cal L}^{\vep,t}_{2,l} (x_1,E_{(2)}^{(r)} \vp) ,\, r=1,2\, , $
we expand the respective test functions as follows, employing
 (\ref{Rna}) and using the notations (\ref{coeff}), (\ref{2ins})
$$
 {\cal L}^{\vep,t}_{2,l} (x_1,E_{(2)}^{(1)} \vp) = 
 \vp(x_1) f^{\mu,\vep,t}_l(x_1)\ \om_{\mu}(x_1) 
 + 2\, b^{\,\mu \nu,\vep,t}_l(x_1)\, \om_{\mu}(x_1) 
    \, (\nabla _{\nu} \vp)(x_1)$$
\eq
  + \int_{x_2} F^{(1)}_{(12)} {\cal L}^{\vep,t}_{2,l} (x_1,x_2)
  \int_{0}^{1} d\rho\, \frac{ (1-\rho)}{d^{\,2}(x_1, x_2)}\,  
   {\dot X}^{\mu}(\rho)\, {\dot X}^{\nu}(\rho) 
   (\nabla_\mu \nabla_\nu  \vp)(X(\rho))\, ,
\label{ein1}
\eqe  
$$
 {\cal L}^{\vep,t}_{2,l} (x_1,E_{(2)}^{(2)} \vp) = 
 - 2 \,\vp(x_1)\, b^{\,\mu \nu,\vep,t}_l(x_1)\,
     \om^{(2)}_{\mu \nu}(x_1) $$
\eq
  + \int_{x_2} F^{(2)}_{(12)} {\cal L}^{\vep,t}_{2,l} (x_1,x_2)
  \int_{0}^{1} d\rho\, \frac{1}{d(x_1, x_2)} \, 
   {\dot X}^{\mu}(\rho)\, 
   (\nabla_\mu \vp)(X(\rho))\, .
\label{ein2}
\eqe
The local, i.e. relevant terms have already been dealt with
in (\ref{b2pt1}), (\ref{b2pt}), and the remainders are treated
as $\, {\ell}^{\vep,t}_{2,l}\ (x_1, \vp)$ ; one obtains
\[
 |\,  {\cal L}^{\vep,t}_{2,l} (x_1,E_{(2)}^{(1)} \vp_2)|
\ \le\ |\,\om^{(1)}(x_1)|\ (t, \tau)^{-1/2}\ 
{\cal P}_{l-1}\log(t,\tau)^{-1}\  {\cal F}_{2,l}(t,\tau)\ ,
\]
\[
 |\,  {\cal L}^{\vep,t}_{2,l} (x_1,E_{(2)}^{(2)}\vp_2)|
\ \le\ |\om^{(2)}(x_1)|\ 
{\cal P}_{l-1}\log (t,\tau)^{-1}\   {\cal F}_{2,l}(t,\tau)\ .
\]
Finally, we realize that ${\cal L}^{\vep,t}_{2,l}(x_1,\vp^{(2)}_2)$ 
equals the r.h.s. of (\ref{2l}) without its first term.
Proceeding again similarly as before  - see 
(\ref{rel}), (\ref{bn2}) and (\ref{tgta}) - provides
\[
|{\cal L}^{\vep,t}_{2,l}(x_1,\vp^{(2)}_2)| 
\ \le\
(\frac{t}{\tau})^{1/2}\
 (t,\tau)^{-1}\ {\cal P}_{l-1}\log (t,\tau)^{-1}\ 
{\cal F}_{2,l}(t,\tau) \ . 
\]
This ends the proof of Proposition 1. \qed
\\

\noindent
The behaviour of the CAS upon removing the UV cutoff, i.e.
$ \vep \searrow 0 $, follows from\\
\noindent
{\bf Proposition 2}:\\
{\it Let $\vep $ be (sufficiently) small. 
With the notations, conventions and the same class of 
renormalization conditions as in Proposition 1 we have the bounds}
\begin{eqnarray}
|\,\pa_{\vep}\, {\cal L}^{\vep,t}_{n,l} (x_1,\vp_{\tau_{2,s},y_{2,s}})|
 & \le & \vep^{-\frac{1}{2}} \,\,
 {\cal P}_l\log{\vep}^{-1}\,\, \, t^{\frac{n-5}{2}} \,\,
{\cal F}_{s,l}(t,\tau)  
\label{uv1} \\
|\,\pa_{\vep}\, {\cal L}^{\vep,t}_{n,l}
    (x_1, E_{(i)}^{(r)}\vp_{\tau_{2,s},y_{2,s}})|
 & \le & \vep^{-\frac{1}{2}}\,\,  
{\cal P}_l\log{\vep}^{-1}\,\,|\,\om^{(r)}(x_1) |\,
      t^{\frac{n+r-5}{2}}\,\, {\cal F}_{s,l}(t,\tau)  
\label{uv3} \\
|\,\pa_{\vep}\, {\cal L}^{\vep,t}_{n,l} (x_1,\vp^{(j)}_{\tau_{2,s},
y_{2,s}})|\, & \le & \vep^{-\frac{1}{2}}  \,\,
 {\cal P}_l\log{\vep}^{-1}\,\, 
 t^{\frac{n-4}{2}}\,\, \tau_j^{-\frac{1}{2}}\,\,
{\cal F}_{s,l}(t,\tau) 
\label{uv4}\\
|\,\pa_{\vep}\, F_{(12)}^{(0)}
{\cal L}^{\vep,t}_{n,l} (x_1, x_2, \vp_{\tau_{2,s},y_{2,s}})|
 & \le &\vep^{-\frac{1}{2}} \,\,
 {\cal P}_l\log{\vep}^{-1}\,\, t^{\frac{n-2}{2}}\,\, 
{\cal F}_{s,l}^{(12)}(t,\tau)  
\label{uv5} \\
|\,\pa_{\vep}\, {\cal L}^{\vep,t}_{2,l}(x_1,\vp_{\tau, y})|
 & \le & \vep^{-\frac{1}{2}}\,\, 
{\cal P}_{l-1}\log{\vep}^{-1}\,\, (t,\tau)^{-\frac{3}{2}}\,\,
{\cal F}_{2,l}(t,\tau)
\label{uv2} \\ 
|\,\pa_{\vep}\, F_{(12)}^{(0)}
{\cal L}^{\vep,t}_{2,l} (x_1, x_2)|
 & \le & \vep^{-\frac{1}{2}} \,\,
 {\cal P}_{l-1}\log{\vep}^{-1}\,\, \,
{\cal F}_{2,l}^{(12)}(t)\ .   
\label{uv6} 
\end{eqnarray}
{\it Proof:} We apply the method developed in the previous
proof. The bound (\ref{uv1}) obviously holds
 in the starting case\, $n=4, l=0\,$.
 Because of the bare interaction (\ref{nawig})
 the FE (\ref{fequvep}) is used if $n+r>4,$ where the difference
test function in (\ref{uv4}) and the modified insertion in
(\ref{uv5}),(\ref{uv6}) count as $r=1$ and $r=3$, respectively.
Regarding the r.h.s. of (\ref{fequvep}) we note that the first
and second term do not contribute to the cases considered, and
the third one only if $n=2,4,6.$\\
Proceeding inductively as in A, B) and C)
of the previous proof, and using the bounds of Proposition 1,
reproduces (\ref{uv1}) for $n>4$ and
(\ref{uv3})-(\ref{uv5}) and (\ref{uv6}).\\
The FE (\ref{fequvep2}) provides bounds on the relevant parts
of the cases $n+r \leq 4$. As the renormalization conditions
(\ref{renbed}), (\ref{renbed4}), (\ref{vepdep}) depend at most weakly on
 $\vep$, we obtain inductively    
\begin{eqnarray}
%\eq
|\,\pa_{\vep}\, c^{\vep,t}_{l}(x_1)\,| \le \,
\vep^{-\frac{1}{2}}\,\, {\cal P}_{l-1}\log {\vep}^{-1}
  \,\cdot t^{- \frac{1}{2}}   & , &
|\,\pa_{\vep}\, a^{\vep,t}_{l}(x_1)\,| \le \,
\vep^{-\frac{1}{2}}\, {\cal P}_{l-1}\log {\vep}^{-1}
  \cdot t^{- \frac{3}{2}}  
\label{uvr1}\\
%\eqe
|\,\pa_{\vep}\, f^{\mu,\vep,t}_{l}(x_1)\,\om_\mu (x_1) |
& \le & \nom \,\,\vep^{-\frac{1}{2}}\,\,
 {\cal P}_{l-1}\log {\vep}^{-1}
  \,\cdot t^{-1}  
 \label{uvr2}\\
|\,\pa_{\vep}\, b^{\mu\nu,\vep,t}_{l}(x_1)\,\om_{\mu \nu}^{(2)} (x_1)\ |
 & \le &  |\, \om^{(2)}(x_1)\,| \,\,
\vep^{-\frac{1}{2}}\,\, {\cal P}_{l-1}\log {\vep}^{-1}
  \,\cdot t^{- \frac{1}{2}}  \ .
\label{uvr3}
\end{eqnarray}
With the aid of the decomposition (\ref{4l}), the bound (\ref{uv1}) for 
 $n=4$ follows from (\ref{uvr1}) and (\ref{uv4}).
It remains to show (\ref{uv2}). We use the decomposition
(\ref{2l}) and perform similar steps as in D)d). From 
(\ref{re22}) and (\ref{uv6}) we obtain
 \begin{eqnarray}
  |\, \pa_\vep\, {\ell}^{\vep,t}_{2,l}\ (x_1, \vp_{\tau, y})| 
 & \leq & \int_{x_2}  \,| \,\pa_{\vep}\, F_{(12)}^{(0)}
{\cal L}^{\vep,t}_{2,l} (x_1, x_2)\,| 
\   \int_{0}^{1}d\rho \, \frac{ (1-\rho)^2}{2!}\,
  | \,( \nabla^{\,3} \vp_{\tau, y})( X(\rho))\, | \nonumber \\
  & \leq & \vep^{-\frac{1}{2}}\,\, {\cal P}_{l-1}\log {\vep}^{-1}
\,\, 
  \,\tau^{- \frac{3}{2}} \,\int_{x_2} {\cal F}_{2,l}^{(12)}(t)
  \int_{0}^{1} d\rho \, K(\tau_{ \de '}, X(\rho),y)\, \nonumber
\end{eqnarray}
and herefrom, cf. (\ref{f2r}), (\ref{rr}) for $N b\,t <\de '\tau$
$(N=3l-2)\,$,
\eq
 |\, \pa_\vep\, {\ell}^{\vep,t}_{2,l}\ (x_1, \vp_{\tau, y})| 
  \ \le \ \vep^{-\frac{1}{2}}\,\, {\cal P}_{l-1}\log {\vep}^{-1}
\,\,  
\tau^{-3/2}\  K(\tau_{\de}, x_{1},y)\ . 
\label{p2s}
\eqe
From (\ref{uvr1})-(\ref{uvr3}) follows
\eq \label{p2rel}
|\,\pa_\vep \, {\cal L}^{\vep,t}_{2,l}(x_1,\, \vp_{\tau, y})_{rel} | < 
\vep^{-\frac{1}{2}}\,\, {\cal P}_{l-1}\log {\vep}^{-1}
\,\cdot \frac{1}{t^{\frac{1}{2}}} 
 \bigg(\, \frac{1}{t}  + \frac{1}{(t \tau)^{\frac{1}{2}}}
 + \frac{1}{\tau}\, \bigg)\
 K(\tau_{\de}, x_1,y)\, .
\eqe
On account of (\ref{2l}) the bounds (\ref{p2s}), (\ref{p2rel})
 establish (\ref{uv2}) for $ N b\,t < \de ' \tau $.\\
To obtain an extension of the bound (\ref{p2s})
 to  $N b\,t \geq \de '\tau$
we again resort to the decomposition (\ref{2l}), yielding
$$
\pa_t\, \pa_\vep\, {\ell}^{\vep,t}_{2,\,l}(x_1, \vp)
 \, = \,\pa_t \,\pa_\vep \, {\cal L}^{\vep,t}_{2,\,l}(x_1,\, \vp)
$$
\eq \label{p2d}
 -\, \pa_t \,\pa_{\vep}\, a^{\vep,t}_{l}(x_1)\vp (x_1) 
+\, \pa_t \,\pa_{\vep}\, f^{\mu,\vep,t}_{l}(x_1)\,\om_\mu (x_1)
+\, \pa_t\,\pa_{\vep}\,
 b^{\mu\nu,\vep,t}_{l}(x_1)\,\om_{\mu \nu}^{(2)} (x_1)\, , 
\eqe
with $\, \om_{\mu}(x) = \nabla_{\mu} \vp(x), \, \, 
\om^{(2)}_{\mu \nu}(x) =  \nabla_{\mu} \nabla_{\nu} \vp(x),
 \, \, \vp (x) = K(\tau, x, y) $. Employing on the
r.h.s. of (\ref{p2d}) in the various terms the corresponding
FE (\ref{fequ}) derived w.r.t. $\vep$ 
and then making use of bounds of Proposition 1 and
of Proposition 2 already established inductively, leads
with now familiar steps to
$$
|\,\pa_t\, \pa_\vep\, {\ell}^{\vep,t}_{2,\,l}(x_1, \vp) | 
 \leq \ \vep^{-\frac{1}{2}}\,\, {\cal P}_{l-1}\log {\vep}^{-1}
\,\cdot \Big [ 
 \, ( t,\tau)^{-\frac{5}{2}}
 \,\,{\cal F}_{2,\,l}(t,\tau)
$$
\eq 
\label{dbp2}
  +\, \Big ( \, t^{-\frac{5}{2}} + \tau^{-\frac{1}{2}}\, t^{-2}
 + \tau^{-1}\, t^{-\frac{3}{2}}    \, \Big )
 \, K(\tau_{\de}, x_1,y)\,\Big ] \, .
\eqe
On integrating $\,\pa_t\, \pa_\vep\, {\ell}^{\vep,t}_{2,l}\ (x_1, \vp)\,$
from $ t = \vep $ (small) with vanishing initial condition
 up to $ t \geq \de '\tau/bN $ the integral has to be split at
 $ t = \de '\tau/bN $. A bound on the
 lower part of the integral is given 
 by (\ref{p2s}). The upper part of the
 integral can be bounded using  (\ref{dbp2})
 observing both $ \tau >  t $ and $ \tau < t$,
and majorizing constants. Combining both contributions
yields for   $ t \geq \de '\tau/bN $ 
$$
|\, \pa_\vep\, {\ell}^{\vep,t}_{2,\,l}(x_1, \vp) | 
 \leq \ \vep^{-\frac{1}{2}}\,\, {\cal P}_{l-1}\log {\vep}^{-1}
\,\cdot
 \,\bigg ( \frac{b N}{\de ' \tau}\bigg )^{\frac{3}{2}}
$$
\eq \label{bp2l} 
 \cdot\, \bigg [ \,{\cal F}_{2,\,l}(t,\tau)
 +\, \bigg ( \,1 + \Big (\,\frac{\de '}{b N}\,\Big )^{\frac{1}{2}}
 + \frac{\de '}{b N} \, \bigg ) \, K(\tau_{\de}, x_1,y)\,\bigg ] \, .
\eqe
Taking into account  once more the decomposition (\ref{2l}),
the bound (\ref{p2rel}) on the relevant part
together with the bounds (\ref{p2s}), (\ref{bp2l})
on the remainder reproduce (\ref{uv2}). Thus
the proof of Proposition 2 is complete. \qed
\\

\noindent
From (\ref{uv1}), (\ref{uv2}) follows the integrability at
$\vep = 0$ and hence the existence of finite limits
$$ \lim_{\vep \searrow 0} \,
 {\cal L}^{\vep,t}_{n,\,l}\, (x_1,\vp_{\tau_{2,s},y_{2,s}})\, , 
 \quad n \geq 2 \ .$$
%%%%%%%%%%%%%%%%%%%%%%%%%%%%%%%%%%%%%%%%%%%%%%%%%%%%%%%%%%%%%%%

\noindent
{\bf Proposition 3}:\\
{\it With the notations, conventions and the same class of 
renormalization conditions as in Proposition 1
- up to the fact that the constants in ${\cal P}_l\log\,$
may now also depend on the mass $m\,$ - 
we claim the following bounds for the CAS in the interval
$ \,1 \le t \le \infty$~:
\eq
|\,{\cal L}^{\vep,t}_{n,l} (x_1,\vp_{\tau,y_{2,s}})|\,\le\ 
{\cal P}_l\log \tau^{-1}\ {\cal F}^{\,t}_{s,l}(\tau)
 \ , \quad n \ge 4  
\label{ip1}
\eqe
\eq
|\, {\cal L}^{\vep,t}_{2,l} (x_1,\vp_{{\tau},y})|\,\le\ 
(1,\tau)^{-1}\
{\cal P}_{l-1}\log (1,\tau)^{-1}\   {\cal F}^{\,t}_{2,l}(\tau)\ .
\label{ip2}
\eqe
The definition of ${\cal F}^{\,t}_{s,l}(\tau)\,$ is given in
(\ref{iges}).} 

\noindent
{\it Proof}~:
The bounds  stated in the proposition are proven inductively
using again the standard scheme. The boundary conditions are the
bounds from Proposition 1 taken at $t=1\,$. They obviously satisfy
the bounds (\ref{ip1}), (\ref{ip2}). The FE is treated in the same way
as in parts A1) and A2) of the proof of Proposition 1. The integration
w.r.t $t$ is performed using the fact that ${\cal F}^{\,t}_{s,l}(\tau)\,$ 
is montonically increasing with $t\,$.
As regards part A1) we now  use for $\,t \geq 1\,$ instead of (\ref{rom})
now $\,C_t(z, z') \leq  O(1)\, \exp( - ( m^2 -\de)\,t\, )\,$,
which results from the upper bounds (\ref{hk5}), (\ref{f21}) on the
heat kernel, and obtain upon integration 
\[
 \int_1^{\,t} dt'\ {\cal F}^{\,t'}_{s,l}(\tau)\ e ^{-(m^2 - \de)\,t'}\
 \le \, O(1/m^2) \,{\cal F}^{\,t}_{s,l}(\tau)\ .
\]
As regards A2) the internal line generated, which connects the
two (partial) trees, see (\ref{gensi}), (\ref{2nd}), has the 
weight (\ref{falt}). Integrating, we majorize the weights of the
other internal lines by their values at $\, t \, $ and use for
(\ref{falt})  
$$ 
 \int_1^{\,t} dt'\,C_{t'}(z', z'') \,\leq \,C_{\underline{t}}\,(z', z'') 
 +  \int_1^{\,t} dt'\,C_{t'}(z', z'')  $$
valid for any $\, 0 < \underline{t} \leq 1 \, $, thus reproducing
the weight factor $\,{\cal F}^{\,t}_{s,l}(\tau)\,$, (\ref{iges}),
in this case, too.   \qed  

Note that the renormalization conditions at $\,t=1\,$ are in one to one 
relation with the values of the corresponding relevant terms
at $\,t=\infty\,$, which have been shown to be finite for $m^2 >0\,$ in
according to Proposition 3. Therefore renormalization conditions 
at $\,t=1\,$ are tantamount to renormalization
 conditions at $\,t=\infty\,$.

We want to close this section with some comments on the test functions
considered and on possible extensions of the class of test functions.
We stay with some informal remarks here, we did not rigorously
analyse the problem of what is a "natural large" class of test functions.   
First note that our test functions can be arbitrarily well localized 
around any point of the manifold. This is an essential criterion for their
viability from the physical point of view. 
Secondly the class of test functions
can be extended  by linearity (\ref{lin1}). 
Since our bounds are in terms of the weight factors decaying with the
tree distance between the points $x_1,y_2,\ldots,y_s\,$ it is quite
evident that the functionals ${\cal L}^{\vep,t}_{n,l}(x_1,\vp)$
can also be extended continuously by bounded convergence
to test functions which are  infinite sums 
$\sum_i \la_i \ \vp_{i,\tau_{2,s}^{(i)},y^{(i)}_{2,s}}\,$ 
with $\sum |\la_i| < \infty\,$. To go further one could either
 prove (in a more functional analysis type of approach) 
that our test functions are dense e.g. in the  set of smooth rapidly
decaying functions on $\cal M\,$ w.r.t. a suitable
norm, and that the ${\cal L}^{\vep,t}_{n,l}(x_1,\vp)$ are continuous
w.r.t. this norm. Or one could try to directly extend the previous proof
to more general test functions in a second step. In this case
the crucial part would be to maintain the line of argument presented
in part C), (\ref{tel}) to (\ref{arr2}), of the previous proof.

%%%%%%%%%%%%%%%%%%%%%%%%%%%%%%%%%%%%%%%%%%%%%%%%%%%%%

\section{Scaling transformations and the minimal form of the bare action }

In this section we want to show that the theory can be renormalized
starting from a bare (inter)action of the form (\ref{nawi}). 
This requires that we do not introduce any position dependent quantity  
in the theory which is not intrinsic to $\, (\mathcal{M}, g ) \, $.
Thus we only consider position independent coupling $\la$,
and renormalization conditions in terms of intrinsic geometric
quantities. We then introduce scaling tranformations 
of the following kind~:\\ 
\noindent
For a four-dimensional Riemannian manifold $\, (\mathcal{M}, g ) \, $
we scale its metric by a constant conformal factor, [NePa],
\begin{equation} \label{sc1}
\rho \in \mathbf{R}_+ : \qquad
   g_{\mu \nu}(x) \rightarrow \rho ^2 \,g_{\mu \nu}(x)\,,\quad
\mbox{shortly }\   g \rightarrow \rho ^2 \,g\ .
\end{equation}
This leads to corresponding changes of geometrical quantities
$$ g^{\mu \nu} \rightarrow \rho^{ - 2}g^{\mu \nu},\quad
\Delta  \rightarrow \rho^{ - 2}\Delta, \quad
|\,g|^{1/2} \rightarrow \rho ^4 |\,g|^{1/2},\quad \ti \de
\rightarrow \rho ^{-4} \,  \ti \de$$
\begin{equation} \label{sc2}
d(x,y) \rightarrow \rho\, d(x,y) , \quad
 \sigma(x,y)^\mu \rightarrow \sigma(x,y)^\mu 
 \end{equation}
$$\Gamma_{\mu \nu}^{\lambda}
 \rightarrow \Gamma_{\mu \nu}^{\lambda}\,, \quad
\nabla_{\mu} \rightarrow \nabla_{\mu} \,, \quad
R^{\lambda}_{\, \,\mu \nu \sigma} \rightarrow
 R^{\lambda}_{\, \,\mu \nu \sigma}\,, \quad
 R_{\mu \nu} \rightarrow R_{\mu \nu}\,, 
   \quad  R \rightarrow \rho^{ - 2} R\,.  $$
  Moreover, the heat kernel $\, K (t, x, y\,; g) \, $
 satisfies the scaling relation
  \begin{equation} \label{sc3}
  K(t, x, y \,;\, g)\, = \,
\rho^4\, K ( \,\rho^2 t, x, y\, ;\,\rho^2 g)  \, ,
\end{equation}
 which follows from its evolution equation 
   $ \,( \partial_t - \Delta_g )  K (t, x, y\, ;\, g) = 0 \, $
  together with stochastic completeness (\ref{hk3}).
 As a consequence the regularized free
   propagator (\ref{propa}), $0 < \varepsilon < t \leq \infty $, 
   $$  C^{\,\varepsilon,\, t}(x, y;\,m^2 \,,g)\, =
 \,\int_{\varepsilon}^{\,t}dt '\,
 e^{\,- m^2 t'}\, K(t', x, y;\,g) \, ,$$
satisfies
\begin{equation}\label{sc4}
C^{\,\varepsilon,\, t}(x, y;m^2 \,,g)\, =
 \rho^2 \,C^{\,\rho^2\varepsilon,\,\rho^2 t}
    (x, y;\frac{m^2}{\rho^2} \,,\rho^2 g)\,.   
\end{equation}
Regarding for a moment the action of the \emph{classical
 scalar field theory},
\begin{equation}
 S(\varphi, m^2, \xi, \lambda; g) \, =\frac{1}{2}\,
 \int_x\, 
 \Big (\, \varphi (-\Delta)\varphi
     + m^2 \varphi^2 + \xi\, R(x)\,\varphi^2  
    + 2 \frac{\lambda}{4!}\, \varphi^4 \Big )\, , 
\end{equation}
we observe, that it is invariant if we supplement the
 scaling (\ref{sc1}) of the metric by the transformations
\begin{equation} 
\varphi(x) \rightarrow \rho^{-1}\varphi(x)\,,\quad
 m^2 \rightarrow \rho^{-2} m^2 \, ,\quad \xi \rightarrow \xi \, ,
  \quad \lambda \rightarrow \lambda \ .
\end{equation}
 We now consider the perturbative expansion of a regularized
$\lambda \phi^4 $- theory
 without counter terms, i.e. in (\ref{funcin}) we have
  \, $ L^{\vep, \vep}( \phi) = \lambda \int dV(x)\, \phi ^4 (x) \,$.  
 A Feynman diagram contributing to an $n$-point CAS having\, $v$\,
 four-vertices and $\, I\, $ internal lines obeys the topological
 relation $  4 v = n + 2 I .$ \, This together with the scaling property
  (\ref{sc4}) of the propagator implies for an $n$-point function
folded with a test function
 $ \varphi = \varphi(x_2, \dots, x_n)$
\begin{equation} \label{sc5}
\mathcal{L}^{\,\varepsilon,\,t}_{n, \,l}
 (x_1, \varphi\, ; m^2, \lambda, g) = \rho^{4-n}
 \,\mathcal{L}^{\,\rho^2\varepsilon,\,\rho^2 t }_{n, \,l}
 (x_1,\varphi \,; \frac{m^2}{\rho^2}\,, \lambda \,,\rho^2 g)\ .   
\end{equation}
In the renormalization proof the CAS were constructed by imposing
renormalization conditions for the relevant terms, see (\ref{renbed}), 
(\ref{renbed4}), and by requiring
the irrelevant terms to vanish at scale $\vep$, see
(\ref{bo1})-(\ref{bo3}). 
As noted the renormalization conditions will now be supposed to be expressed
in terms of intrinsic quantities, and they will be supposed to satisfy
scaling  (both statements are true for vanishing  renormalization 
conditions).   Because of the behaviour of $\, \sigma(x,y)^\mu \, $ 
under scaling, (\ref{sc2}), this means
\begin{eqnarray}
  a^{\varepsilon, \,\infty}_l ( x ; m^2, g) & = & \rho^2 \,
   a^{\,\rho^2 \varepsilon,\, \infty}_l ( x ; \rho^{-2}\, m^2 , \rho^2 \,g)
          \label{sc7}  \\
   f^{ \,\mu, \,\varepsilon, \,\infty}_l ( x ; m^2, g) & = & \rho^2 \,
   f^{\,\mu,\,\rho^2 \varepsilon,\, \infty}_l ( x ; \rho^{-2}\, m^2,
       \rho^2 \,g) \label{sc8}  \\
    b^{ \,\mu \nu, \, \varepsilon, \,\infty}_l ( x ; m^2, g) & = &
    \rho^2 \,
   b^{\,\mu \nu ,\,\rho^2 \varepsilon,\, \infty}_l
  ( x ; \rho^{-2}\, m^2, \rho^2 \,g)  \label{sc9}  \\
   c^{\varepsilon, \,\infty}_l ( x ; m^2, g) & = & 
   c^{\,\rho^2 \varepsilon,\, \infty}_l ( x ; \rho^{-2}\, m^2,
         \rho^2 \,g) \ . \label{sc10} 
\end{eqnarray}
For the standard case of $\vep$-independent renormalization conditions 
the scaling of $\vep$ can of course be ignored.
At the tree level the relation (\ref{sc5}) holds  as shown above.
Using the FE with the standard inductive scheme it then follows that\\
\centerline{{\it  (\ref{sc5}) holds in the case of renormalization 
conditions satisfying (\ref{sc7})-(\ref{sc10}).}}
Renormalization conditions imposed at some scale $t_R < \infty\,$
are in one to one relation to those imposed at  $t= \infty\,$, and
the local terms $ a^{\,\varepsilon,\, t_R}_l\,$ etc. 
can be viewed either as renormalization conditions imposed at this
scale or as resulting from integrating the FE over $[t_R,\infty)\,$
with  renormalization conditions imposed at $\infty\,$. From this
fact and (\ref{sc5})  one 
deduces that the relations corresponding to (\ref{sc7})-(\ref{sc10}) 
for renormalization conditions imposed at finite $t_R\,$ are
\eq
 a^{\varepsilon, \,t_R}_l ( x ; m^2, g)\,  = \, \rho^2 \,
   a^{\,\rho^2 \varepsilon,\, \rho^2 t_R}_l 
( x ; \rho^{-2}\, m^2 , \rho^2 \,g) \quad \mbox{etc.}
          \label{ss11}  
\eqe

\noindent
In the subsequent analysis of the counter terms it will be helpful
to first analyse the {\it massless} theory for $t$ in the interval
$[\vep,T]\,$ to eliminate one of the parameters subject to scaling.  
While restricting to $[\vep,T]\,$, 
the less singular corrections stemming from the
massiveness (see (\ref{dev}) below) 
can be dealt with afterwards. 
The same can then be done (trivially) 
for the finite contributions coming
from integrating the FE of the massive theory over $[T,\infty)\,$.\\
For the massless theory we introduce the following notation~: we denote
\[
a_l^{\vep,t}(x;g) \to  a_{l,t_R}^{\vep,t}(x;g)\,,\quad \mbox{ etc.}  
\]
to explicitly introduce all parameters subject to scaling,
including the scale of the renormalization point $t_R\,$.
Furthermore we will introduce the sequence of scales 
\[
t_n~:= \ka ^{-n}\, t_R,\ \ \kappa>1\,,\quad  1 \le n \le N\,, \ 
\mbox{ such that } \ \vep =t_N\ .
\]
Then we use the shorthands  
\eq
a_{l,t_R}^{n}(x;g)~:= a_{l,t_R}^{\vep,t_n}(x;g)
  \quad \mbox{ etc.}\ , \label{n1}
\eqe
and for the renormalization constants at $t=t_R$
\eq
a_l^{t_R}(x;g)~:= a_{l,t_R}^{\vep,t_R}(x;g)
  \quad \mbox{ etc.}  \label{n2}
\eqe
As a consequence of the properties of the heat kernel, 
the  terms $ a_{l,t_R}^{n}(x;g)\,$ etc. 
are smooth scalars on the manifold. For the manifolds considered
(of sectional curvature bounded above and below, as defined in Sect.2\,), 
we have proven bounds which are uniform in the curvature
since our bounds on the heat kernel are uniform in this case.
The same holds for their (low order) derivatives 
$(t \De)^s  a_{l,t_R}^{n}(x;g)\,$ 
etc., since we obtain the same bounds for these derivatives due to 
(\ref{D}). We can therefore decompose these terms according 
to their tensorial
   character into individual contributions from curvature,
   respecting the scaling property, such that in this decomposition
   there will only appear terms depending smoothly on the geometric
   quantities. This gives
 \begin{align}
 a_{l,t_R}^{n}(x;g) = & \,\,
 \alpha_{l,t_R}^{n}
   + R (x) \,\xi_{l,t_R}^{n} 
+ \de a_{l,t_R}^{n}( x ; g) \label{sc11} \\  
  f^{ \,\mu,\, n}_{l,t_R} ( x ; g)  = &  \,\, 0 + 
         \de  f^{ \,\mu, \,n}_{l,t_R} ( x ;  g)
     \label{sc12} \\    
  b^{ \,\mu \nu, \,n}_{l,t_R} ( x ; g) = &
     \,\, g^{\mu \nu}(x) \,\beta_{l,t_R}^{n}
 + \de b^{ \,\mu \nu,\,n}_{l,t_R}(x;g)\label{sc13}\\  
  c_{l,t_R}^{n}(x;g) = & \,\,
             \gamma_{l,t_R}^{n}  
+ \de   c_{l,t_R}^{n}(x;g)\ . \label{sc14}      
 \end{align}
The zero written
in (\ref{sc12}) reminds us that this term vanishes
identically in the case of constant curvature. 
The remainder terms in this decomposition may be analysed further
 \begin{align}
\de a_{l,t_R}^{n}( x ; g)  = & \,\,
 t_R\, \Big( \Delta R(x)\, h^{(1,n)}_l + R^{\,2} (x)\, h^{(1',n)}_l
 +  R^{\,\mu \nu}(x) R_{\,\mu \nu}(x)\, h^{(1'',n)}_l \nonumber\\ 
 & \,\, + R^{\,\mu \nu\la \si }(x) R_{\,\mu \nu\la \si}(x)\,
    h^{(1''',n)}_l \Big ) 
  \,\,  +    \cdots  \label{dea} \\  
\de  f^{ \,\mu, \,n}_{l,t_R} ( x ;g) = & \,\,
t_R\, g^{\,\mu \nu}(x) \, R_{  , \, \nu}(x) \,
 h^{(2,n)}_l  + \cdots \label{def} \\  
\de b^{ \,\mu \nu, \, n}_{l,t_R} ( x ;  g) = & \,\,
t_R\left(  R^{\,\mu \nu}(x) \, 
       h^{(3,n)}_l +  g^{\,\mu \nu}(x) R(x)  h^{(3',n)}_l \right)\,+ 
      \cdots  \label{deb}\\   
 \de   c^{n}_{l,t_R} (x ; g)  = &  \,\, 
t_R\, R (x) \,  h^{(4,n)}_l + \cdots \ \, .\label{dec}
 \end{align}
All the $h$-functions in this decomposition have mass dimension zero
and are therefore independent of $t_R$ which is the only scale.
The dots indicate  terms of   higher scaling dimension
in the expansion w.r.t. curvature terms.
We then\\
{\it assume that these expansions are
asymptotic \footnote{asymptoticity is obviously
 required up to second order in $\rho ^2\,$  only.}, 
in the sense that the remainders satisfy}
\eq  
|\,\de a^{n}_{l,t_R} ( x ;\rho ^2 \,g)|\,,\
|\,\om_{\mu}(x) \,\de  f^{ \,\mu, \,n}_{l,t_R} ( x ; \rho ^2 g)|\,,\
|\,\om^{(2)}_{\mu\nu}(x)\, \de b^{ \,\mu \nu, \, n}_{l,t_R} ( x ;\rho ^2
g)| 
\ \le\ O(\rho^{-4})\ ,
\label{desc1}
\eqe
\eq
 |\, \de   c^{n}_{l,t_R} ( x ; \rho ^2 g) | \ \le \  O(\rho ^{-2})\ .
\label{desc2}
\eqe
Here $ n\,$ and $t_R\, $ are (of course) kept fixed and furthermore, 
the rank 1 resp. rank 2 cotensor fields 
$\,\om_{\mu}(x)\,,\ \om^{(2)}_{\mu\nu}(x)\,$ are assumed to stay 
invariant under scaling $g\to \rho ^2 \,g\,$. The bounds are
in agreement with the leading terms written in (\ref{dea})-(\ref{dec}).
This assumption appears plausible and is often taken for granted,
 see e.g. [HoWa3]. Its proof requires a more thorough analysis of the
heat kernel and its convolutions than is given here.

\noindent
{\bf Proposition 4}~:\\
{\it Assuming (\ref{desc1}),(\ref{desc2}), then for
 position independent coupling $\la$ 
there exist renormalization conditions 
of the form (\ref{renbed}, \ref{renbed4})
such that
the bare action takes the simple form (\ref{nawi}), 
this means that for $l \ge 1$
\eq
   L_l^{\vep}(\vp) = \ {1 \over 2}\,\int_x\ \{ (\,\al_l^{\vep}
 +\,\xi_l^{\vep}\  R(x))\,\vp^2(x)\, -\,
   b_l^{\vep}\,\vp (x) \De\vp (x)\, +\,
    {2 \over 4!}\, c_l ^{\vep}\, \vp^4(x)\} 
\label{nawi1}
\eqe 
with the following bounds
\eq
|\,\al^{\vep}_{l}\,| \le \  \frac{1}{ \vep}\ {\cal P}_{l-1}\log
\frac{1}{ \vep}\ , \quad 
|\,\xi_l^{\vep}\,| \le \  {\cal P}_{l}\log
\frac{1}{ \vep}\ , \quad 
|\,b_l^{\vep}\,| \le \  {\cal P}_{l-1}\log
\frac{1}{ \vep}\ , \quad 
|\,c_l^{\vep}\,| \le \  {\cal P}_{l}\log
\frac{1}{ \vep}\ .
\label{cac}
\eqe
}

\noindent
{\it Proof~:}\\ 
We first note that Proposition 1 can be proven in complete analogy
when imposing renormalization conditions of the form 
(\ref{renbed}), (\ref{renbed4})
at scale $t_R = T>0\,$ for the massless theory.
The scale  $T\,$ is the one up to which we have precise control on the
heat kernel, cf. (\ref{hk10}),  
and it is thus related to the geometry of $\cal M\,$. 
Furthermore we can expand for $\vep \le t \le T\,$
\eq
\,{\cal L}^{\vep,t}_{n,l} (m^2;x_1,\vp_{\tau,y_{2,s}})\ =\
\,{\cal L}^{\vep,t}_{n,l} (0;x_1,\vp_{\tau,y_{2,s}})\ +\
m^2\,\pa_{m^2}{\cal L}^{\vep,t}_{n,l} (0;x_1,\vp_{\tau,y_{2,s}})
\label{dev}
\eqe
\[
 +\ 
m^4\,\int_{0}^{1} d\la \ (1-\la)\,\pa_{m^2}^2{\cal L}^{\vep,t}_{n,l} (\la
m^2;x_1,\vp_{\tau,y_{2,s}})\ .
\]
We first analyse the massless theory and then comment on the 
derivative terms.

We use the notation (\ref{n1}), (\ref{n2}). The theory is specified
through renormalization conditions of the form  (\ref{renbed}), 
(\ref{renbed4}) imposed at scale $t_R=T$~:
\eq
a^{T}_l(x;g)=0\, ,\
f^{\mu, T}_l(x;g)=0\, ,\
b^{\mu\nu,T }_l(x;g)=0\, ,\
c ^T_l(x;g)=0\ ,
\label{renscal}
\eqe
together with  boundary conditions of the type
(\ref{bo1})-(\ref{bo3}) at scale $\vep =\ka ^{-N} T$ 
for $l'\le l\,$. Our aim is to analyse the bare action. 
From Proposition 1 we obtain for $\, l > 0\,$ the bounds
\begin{align}
|\, a_{l, T}^{n}(x;g)| \leq & \,\, O(1)\ \ka ^{n}\,\,n^{l-1}  
   \label{pp1} \\  
|\,f^{ \,\mu,\, n}_{l,T} ( x ; g)\,\om_{\mu}(x)|
   \leq &  \,\, O(1)\  |\,\om(x)|\,\, \ka ^{\frac{n}{2}}\,\, n^{l-1}   
          \label{pp2} \\    
|\, b^{ \,\mu \nu, \,n}_{l,T}( x ; g)\, \om^{(2)}_{\mu \nu} (x)| 
  \leq & \,\, O(1) \ |\, \om^{(2)}(x)|\,\,  n^{l-1} \label{pp3}\\  
|\, c_{l,T}^{n}(x;g) | \leq & \,\, O(1) \  n^l \, . \label{pp4}      
 \end{align}

In the sequel we present the detailed argument for
the relevant term $a(x;g)\,$, whereas the analogous treatment
of the other ones is stated in shortened form.
 In view of the decomposition (\ref{sc11}) 
 we want to prove inductively in $n$ \footnote{More precisely induction
is in $(l,n)\,$ in the order $(l,1),(l,2),\ldots,(l,N),(l+1,1),\ldots$,
but the step   $(l,N)\to (l+1,1)$ is trivial.}
\eq
|\al_{l,T}^n| \le O(1) \sum_{n '=1}^{n} \ka ^{n'}\ {n'} ^{\,l-1}\,,\
|\xi_{l,T}^n| \le O(1) \sum_{n '=1}^{n} {n'}^{\,l-1}
\,,\
|\de a_{l,T}^n(x;g)| \le  O(1) \sum_{n '=1}^{n} 
\ka^{-n'}{n'}^{l-1}\ .
\label{ind}
\eqe
First note that the uniqueness of the solutions
of the FE implies that 
 the relevant term $\,a_{l,T}^{n+1}(x,g)\,$ satisfies
\eq
a_{l,T}^{n+1}(x;g)= {\hat a}_{l,\kappa ^{-n} T}^1(x;g)
\label{hut}
\eqe
where  $ {\hat a}_{l,\kappa ^{-n} T}^1(x;g)\,$
is defined to be the corresponding relevant
term at scale $\kappa ^{-(n+1)} T\,$
for the theory renormalized at scale $\kappa ^{-n} T$,
with renormalization conditions of the following form
\eq
{\hat a}_{l}^{\kappa ^{-n} T}(x;g)= a_{l,T}^{n}(x;g) 
\ \ \mbox{ (analogously for the $f\,,\ b \,,\ c$-terms)}\ . 
\label{rhut}
\eqe
This just means that we take renormalization conditions 
at scale $T$, integrate down to $\kappa ^{-n} T\,$,
and take the values we arrive at for the local terms, as
renormalization conditions at the scale  $\kappa ^{-n} T\,$.  
By the uniqueness statement
we obtain the same Schwinger functions
as when imposing $ a_{l}^{T}(x;g)\,$ etc. at scale $T\,$.\\
From the scaling relations, cf. (\ref{ss11}), we have
\eq
a_{l,T}^n(x; g)\ =\ \ka^{n}\  a_{l,\kappa ^{n}T}^n(x;\ka^{n}g)
\ =\  
\ka^{n}\  {\hat a}_{l}^{T}(x;\ka^{n}g)\, , 
\label{shutr}
\eqe
\eq
a_{l,T}^{n+1}(x;g)\ =\ \ka^{n}\  a_{l,\kappa ^{n}T}^{n+1}(x;\ka^{n}g)
\ =\ 
\ka^{n}\  {\hat a}_{l, T}^1(x;\ka^{n}g) \ .
\label{shut}
\eqe
In the case of $ c_{l,T}^{n} (x;g)\ $ such relations hold
without the external factor $\, \ka^n \,$. Moreover,
\begin{align}
 b^{ \,\mu \nu, \,n}_{l,T}( x ; g)\, \om^{\,(2)}_{\mu \nu} (x)\, = &
 \,\,\ka^{n}\ {\hat b}^{\,\mu\nu,T }_{l}(x;\ka^n g )\,
 \om^{\,(2)}_{\mu \nu} (x)\, , \label{shutrb} \\
 b^{ \,\mu \nu, \,n+1}_{l,T}( x ; g)\, \om^{\,(2)}_{\mu \nu} (x)\, = &
 \,\,\ka^n \,\, {\hat b}^{\,\mu\nu,1 }_{l, T}(x;\ka^n g )\,
   \om^{\,(2)}_{\mu \nu} (x)\ ,
  \label{shutb} 
\end{align}
and the analogue for $\,f^\mu \,$ is obtained replacing
$\, b^{\mu \nu}\,$ by $\, f^\mu \,$ and $\, \om^{\,(2)}_{\mu \nu} (x)\,$
by $\, \om_{\mu} (x)$. 
Using (\ref{pp1})-(\ref{pp4}) and (\ref{shutr})-(\ref{shutb}), we then obtain
\eq
|\,{\hat a}^{T}_{l}(x;\kappa ^{n}g) |\ \le\  O(1)\ n^{l-1}\, ,\quad
|\,{\hat f}^{\,\mu, T}_{l}(x;\ka ^{n} g)\, \om_{ \mu}(x) |
\ \le \ O(1)\ |\,  \om(x)|_{\,\kappa ^{n}g} \,\,  n^{l-1}  \,, 
\label{ren0}
\eqe
\eq
 |\,{\hat b}^{\mu\nu,T }_{l}(x;\ka ^{n} g)\, \om^{\,(2)}_{\mu \nu}(x)|
  \ \le\  O(1)\ |\, \om^{\,(2)}(x)|_{\,\kappa ^{n}g} \,\,n^{l-1}\, , 
   \quad |\,{\hat c}^T_l(x;\kappa ^{n}g) |
   \ \le\ O(1)\  n^{l} 
\label{ren1}
\eqe
where we denoted by
$\, | \cdot |_{\,\kappa ^{n}g} \,$ the norm (\ref{nor}) generated
by $\, \kappa ^{n}g\, $. 

We now consider more general massless Schwinger functions 
$\,\hat {\cal L}^{\kappa ^{- 1}T,t}_{p,l}(x_1,\vp_{{\tau},y_{2,s}};\ti g)\,$
resulting from a metric $\ti g\,$ of the class defined in 
Section 2 \footnote{$\ti g =\kappa ^n g\,$ certainly belongs to this 
class if $g$ does}  
and satisfying renormalization conditions
of the form (\ref{ren0}), (\ref{ren1})
at loop orders $l' <l\,$. At loop order $l\,$ we first assume vanishing
renormalization conditions. Afterwards the contribution
coming from renormalization conditions
at loop order $l$,  bounded as in
(\ref{ren0}),(\ref{ren1}) will be added to the result obtained.
Integrating the flow equations for these Schwinger functions
within the interval $[\kappa ^{- 1}T,\,T]\,$,
one verifies with the aid of the usual 
inductive scheme and analogously as in Proposition 1, 
for $t \in [\kappa ^{- 1}T,\,T]\,$, the bounds 
\eq
|\,\hat {\cal L}^{\kappa ^{- 1}T,t}_{p,l} (x_1,\vp_{\tau,y_{2,s}};\ti
g)|
\,\le\ 
O(1)\ n^l\ {\cal F}_{s,l}(t,\tau) \ , \quad p \ge 6  
\label{le1}
\eqe
\eq
|\,\hat {\cal L}^{\kappa ^{- 1}T,t}_{p,l} (x_1,\vp_{\tau,y_{2,s}};\ti
g)|
\,\le\ 
O(1)\ n^{l-1}\ {\cal F}_{s,l}(t,\tau) \ , \quad  p \le 4  
\ .
\label{le2}
\eqe
These bounds are dictated by the size of the boundary conditions
for $l' < l\,$ which enter on the r.h.s. of the FE.
In fact one  realizes that the factors of $n^{l'}\,$ appearing in the
bound on the r.h.s. can be factored out
and majorized by  $n^{l}\,$ resp. $n^{l-1}\,$. 
The remainder is then inductively bounded   (uniformly in $n$)
by  the ${\cal F}_{s,l}(t,\tau) \,$-factors times a $(p,l)$-dependent
constant. 
For the relevant terms these bounds imply
\eq
|\,{\hat a}^{1,0}_{l,T}(x;\ti g)|\ \le\ O(1)\  n^{l-1}\, ,\quad
|\,{\hat f}^{\mu, 1,0}_{l,T}(x;\ti g)\, \om_{ \mu}(x) | 
\ \le\  O(1)\ |\, \om(x)|_{\,\ti g}\  n^{l-1}  \, ,\
\label{ctt0}
\eqe
\[
|\,{\hat b}^{\mu\nu,1,0 }_{l,T}(x;\ti g)\, \om^{\,(2)}_{\mu \nu}(x)| 
\ \le\  O(1)\ |\, \om^{\,(2)}(x)|_{\,\ti g}\  n^{l-1}\, , \quad
|\,{\hat c}^{1,0}_{l,T}(x;\ti g)|\ \le \ O(1)\ n^{l-1} \ .
\]
Here the upper index $0$ indicates that we were calculating
with vanishing  renormalization conditions
at loop order $l\,$.
On decomposing as in (\ref{sc11})
\eq
{\hat a}_{l,T}^{1,0}(x;\ti g)= {\hat \al}_{l,T}^{1,0} 
+ \ti  R (x) \,{\hat \xi}_{l,T}^{1,0}
+ {\de \hat a}_{l,T}^{1,0}(x;\ti g) 
\label{ax}
\eqe
 we then obtain from (\ref{ctt0}) by linear independence
\eq 
|\, {\hat \al}_{l,T}^{1,0}\,|\,,\
|\, {\hat \xi}^{1,0}_{l,T}\,|\,,\
|\, {\de \hat a}_{l,T}^{1,0}(x;\ti g)\,| \ \le O(1)\   n^{l-1}\ .
\eqe
Specializing to $\ti g =\ka ^{n} g\,$ in (\ref{ax}) yields
\eq
{\hat a}_{l,T}^{1,0}(x;\ka ^n g)= {\hat \al}_{l,T}^{1,0} 
+ \ka ^{-n}   R (x) \,{\hat \xi}_{l,T}^{1,0}
+ {\de \hat a}_{l,T}^{1,0}(x;\ka ^n  g) 
\label{axn}
\eqe
where our smoothness assumption  (\ref{desc1}) on 
${\de \hat a}_{l,T}^{1,0}(x;\ka ^n  g)\,$ implies
\eq
 |\, {\de \hat a}_{l,T}^{1,0}(x;\ka ^n  g)| \ \le\ O(1)\ 
\ka ^{-2n}\ n^{l-1}\ .
\label{deas}
\eqe 
Upon scaling according to (\ref{shut}) we then obtain 
\eq
{a}_{l,T}^{n+1,0}(x;g)= {\al}_{l,T}^{n+1,0} 
+  R (x) \,\xi_{l,T}^{n+1,0}
+ \de  a_{l,T}^{n+1,0}(x;g) 
\label{axn1}
\eqe
with the bounds
\eq
| {\al}_{l,T}^{n+1,0}| \le O(1)\  \ka^n \ n^{l-1}\,,\ \
| {\xi}_{l,T}^{n+1,0}| \le O(1) \ n^{l-1}\,,\ \
| {\de a}_{l,T}^{n+1,0}(x;g)| 
   \le O(1) \,\frac{n^{l-1}}{\ka^n}\ .
\label{axnb}
\eqe
Adding the contributions from the renormalization 
condition obeying the inductive bounds from (\ref{ind}),
 then yields
\[
|\,\al_{l,T}^{n+1}\,| \le  O(1) \, \sum_{n '=1}^{n+1} \ka ^{n'}
{n '}^{\,l-1}
 \le  \,  O(1) \  \ka ^{n+1} ({n+1})^{\,l-1}\ ,
\]
\[
|\, { \xi}_{l,T}^{n+1}\,| \le   O(1) \,  
\sum_{n '=1}^{n+1} {n'} ^{\,l-1} 
\le \, O(1) \, (n+1)^{l}\ ,
\]
\[
|\, {\de  a}_{l,T}^{n+1}(x; g)\,| \le  \,  O(1) \, \sum_{n '=1}^{n+1}
 \ka ^{-n'}{n'}^{\,l-1}\le    O(1) \ ,
\]
thus establishing the bounds (\ref{ind}) by induction.
The statement for $n+1 =N$ implies Proposition 4, noting in
particular that the last
inequality allows for eliminating the term $\de a_{l,T}^N(x;g)\,$
by a {\it finite} change of the corresponding renormalization 
condition at scale  $T\,$. 

The other relevant terms are dealt with 
analogously. Regarding
$\, c_l\,$ we obtain in place of (\ref{axn1}), (\ref{axnb}) 
\eq
{c}_{l,T}^{n+1,0}(x;g)= {\gamma}_{l,T}^{n+1,0} 
 + \de  c_{l,T}^{n+1,0}(x;g)\,,
\eqe
\eq
| {\gamma}_{l,T}^{n+1,0}| \le O(1)\ n^{l-1}\,,\ \
| {\de c}_{l,T}^{n+1,0}(x;g)|  \le O(1) \,\ka^{- n}\, n^{l-1}\ .
\eqe
As for $ {b}^{\, \mu \nu}_{l}\,$, decomposing as in (\ref{sc13})
\eq \label{bx}
 {\hat b}^{\, \mu \nu, \,1, 0}_{l, T} ( x ;\ti g) = 
     \,\,\ti g^{\,\mu \nu}(x) \,{\hat \beta}_{l,T}^{\,1, 0}
 + \de {\hat b}^{ \,\mu \nu,\,1, 0}_{l,T}(x;\ti g)
\eqe 
we get from (\ref{ctt0}) 
\eq \label{bxb}
|\,\ti g^{\,\mu \nu}(x)\,\om^{\, (2)}_{\mu \nu}(x)\,
{\hat \beta}_{l,T}^{\,1, 0} |\,, \ \,
|\,\om^{\, (2)}_{\mu \nu}(x)\,
 \de {\hat b}^{ \,\mu \nu,\,1, 0}_{l,T}(x;\ti g)|\
 \leq \ O(1)\ |\,\om^{\, (2)}(x)|_{\ti g}\  n^{l-1}\ .
\eqe
The second bound implies for ${\ti g}\,=\,g$
\[
|\,\om^{\, (2)}_{\mu \nu}(x)\,
 \de {\hat b}^{ \,\mu \nu,\,1, 0}_{l,T}(x;g)|\
 \leq \ O(1)\ |\,\om^{\, (2)}(x)|_{ g}\  n^{l-1}\ ,
\]
and using (\ref{desc1}) then
provides 
\eq \label{bxbsc}
 |\,\om^{\, (2)}_{\mu \nu}(x)\,
 \de {\hat b}^{ \,\mu \nu,\,1, 0}_{l,T}(x;\ka^n g)|
\ \leq \ O(1)\ |\,\om^{\, (2)}(x)|_g\  \ka^{-2n}\  n^{l-1} \ .
\eqe
Upon scaling, (\ref{shutb}), 
and observing $\, \ka^n \,|\,\om^{\, (2)}(x)|_{\, \ti g} =
| \, \om^{\,(2)}(x)|_g \,$, 
we obtain from (\ref{bx})-(\ref{bxbsc})
\eq
|\,g^{\,\mu \nu}(x)\,\om^{\, (2)}_{\mu \nu}(x)\,
  \beta_{l,T}^{\,n+1, 0} | \, \,
\ \leq\ O(1)\ |\,\om^{\, (2)}(x)|_g\ n^{l-1}\,,
\eqe
\eq \label{bdb}
 |\,\om^{\, (2)}_{\mu \nu}(x)\,
 \de  b^{ \,\mu \nu,\,n+1, 0}_{l,T}(x; g)|
\ \leq \ O(1)\ |\,\om^{\, (2)}(x)|_g\ \ka^{-n}\, n^{l-1}\ .
\eqe
Finally, proceeding similarly we find
\eq \label{bdf}
 |\,\om_{\mu}(x)\,
 \de f^{ \,\mu,\,n+1, 0}_{l,T}(x; g)|
\ \leq\  O(1)\ |\,\om(x)|_g\ \ka^{-n}\, n^{l-1}\,.
\eqe
Since (\ref{bdb}), (\ref{bdf}) hold with general $\,\om^{\,(2)} \,$ 
and $\,\om \, $, respectively, the bounds extend to the
 individual tensorial components.

\noindent
The proof of Proposition 4 is finished  through the following
remarks~:\\  
i) To go back to the massive theory we have to add the two derivative terms
from (\ref{dev}). An $m^2$-derivative acting on the propagator 
produces an additional factor of $t$. As a consequence of this 
we get the bounds
\eq
|\pa ^{s}_{m^2}{\cal L}^{\vep,t}_{n,l} (x_1, \vp_s)|\,\le\ 
  t^{\frac{n+2s-4}{2}}\ {\cal P}_l\log  (t,\tau)^{-1} \
 {\cal F}_{s,l }(t,\tau)\ .
\eqe
This implies that for $\,s\ge 1\,$ there is only one relevant term
\[
\int_{x_2} \pa _{m^2}{\cal L}^{\vep,t}_{2,l} (x_1, x_2)
\]
which by the previous statement is logarithmically bounded.
Applying the expansion (\ref{sc11}) to this term, 
all terms produced can be absorbed -respecting the bounds- 
in the terms already present
in the massless theory. So the previous result is maintained.\\[.1cm]
ii)  We restore the massive theory at scale $T\,$ 
by adding the contributions from the last two terms on the r.h.s. 
of (\ref{dev}). According to Proposition 3, 
renormalization conditions at scale $T\,$  
can then be translated into renormalization conditions at scale $t \to
\infty$ for the massive theory. 
\qed

\noindent
{\bf Acknowledgement~:}\\
The authors are indebted to the referee for
careful study of the paper and for demanding clarification of two items.

%%%%%%%%%%%%%%%%%%%%%%%%%%%%%%%%%%%%%%%%%%%%%%%%%%%%%%%%%%%%%%%
\renewcommand{\theequation}{\thesection.\arabic{equation}}
\setcounter{equation}{0}
\begin{appendix}      
\section{Some Notions from Riemannian Geometry}
Here, we briefly recall some basic properties of Riemannian manifolds
pertinent to the main text and thereby introduce the definitions and
conventions used. 
For a detailed exposition we refer to [Wil].
We consider a connected four-dimensional smooth manifold $\cal M$.
A Riemannian metric on ${\cal M}$ is a tensor field $g$ of type (0,2)  
(more technically: a section of $ \otimes^2\, \mathcal{T}^{\,*} {\cal M}$,
where $ \mathcal{T}^{\,*} {\cal M}$ is the cotangent bundle ) which
associates to each point $ p \in {\cal M}$ a positive-definite inner
product on  $ \mathcal{T}_p {\cal M}$, the tangent space to ${\cal M}$ at $p$. 
Given a chart with local coordinates $ x = (x^1, x^2, x^3, x^4 ) \in
\mathbf{R}^4 $, and denoting by 
$ \partial _{\mu} := \partial /\partial x^{\mu}$
and by $ dx^{\mu}, \mu = 1,2,3,4 $, the corresponding coordinate
vector and covector fields, respectively, the Riemannian metric tensor
has the form
\begin{equation} \label{a1}
g \, = \, g_{\mu \nu } (x) \,dx^\mu \otimes dx^\nu \, ,
                   \qquad   g_{\mu \nu } (x) \, = \,   g(\partial_{\mu}, 
                  \partial _{\nu}) .
\end{equation}
At each point $x$ the components $ g_{\mu \nu } (x) $ form the entries
of a symmetric positive-definite matrix. In (\ref{a1}) and henceforth the 
summation convention is implied. Moreover, with
\begin{equation} \label{a2}
  g^{ \lambda \mu} (x) g_{ \mu \nu}(x) := \delta _{\nu}^{\lambda} \, , 
    \qquad |g(x)| \equiv {\rm {det} }\big(  g_{\mu \nu } (x) \big) 
\end{equation} 
the Riemannian volume element reads
\begin{equation} \label{a3}
    dV(x) \, = \, |g(x)|^{\frac{1}{2}}\, dx^1 dx^2 dx^3 dx^4 \, ,
\end{equation}
and the Laplace-Beltrami operator acting on a scalar field is defined by
\begin{equation} \label{a4}
 \Delta \phi (x) \, = \, |g(x)|^{- \frac{1}{2}} \partial _\mu \,
                    g^{\mu \nu}(x)  |g(x)|^{\frac{1}{2}}
 \,\partial _\nu \,\phi(x) \, .
\end{equation}
The Levi-Civita connection $ \nabla $ of the Riemannian metric $g$
leads to the covariant derivative of the coordinate vector fields
\begin{equation} \label{a5}
\nabla_{\partial _\nu}  \partial _\mu \, =
                          \, \Gamma _{\mu \nu}^{\lambda} (x) \, 
\partial _{\lambda}
\end{equation}
with the Christoffel symbols
\begin{equation}\label{a6}
\Gamma _{\mu \nu}^{\lambda}(x) \, = \, \frac{1}{2} g^{\lambda \varrho  }
 \big( \partial _\mu \, g_{\varrho  \nu} + \partial_\nu \, g_{\varrho \mu}
         -\partial_\varrho \, g_{\mu \nu} \big )
 \, = \, \Gamma_{\nu \mu}^{\lambda } \, .
\end{equation}
The Riemannian curvature tensor $ R $ of the connection $ \nabla $ maps
the triple of vector fields $ X,Y,Z $ to the vector field
\footnote{There is obviously a freedom in choosing an overall sign,
    which has to be observed, similarly in the case of the Ricci tensor.}
\begin{equation} \label{a7}
R (X,Y ) Z \, = \, \big( \nabla_X \nabla_Y - \nabla_Y \nabla_X 
                      - \nabla_{[X, Y]} \, \big ) Z \, .
\end{equation}
In  local coordinates with $ X = X^{\mu} (x) \, \partial_{\mu} $ and 
similarly for $ Y,Z $ the curvature tensor has the form  
\begin{equation} \label{a8}
  R(X,Y)Z \, = \, R_{\, \, \,  \sigma \mu \, \nu}^{\varrho } 
\, Z^{\sigma} X^{\mu} 
                             \, Y^{\nu} \, \partial_{\varrho }
\end{equation}
with components
\begin{equation}\label{a9}
 R_{\, \, \,  \sigma \mu \, \nu}^{\varrho } (x) \,
                        =  \,\partial_\mu \, \Gamma_{\sigma \nu}^{\varrho }
                 - \, \partial_\nu \, \Gamma_{\sigma \mu}^{\varrho }
         + \Gamma_{\lambda \mu}^{\varrho }\, \Gamma_{\sigma \nu}^{\lambda} 
       - \Gamma_{\lambda \nu}^{\varrho } \,\Gamma_{\sigma \mu}^{\lambda} \, .
 \end{equation} 
The components of the Ricci tensor follow by internal contraction as
\begin{equation}\label{a10}
   R_{\sigma  \nu} (x) \, := \, R_{\, \, \, \sigma \mu \nu}^{\mu} (x) \, ,
\end{equation}
and the Ricci curvature at the point $p$ with local coordinates $x$ in the
direction of the tangent vector $ v \in \mathcal{T}_p {\cal M}$ is defined by
 \begin{equation}
     Ric_p(v) \, := 
\, \frac{R_{\sigma \nu} (x)\, v^{\sigma} v^{\nu}}
                                {g_{\sigma \nu} (x) \, v^{\sigma} v^{\nu}} \, .
\end{equation}
Moreover, the \emph{scalar curvature} is given by
\begin{equation}
      R(x) \, := \, g^{\sigma \nu}(x) \, R_{\sigma \nu} (x) \, . 
\end{equation} 
Let  $v,w \in \mathcal{T}_p \,\mathcal{M}  $ span the two-dimensional 
subspace $ S $. Then the
\emph{sectional curvature} of $\mathcal{M}$ at the point $p$ along 
the section $S$ 
is defined as
\begin{equation} 
   Sec_p (v,w) := \,  - \, 
\frac{ g_p ( R_p (v, w)v, w)}{g_p (v,v) g_p (w,w) - g_p (v,w) ^{\,2} } \, .
\end{equation}
It depends only on the section  $S$, not on the spanning vectors $ v,w $.
Given in $ \mathcal{T}_p \,\mathcal{M} $ an orthonormal basis 
$\xi_{(r) }, r=1,.. ,4, $ with 
components $\{\xi^{\mu}_{(r)} \}$ implies
\begin{equation}
g^{\mu \nu} (x) \, = \, \sum_{r=1}^{4}\xi^{\mu}_{(r)}\,\xi^{\nu}_{(r)}
\end{equation}
and leads to sectional curvatures, $r \not= s,$ 
\begin{equation}
  Sec_p\, (\xi_{(r) },\xi_{(s) }) = R_{\sigma \alpha \mu \nu}(x) \,
       \xi^{\sigma}_{(r)}\,\xi^{\alpha}_{(s)} 
    \, \xi^{\mu}_{(r)}\,\xi^{\nu}_{(s)} \, .
\end{equation}
 Herefrom it follows that
\begin{eqnarray}
 Ric_p ( \xi_{(s) }) & = & \sum_{r, \, r \not= s}  
Sec_p\, (\xi_{(r) },\xi_{(s) }) \, ,\\
  R(x) & = & 2 \sum_{r<\, s}  Sec_p\, (\xi_{(r) },\xi_{(s) }) \, .
  \end{eqnarray}

The geodesics passing through a point $ p \in {\cal M}$  can in general
only be defined for values of the (affine) parameter confined to a
finite interval. They generate a map from an open domain of the
tangent space into the manifold, called the exponential map,
$ \exp : \Omega \subset \mathcal{T}_p {\cal M}\rightarrow {\cal M}$. 
Its inverse are the Riemannian normal coordinates. A manifold is
geodesically complete, if this parameter interval everywhere
extends to $\mathbf{R} $, and hence $ \Omega = \mathcal{T}_p {\cal M}$,
for all $p \in {\cal M}$. For points $ p,q \in {\cal M}$ 
 the distance function $ d(p,q) = d(q,p) $ is defined by
 $ d(p,q) \, = \, \inf_{\alpha} L(\alpha) $ ,
where $\alpha $ runs over all $ C^1$ curve segments 
 joining $p$ to $q$, i.e. $ \alpha : [a,b\,] \rightarrow {\cal M}, \,
 \alpha (a)=p, \, \alpha (b) = q $, and its arc length given by
 \begin{equation} \label{a12i}
   L(\alpha) \, = \, \int_{a}^{b} dt \,
   \Big (\, g_{\alpha (t)} \big (\dot{\alpha} (t), 
\dot{\alpha} (t) \big )\Big )^{\frac{1}{2}} \, .
\end{equation}
If $ p$ is sufficiently close to $q$ there is always  a unique geodesic
determining $ d(p, q)\,$.
 Regarding a geodesic ball in $ {\cal M}$ with
center $ p $ and with radius $r$ ,
\begin{equation}\label{a13}
\mathcal{B}(p, r) = \{ q \in {\cal M}|\,  d (p,q) < r \} \, ,
\end{equation}
its Riemannian volume is denoted by
\begin{equation}\label{a14}
        | \mathcal{B}(p, r) | \, = \, \int_{\mathcal{B}} dV \, .
\end{equation}
For $\,x,y \in \cal M\,$ we introduce the bi-tensor of type scalar-vector
\eq
\si(x,y)^{\mu}~:=\ \frac12 g^{\mu \nu}(y)\, \frac{\pa}{\pa y^{\nu}}\
d^{\,2}(x,y)
\label{sig}
\eqe
which satisfies
\eq
\si(x,y)^{\mu}\ \si(x,y)^{\nu}\ g_{\mu \nu}(y)\ =\ 
d^{\,2}(x,y)\ .
\label{nsi}
\eqe

In the renormalization proof we  need covariant Taylor expansion 
formulae in the Schl\"omilch form, i.e.  with integrated
remainders, which are obtained as follows: \footnote{We give the
  complete argument since we only found part of it
in the literature [BaVi].} 
Given a complete Riemannian 
manifold $ \big( \mathcal{M}, g \big) $, and 
a chart $ \big( \mathcal{U}, x \big) $ with local coordinates $x$,
a geodesic $x(s)$ parametrized by its arc length $s$ satisfies
\begin{eqnarray}
\label{geod}
{\ddot x}^{\lambda} (s) + \Gamma ^{\lambda}_{\mu \nu} (x(s))
         \, {\dot x}^\mu (s) \,{\dot x}^\nu (s) & = & 0 \,, \\
    g_{\mu \nu} (x(s)) \, {\dot x}^\mu (s) \,{\dot x}^\nu (s) & = & 1\, .
\label{geod2}
\end{eqnarray}   
Let $ f \in C^{\infty} (\mathcal{M}) $ and $ F(s) := f (x(s)) $, then 
\begin{equation}
\Big (\frac{d}{ds}\Big )^n \, F(s) = 
     \big( \nabla_{\nu_n} \cdots  \nabla_{\nu_1} f \big) (x(s)) \,
          {\dot x}^{\nu_1} (s) \cdots {\dot x}^{\nu_n} (s) \, .
\label{F}
\end{equation} 
The proof is by induction, using (\ref{geod}).

We consider the geodesic segment with initial point $x_0 = x(0) $ and
end point $ x = x(s) $, hence $ d(x, x_0) = s\, $.
With (\ref{sig}) we then have the relation, 
see e.g. [Wil, sect.\,6.3 ],
\begin{equation}  
\sigma (x, x_0) ^\nu \, = \,  - s \,{ \dot x}^{\nu} (0) \, .
\label{sip}
\end{equation}
From the Taylor formula with remainder 
\begin{equation}
F(s)  =  F(0) + \sum_{l=1}^{n} \frac{s^l}{l!}\, F^{(l)} (0) + R_n 
\, , \,\quad
 R_n  =  \int_{0}^{s} dr\, \frac{ (s-r)^n}{n!} \, F^{(n+1)} (r) \,,
\end{equation} 
we obtain, using (\ref{F}), (\ref{sip}),
\begin{eqnarray}
\label{sloe}
f(x) = f(x_0) &+& \sum_{l=1}^{n} \frac{(-1)^l}{l!} \,\,
          \sigma(x,x_0)^{\nu_l} \cdots \sigma(x,x_0)^{\nu_1}
      \big( \nabla_{\nu_l} \cdots  \nabla_{\nu_1} f \big) (x_0) \, 
\nonumber \\
        & + & R_n \, ,
\end{eqnarray}    
\begin{equation}
  R_n(x,x_0)  =  \int_{0}^{\,d(x,x_0)} dr\, \frac{ (d(x,x_0)-r)^n}{n!} \,\,
       {\dot x}^{\nu_{n+1}} (r) \cdots {\dot x}^{\nu_1} (r) 
        \big( \nabla_{\nu_{n+1}} \cdots  \nabla_{\nu_1} f \big) (x(r)) \, . 
\label{Rn}
\end{equation}
Between fixed $ x, x_0 $ we can reparametrize the geodesic
segment \,$ x(r) = X(\rho)$ , with\\
 $ r = d(x_0, x)\rho\, ,\,0 \leq \rho \leq 1$, implying \,
$ g_{\mu \nu}(X(\rho))\,{ \dot X}^{\mu}(\rho){\dot X}^{\nu}(\rho)
 = d^{\,2}(x_0, x)$. Then
\begin{equation} \label{Rna}
 R_n(x,x_0)  =  \int_{0}^{1} d\rho\,
         \frac{(1 - \rho )^{\,n}}{n! } \,\,
 {\dot X}^{\nu_{n+1}} (\rho) \cdots {\dot X}^{\nu_1} (\rho) 
  \big( \nabla_{\nu_{n+1}} \cdots  \nabla_{\nu_1} f \big) (X(\rho)) \, .
\end{equation} 
 In the remainder $R_n $ the contraction of a tensor of type $(n+1,0)$ 
with a tensor
of type $(0,n+1)$ can be viewed via the (inverse) Riemannian metric as 
the scalar product of two tensors of type $(0,n+1) $.
 To bound $|R_n(x, x_0)|$ , Cauchy's inequality is used
 observing (\ref{geod2}),
\begin{equation}\label{Rnb}
      | R_n(x,x_0) | \leq 
\end{equation} 
$$ \int_{0}^{\, d(x,x_0)} dr\, 
\frac{ (d(x,x_0)-r)^n}{n!}\,\,|( \nabla^{n+1} f) (x(r))|\,
 = d^{\, n+1}(x_0, x) \int_{0}^{1} d\rho\,
         \frac{(1 - \rho )^{\,n}}{n! } \,\,
      |( \nabla^{n+1} f) (X(\rho))|\,,$$
where the norm square is given by
\begin{eqnarray}
|( \nabla^{n+1} f)(x)|^{\,2} =  
\qquad \qquad \qquad \qquad\qquad \nonumber \\ 
  \big( \nabla_{\mu_{n+1}} \cdots  \nabla_{\mu_1} f \big) (x) \, 
  g^{\mu_{n+1} \nu_{n+1}}(x) \cdots g^{\mu_1 \nu_1}(x)
  \big( \nabla_{\nu_{n+1}} \cdots  \nabla_{\nu_1} f \big) (x) \, . 
\label{nor}  
\end{eqnarray}  
Majorising in (\ref{Rnb})
 the norm on the geodesic segment $ \gamma $ between 
$x_0$ and $x$ yields the bound
\begin{equation}
   |\,  R_n (x,x_0) | \leq \frac{d^{\,n+1}(x, x_0)}{(n+1)!}\,\, 
\sup_{y \in \gamma} 
                  |( \nabla^{n+1} f) (y)|\, .
\label{majo}
\end{equation}            
\end{appendix}

 \section*{References}

\begin{itemize}
\item[[BaVi]] A.O. Barvinsky, G.A. Vilkovisky, The generalized
  Schwinger-De Witt technique in gauge theories and quantum gravity,
Phys.Rep.{\bf 119} (1985) 1-74.
\item[[BEM]] J. Bros, H. Epstein and U. Moschella, Towards a 
 General Theory of Quantized
  Fields on the Anti-de Sitter Space-Time,\\ Commun. Math. Phys.
 \textbf{231} (2002) 481 - 528.
\item[[BFV]] R. Brunetti, K. Fredenhagen and R. Verch,
  The Generally Covariant Locality Principle - A New
  Paradigm for Local Quantum Field Theory, \\
      Commun. Math. Phys. {\bf 237} (2003) 31 - 68.
\item[[BiDa]] N.D. Birrell and P.C.W. Davies, \emph{Quantum
   Fields in Curved Space},\\ Cambridge University Press, 1982.
\item[[BiFr]] L. Birke and J. Fr\"ohlich,\, KMS, etc ,
     Rev. Math. Phys. \textbf{14} (2002) 829 - 873. 
\item[[Bir]] N.D. Birrell, Momentum space renormalization
  of $\la\, \phi^4$ in curved space-time,\\ J. Phys. {\bf A13}
  (1980) 569-584. 
\item[[BPP]] T.S. Bunch, P. Panangaden and L. Parker, On
  renormalization of $\la\, \phi^4$ field theory 
   in curved space-time: I,  J. Phys. {\bf A13} (1980) 901-918.  
\item[[BrFr]] R. Brunetti and K. Fredenhagen, Microlocal
  Analysis and Interacting Quantum Field Theories: Renormalization
   on Physical Backgrounds,\\
      Commun. Math. Phys. {\bf 208} (2000) 623-661.
\item[[Bun1]] T.S. Bunch, Local Momentum Space and Two-loop
   Renormalization  of $\la\, \phi^4$ Field Theory in
   Curved Space-Time, Gen. Rel. Grav. {\bf 13} (1981) 711-723.
\item[[Bun2]] T.S. Bunch, BPHZ Renormalization  of $\la\, \phi^4$ 
    Field Theory in Curved Space-Time,
          Ann. Phys. (N.Y.) {\bf 131} (1981) 118-148. 
\item[[BuPn]] T.S. Bunch and P. Panangaden,
  On renormalization of $\la\, \phi^4$ field theory 
   in curved space-time: II,  J. Phys. {\bf A13} (1980) 919-932. 
\item[[BuPr]] T.S. Bunch and L. Parker, Feynman propagator
   in curved space-time: A momentum-space representation, 
   Phys. Rev. {\bf D20} (1979) 2499-2510. 
\item[[Cha]] I. Chavel,\emph{ Riemannian Geometry: A Modern
    Introduction},\\  Cambridge University Press, 1993.
\item[[CLY]] S. Y. Cheng, P. Li and S. T. Yau,
  On the upper estimate of the
     heat kernel of a complete Riemannian manifold,
  Am. J. of Math. {\bf 103} (1981), 1021-1063.  
\item[[Dav1]] E. B. Davies,\emph{ Heat kernels and spectral theory},
     Cambridge University Press 1989.
\item[[Dav2]] E. B. Davies,  Gaussian upper bounds for the heat kernels
           of some second order operators on Riemannian manifolds,
          J. Funct. Anal {\bf 80} (1988), 16-32.  
\item[[Dav3]] E. B. Davies, 
Pointwise bounds on the space and time derivatives of heat
kernels, J. Operator Theory {\bf 21} (1989), 367-378.  
\item[[Gri]] A. Grigor'yan, Estimates of heat kernels on 
     Riemannian manifolds, in \emph{ Spectral Theory and Geometry},
London Math. Soc. Lecture Notes {\bf 273} (1999), 140-225,
    ed. E.B. Davies and Yu. Safarov, Cambridge Univ.Press.  
\item[[HoWa1]] S. Hollands and R.M. Wald, Local Wick polynomials
  and time ordered products of quantum fields in curved
  spacetime, Commun. Math. Phys. {\bf 223} (2001) 289-326.
\item[[HoWa2]] S. Hollands and R.M. Wald, Existence of
   Local Covariant Time Ordered Products of Quantum Fields in
     Curved Spacetime,\\ Commun. Math. Phys. {\bf 231} (2002) 309-345.
\item[[HoWa3]] S. Hollands and R.M. Wald, 
       On the Renormalization Group in Curved Spacetime,\\ 
      Commun. Math. Phys. {\bf 237} (2003) 123-160.
\item[[KKS]] G. Keller, Ch. Kopper and M. Salmhofer,
Perturbative renormalization and effective Lagrangians in $\Phi_4^4$,
Helv. Phys. Acta {\bf 156}  (1992), 32-52. 
\item[[Kop1]] Ch. Kopper,\emph{ Renormierungstheorie mit 
    Flussgleichungen},\\ Shaker Verlag Aachen 1998.
\item[[Kop2]]  Ch. Kopper, Renormalization Theory based on Flow equations,\\
lecture in honour of Jacques Bros, hep-th 0508143, to appear in \\
Progress in Mathematics, Birkh\"auser 2006.
\item[[LiYa]] P. Li and S. T. Yau, On the parabolic kernel of the
     Schr\"odinger operator,\\ Acta Math. {\bf 156}  (1986), 153-201. 
\item[[L\"u]] M. L\"uscher, Dimensional Regularization in the
   Presence of Large Background Fields,
       Ann. Phys. (N.Y.) {\bf 142} (1982) 359-392. 
\item[[M\"u]] V.F. M\"uller, Perturbative Renormalization by Flow
  Equations,\\ Rev. Math. Phys.  {\bf 15} (2003) 491-557.
\item[[NePa]] B.L. Nelson and P. Panangaden, Scaling behavior
     of interacting quantum fields in curved spacetime,
   Phys. Rev. {\bf D25} (1982) 1019-1027.   
\item[[Pol]] J. Polchinski, Renormalization and Effective Lagrangians,
   Nucl. Phys. {\bf B231} (1984) 269-295. 
\item[[Sal]] M. Salmhofer, \emph{Renormalization - An Introduction},
       Springer-Verlag 1998.
\item[[SoZh]] P. Souplet and  Q. Zhang, Sharp gradient estimate and
    Yau's Liouville theorem for the heat equation on
     noncompact manifolds, arXiv:math.DG/0502079, to appear in
Bull. London Math. Soc. 2006 
\item[[Tay]] M.E. Taylor, \emph{Partial Differential Equations I},
     AMS 115, Springer-Verlag, 1996
\item[[Var]] N.Th. Varopoulos, Small time Gaussian estimates of
     heat diffusion kernel. I. The semigroup technique, Bull.
      Sc. math., 2 s\'erie, 113 (1989) 253-277. 
\item[[Wil]] T.J. Willmore, \emph{Riemannian Geometry},
                         Oxford University Press, 1996

\end{itemize}

\end{document}